\documentclass[twocolumn,english,footinbib,bibnotes,superscriptaddress]{revtex4-1}

\usepackage[usenames,dvipsnames]{color}
\usepackage{graphicx}
\usepackage{dcolumn}
\usepackage{amsmath}
\usepackage{amssymb}
\usepackage{times}
\usepackage{hyperref}
\hypersetup{colorlinks=false}

\usepackage{multirow}

\newcommand{\mr}[1]{\multirow{2}{*}{#1}}

\newcommand{\bracket}[1]{\left( #1 \right)}

\graphicspath{{figures/}}

\begin{document}

\title{Topological semimetals and insulators in three-dimensional honeycomb materials}

\author{Dennis Wawrzik}
\affiliation{Institute for Theoretical Physics, University of Cologne, 50937 Cologne, Germany}
\author{David Lindner}
\affiliation{Institute for Theoretical Physics, University of Cologne, 50937 Cologne, Germany}
\author{Maria Hermanns}
\affiliation{Institute for Theoretical Physics, University of Cologne, 50937 Cologne, Germany}
\affiliation{Department of Physics, University of Gothenburg, SE 412 96 Gothenburg, Sweden}
\author{Simon Trebst}
\affiliation{Institute for Theoretical Physics, University of Cologne, 50937 Cologne, Germany}

\date{\today}

\begin{abstract}
Semimetals, in which conduction and valence bands touch  
but do not form Fermi surfaces, have attracted considerable interest for their anomalous properties
starting with the discovery of Dirac matter in graphene and other two-dimensional honeycomb materials. 
Here we introduce a family of three-dimensional honeycomb systems whose electronic band structures exhibit a variety of topological semimetals with Dirac nodal lines. We show that these nodal lines appear in varying numbers and mutual geometries, depending on the underlying lattice structure. 
They are stabilized, in most cases, by a combination of time-reversal and inversion  symmetries  
and are accompanied by topologically protected ``drumhead" surface states. In the bulk, these nodal line systems exhibit Landau level quantization and flat bands upon applying a magnetic field. In the presence of spin-orbit coupling, these topological semimetals  are found to generically form (strong) topological insulators. 
This comprehensive classification of the electronic band structures of three-dimensional honeycomb systems might serve as guidance for future material synthesis.
\end{abstract}

\maketitle


\section{Introduction}

In Fermi liquid theory, conventional metals are described by the physics in the vicinity of their Fermi surface.
A defining characteristic of a Fermi surface is its codimension of one, which enters, for instance, directly in thermodynamic signatures such as the linear temperature scaling of the specific heat of a metal. 
Since the early days of solid-state band theory, it has also been known that two bands can accidentally touch each other along manifolds of higher codimension (and thus smaller dimensionality) such as lines or even 
singular points in momentum space \cite{Herring1937}. 
What has been appreciated only much more recently \cite{Volovik2003,Volovik2007} is that in the presence of additional discrete symmetries such as time-reversal, inversion or certain lattice symmetries, these band touchings can be pinned to the Fermi energy, resulting in semimetals whose nodal structures have codimension two or higher. This insight has paved the way to the recent theoretical prediction and subsequent experimental discoveries of Weyl semimetals \cite{Wan2011,Armitage2018} and nodal-line semimetals \cite{Burkov2011,Phillips2014,Chiu2014,Bzdusek2016}. Such semimetals exhibit many interesting anomalous properties \cite{Rhim2015,Bzdusek2016} that trace back directly to the non-trivial momentum-space topology of their electronic band structures. As such they are often considered to be gapless cousins of time-reversal invariant topological insulators \cite{Kane2005,Bernevig2006,Moore2007,Fu2007,Roy2009,Koenig2007,Hasan2010,Qi2011}.

\begin{figure*}[t]
	\includegraphics[width=\textwidth]{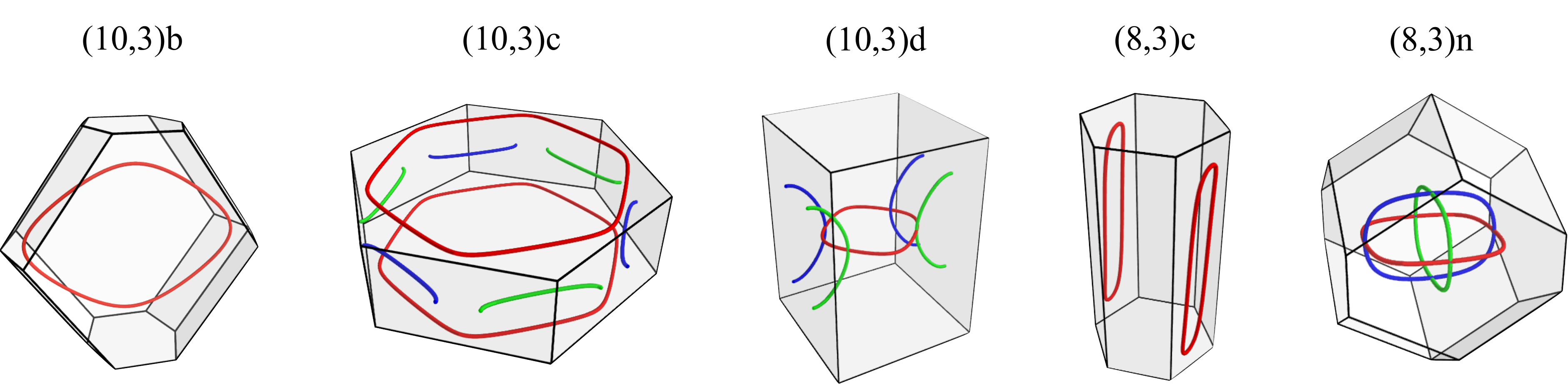}
	\caption{Overview of {\bf nodal line geometries} for a set of three-dimensional generalizations of the
		      honeycomb lattice (specified in further detail in Tab.~\ref{tab:lattice_overview}).
		      Depending on the underlying lattice geometry one or three nodal lines form that are either
		      separated in momentum space [(10,3)c], form nodal chains [(10,3)d] or complex networks
		      with multiple touching points [(8,3)n].}
	\label{Fig:OverviewNodalLines}
\end{figure*}

In this manuscript, we go back to one of the starting points in the exploration of semimetals -- the observation of Dirac points in graphene-like honeycomb materials -- and ask whether there is a systematic way to discover novel, non-trivial band structure phenomena by generalizing the underlying honeycomb lattice geometry to three spatial dimensions.
Considering the most elementary tricoordinated lattices in three dimensions (discussed in more detail below),
we indeed generically find {\em topological} phenomena in their respective band structures.
This includes the formation of Dirac nodal-line semimetals in lieu of a conventional metal (with a Fermi surface) in many of these systems.
Depending on the underlying lattice geometry, we find an odd number of nodal lines that remain separated from one another in momentum space or form nodal chains or more complex networks with multiple crossing points, as illustrated in Fig.~\ref{Fig:OverviewNodalLines}  below. 
The topological nature of these nodal-line semimetals reveals itself in flat-band surface states, also called drumhead states \cite{Burkov2011,Phillips2014,Chiu2014}, 
whose origin can be traced back to a non-trivial winding number associated with each nodal line. 
These nodal-line features are stabilized by discrete symmetries. 
Recent work, aimed at systematically classifying these symmetries, has identified possible combinations of inversion and time-reversal symmetry $(PT)$ \cite{Kim2015,Fang2015,Chan2016}, sublattice/chiral symmetry \cite{Zhao2013}, crystal reflection \cite{Chiu2014}, non-centrosymmetric  \cite{Yamakage2016} and non-symmorphic lattice symmetries \cite{Yang2017} as permissible symmetry sets. Most of our nodal line scenarios are found to comply with the $PT$-protection mechanism, supplemented by sublattice symmetry which pins the nodal line(s) to the Fermi energy.
One exception is found where in the absence of any other symmetries the sublattice symmetry alone suffices to stabilize and pin three nodal lines to the Fermi energy.
These nodal-line systems give rise to additional topological band structure phenomena upon 
applying an external magnetic field or by considering the effect of spin-orbit coupling. 
In the case of an external magnetic field, one finds a Landau level quantization of extended flat bands in the bulk.
In the presence of spin-orbit coupling, the bulk energy spectrum gaps out and the nodal-line semimetals are found to generically transition into (strong) topological insulators.

For some of the three-dimensional honeycomb lattices an alternative scenario plays out with these systems forming conventional metals with a Fermi surface. Upon closer scrutiny, however, some of these systems are found to also exhibit topological bulk and surface features. The non-trivial topology in these metallic systems arises from Weyl or Dirac cones hidden above/below the Fermi energy, which endows each Fermi surface with a non-trivial Weyl charge and results in characteristic Fermi arc surface states.
For the remaining cases of conventional metals, we find that the inclusion of spin-orbit coupling can lead to non-trivial band-structure phenomena, including the formation of a  Dirac semimetal.

\begin{table}[b]
	\centering
	\begin{tabular}{c||c|cc||c}
	\, lattice \, & \,\, tight-binding \,\, & \multicolumn{2}{c||}{\,\, spin-orbit coupling \,\,} & \,\, Kitaev model \,\, \\ 
    	& & &  \\[-2mm]
	\hline \hline
    	& & &  \\[-2mm]
	\,{(10,3)a} \,&  Fermi surface$^*$ & & Fermi surface$^*$  & Fermi surface$^*$  \\
	{(10,3)b} & nodal line & & strong TI 1;~(000) & nodal line \\
	{(10,3)c} & 3 nodal lines & &  strong TI  1;~(001) & 3 nodal lines \\ 
	{(10,3)d} & nodal chain & & strong TI 1;~(000)  & nodal chain \\ 
    	& & &  \\[-2mm]
	\hline
    	& & &  \\[-2mm]
	{(9,3)a} & Fermi surface & &  strong TI 1;~(000) & Weyl nodes \\ 
    	& & &  \\[-2mm]
	\hline
    	& & &  \\[-2mm]
	{(8,3)a} & Fermi surface$^*$  & & Fermi surface$^*$  &  Fermi surface$^*$  \\
	\mr{(8,3)b} & \mr{Fermi surface} & \mr{\Big \{} &  weak TI 0;~(101) & \mr{Weyl nodes} \\
	& & & Dirac nodes  & \\
	\mr{(8,3)c} & \mr{nodal line} & \mr{\Big \{} & strong TI 1;~(010) & \mr{nodal line} \\
	& & & Dirac nodes & \\
	\mr{(8,3)n} & \mr{3 nodal lines} &  \mr{\Big \{} & strong TI 1;~(001)  & \mr{gapped} \\
	& & & Dirac node & \\
	\end{tabular}
	\caption{{\bf Overview of results}. 
			Classification of the nodal structure at the Fermi energy for a number of 3D honeycomb lattices 
			(specified in further detail in Tab.~\ref{tab:lattice_overview}).
			Results for the electronic band structure captured by a tight-binding model are given in the second column. 
			The effect of spin-orbit coupling on the band structure is given in the third column.
			The last column provides a comparison to the physics of {\em Majorana} fermions 
			on the same lattices relevant to a family of three-dimensional Kitaev models \cite{OBrien2016}.
			The asterisk indicates topological metals where a Weyl/Dirac node is encapsulated by the Fermi surface.}
	\label{Tab:Overview}
\end{table}

Taken all together, our manuscript presents an exhaustive classification of the electronic band structure phenomena in three-dimensional honeycomb structures. A summary of our results is given in Tab.~\ref{Tab:Overview}, which indicates for each 3D generalized honeycomb lattice the nodal structure at the Fermi energy both for an elementary tight-binding model and in the presence of spin-orbit coupling. This classification provides context to and generalizations of some earlier results obtained for the 
hyperhoneycomb lattice \cite{Mullen2015,Ezawa2016} and the 
hyperoctagon lattice \cite{Tsuchiizu2016}. 
It also complements a recent classification of Majorana metals in 3D Kitaev spin liquids \cite{OBrien2016,Yamada2017b} pursued for the same set of 3D generalizations of the honeycomb lattice.
We will comment on analogies and differences of the nodal line physics between the electronic systems considered here and the Majorana metals arising in 3D Kitaev magnets in the discussion at the end of the manuscript.


The remainder of the manuscript is structured as follows.
 We start with an introduction of the 3D honeycomb lattices in Section \ref{sec:3DLattices}.
 Section \ref{sec:NodalLineSemiMetals} is devoted to the physics of topological nodal-line semimetals,
 which includes a detailed discussion of the underlying symmetry protection, topological surface states, as
 well as the formation of topological bulk states upon applying an external magnet field or via the inclusion of
 spin-orbit coupling.
 Section \ref{sec:TopologicalMetals} discusses the alternative scenario of a conventional metal forming
 in some of the 3D honeycomb lattices, putting an emphasis on topological bulk and surface 
 phenomena present also in these systems.
 We round off our discussion in Section \ref{sec:Discussion} that also touches on the relevance of our study 
 for future material synthesis. 
 Supplementary information  is provided in the Appendix.


\section{Three-dimensional honeycomb lattices}
\label{sec:3DLattices}

\begin{figure}[t]
	\includegraphics[width=\columnwidth]{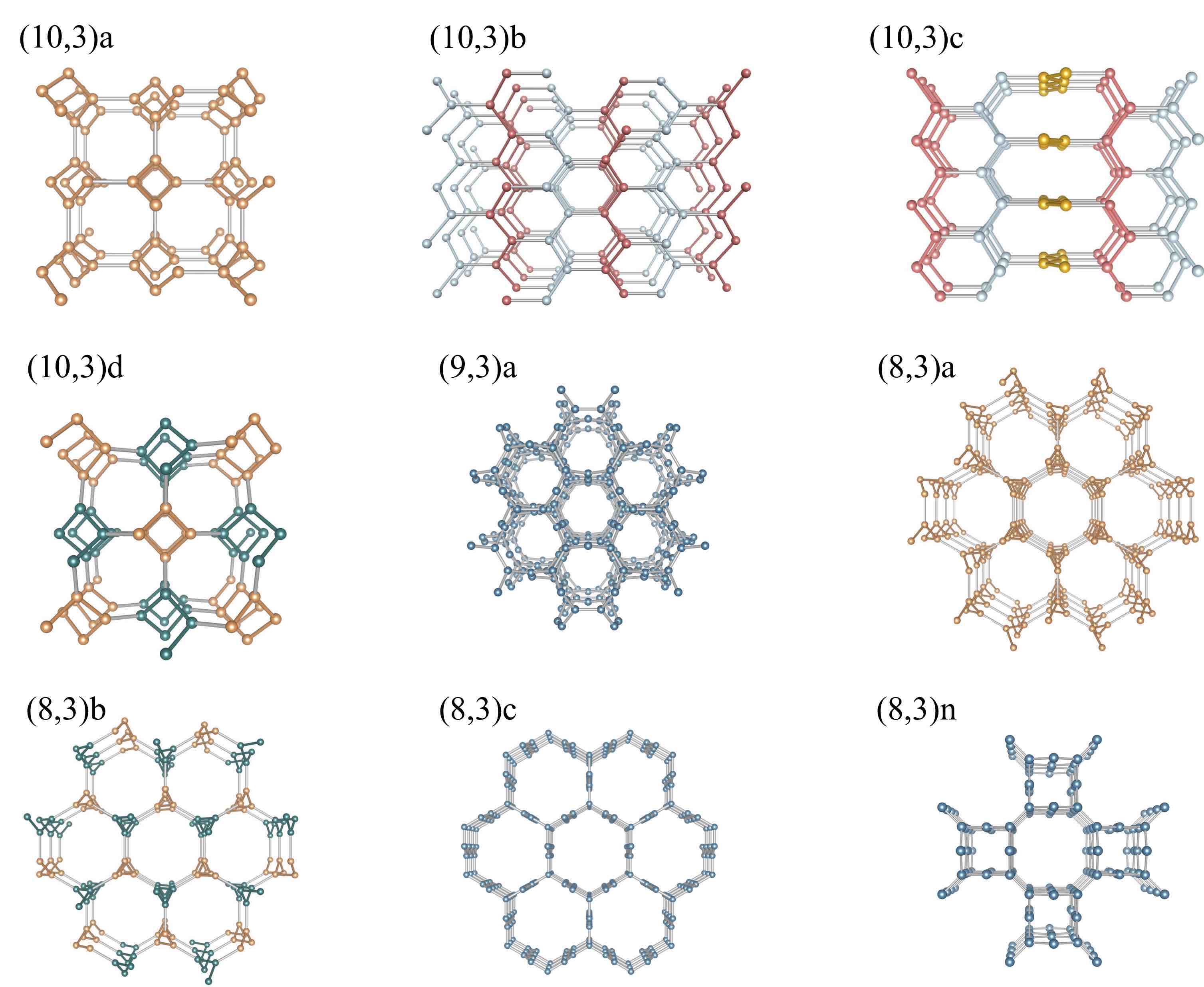}
	\caption{Illustration of the {\bf three-dimensional honeycomb lattices} of Table \ref{tab:lattice_overview}.
			Each lattice, labeled by its Schl\"afli symbol, is shown along one of its high-symmetry directions. To highlight important features of the lattice structures, we colored certain bonds/sites: counter-clockwise rotating spirals are marked in orange, while clockwise-rotating ones are blue. For (10,3)b and c, we visualized the different directions of the zig-zag bonds using different colors.  }
	\label{Fig:lattice_illustration}
\end{figure}

The principal input of our study are three-dimensional generalizations of the honeycomb lattice, i.e.\ lattice structures that, 
like the familiar honeycomb lattice, exhibit only {\em tricoordinated} sites. While such low-coordinated lattice structures are relatively rare in conventional solids, they are common-place in three-dimensional graphene networks \cite{Weng2015}, synthesized as the magnetic sublattice in polymorphs of the iridate Li$_2$IrO$_3$ \cite{Takayama2015,Modic2014} and potentially realized in certain metal-organic compounds \cite{Yamada2017a}.
Here we take a more abstract point of view and consider a set of {\em prototypical} tricoordinated lattice structures that  
contain only elementary loops of identical length. Each one of these lattices is a representative of an entire family of higher harmonics \cite{Modic2014} of tricoordinated lattices (where some loops are expanded at the expense of shortening others in a systematic way).
The variety of these prototypical lattice structures (with equal loop length) has been extensively classified  in pioneering work of A.F.~Wells in the 1970s \cite{Wells1977}. We provide a summary of these lattices in Table \ref{tab:lattice_overview} along with an illustration in Fig.~\ref{Fig:lattice_illustration}. Each lattice is indexed by the Schl\"afli symbol $(p,c)$, which indicates the length of the elementary loop $p$ (polygonality) and coordination $c=3$, followed by a letter. In this notation, the honeycomb lattice would be indexed as (6,3) (not to be followed by a letter as it is the only tricoordinated lattice with loops of length 6). Other familiar lattices include the hyperhoneycomb \cite{Takayama2015} lattice carrying index (10,3)b and the hyper\-octagon \cite{Hermanns2014} lattice (which is also referred to as the Laves graph \cite{Laves1993} or $K_4$ crystal \cite{K4} in the literature) that carries index (10,3)a. Beyond their elementary loop length, the prototypical lattices of Table \ref{tab:lattice_overview} distinguish themselves by the number of sites in the unit cell (varying between 4 and 16) and their fundamental lattice symmetries as indicated in the table. Detailed information on all lattices, including their unit cells, lattice and reciprocal lattice vectors is provided in  Appendix \ref{App:Lattices}.

\begin{table}[t]
  \begin{tabular}{l||c|cc|cc}
    \mr{\, lattice \,} 	& sites in 	& inversion  	& non & \multicolumn{2}{c}{space group}  \\
    	     		& unit cell & symmetry  	& symmorphic &  symbol  & No. \\
    & & & & &  \\[-2mm]
    \hline \hline
    & & & & &  \\[-2mm]
    (10,3)a 	&  4 		& chiral 		& \checkmark & I$4_132$ 		& 214 \\
    (10,3)b 	&  4		& \checkmark 		& \checkmark & Fddd 			& 70 \\
    (10,3)c 	&  6		& chiral		& \checkmark & P$3_112$		& 151 \\
    (10,3)d 	&  8		& \checkmark		& \checkmark & Pnna 		& 52 \\
    & & & & &  \\[-2mm]
    \hline
    & & & & &  \\[-2mm]
    (9,3)a 	& 12	& \checkmark		& --- & R$\bar{3}$m		& 166 \\
    & & & & &  \\[-2mm]
    \hline
    & & & & &  \\[-2mm]
    (8,3)a 	& 6	& chiral		& \checkmark & P$6_222$		& 180 \\
    (8,3)b 	& 6	& \checkmark		& --- & R$\bar{3}$m		& 166 \\
    (8,3)c 	& 8	& \checkmark		& \checkmark & P$6_3$ / mmc 	& 194 \\
    (8,3)n 	& 16& \checkmark & --- & I4 / mmm		& 139 \\
  \end{tabular}
  \caption{Overview of  {\bf three-dimensional honeycomb lattices} with tricoordinated sites.  
  		The first column gives the Schl\"afli symbol $(p,c)$ indicating the length of the elementary loop $p$ (polygonality)
		and coordination $c=3$ followed by a letter.
		For each lattice, we provide the number of sites in the unit cell (second column) 
		along with its basic symmetry properties (third and fourth column) 
		and its space group information (last two columns).
		Further details on the lattice vectors in real and momentum space are provided in Appendix \ref{App:Lattices}.
		}
\label{tab:lattice_overview}
\end{table}


\section{Nodal-line semimetals}
\label{sec:NodalLineSemiMetals}

Probably the most striking topological band structure phenomenon in the three-dimensional honeycomb systems that we consider is the occurrence of Dirac nodal-line semimetals for various lattice geometries. 
As illustrated in Fig.~\ref{Fig:OverviewNodalLines} 
we find multiple scenarios with varying numbers of nodal lines and mutual geometries. In the following we will discuss various aspects of these nodal-line band structures. We will first concentrate on the elementary symmetry mechanism that leads to the formation of these nodal lines in the (nearest-neighbor) tight-binding Hamiltonian and establish that for all but one case it is a combination of time-reversal and inversion symmetry that is at play. The one exception is lattice (10,3)c for which a simple sublattice symmetry suffices. In fact, sublattice symmetry is a crucial symmetry for all lattices as it ensures particle-hole symmetry which in turn pins the nodal line(s) to the Fermi energy.
However, if the system is not perfectly particle-hole symmetric (e.g. upon breaking the sublattice symmetry or by moving away from half filling) the nodal line structure will (at least partially) move away from the Fermi energy. We briefly discuss this instability, which every electronic system clearly exhibits as a sublattice-symmetry breaking next-nearest neighbor hopping is not symmetry-forbidden. We then move to a characterization of the so-called drumhead surface states which accompany the bulk nodal lines and reflect their non-trivial topological character. 

Finally, we turn to a discussion of topological bulk phenomena. First, upon applying an external magnetic field one can stabilize flat bands in the bulk that show a characteristic $\sqrt{n}$ Landau level quantization. Last but not least, we will turn to the effect of spin-orbit coupling on the nodal-line semimetals and show that it typically induces a bulk gap and leads to the formation of topological insulators.

%


\subsection{Symmetry protection}\label{sec:sym}

In order to discuss the symmetry mechanism that leads to the formation of nodal lines in our family of three-dimensional honeycomb systems,
the relevant symmetries are time-reversal symmetry $T$, inversion symmetry $P$, sublattice symmetry $S$, and combinations thereof. Let us first describe the effect of each symmetry individually. 
\paragraph*{Time-reversal symmetry: }
Time-reversal symmetry (TRS) is an anti-unitary symmetry, which for the systems at hand squares to $-1$. 
It relates the energy of an eigenstate with spin-z component $\sigma$ and momentum $\vec k$ to that of a state with spin $-\sigma$ and momentum $-\vec k$. 

\paragraph*{Inversion symmetry:}
All but three of the lattices we consider here are inversion symmetric (see Table \ref{tab:lattice_overview}). 
Inversion symmetry is a unitary symmetry, which (like time-reversal symmetry) relates energies at $\vec k$ and $-\vec k$ to each other. 
Consequently, the combination of inversion and time-reversal --- in the following denoted by $PT$ --- is an anti-unitary symmetry that leaves the momentum invariant. 
Thus, it is a symmetry that is obeyed in the full Brillouin zone. 
As pointed out in Ref.~\onlinecite{Chan2016}, $PT$ can be used to define a $\mathbb{Z}_2$ topological invariant that can protect nodal lines. 

\paragraph*{Sublattice symmetry:}
With the exception of (9,3)a, all the lattices at hand are bipartite. 
Therefore, as long as we only consider nearest-neighbor hopping, the resulting Hamiltonian is symmetric under sublattice symmetry. 
The latter can be used to define a $\mathbb{Z}_2$ topological invariant that in itself can protect nodal lines \cite{Zhao2013}. 
Depending on the lattice details, sublattice symmetry may be implemented in two distinct ways in momentum space. 
For the lattices (10,3)a, (8,3)a and (8,3)b, the sublattice has {\em different} translation vectors than the original lattice. 
To implement this symmetry, one needs to artificially enlarge the unit cell of the lattice, which results in a nontrivial translation in momentum space by half a reciprocal lattice vector, denoted by $\vec k_0$. As a consequence, sublattice symmetry imposes that for each eigenstate with energy $E$ at momentum $\vec k$, there has to be a corresponding one at momentum $-\vec k+\vec k_0$ with energy $-E$. 
In all the other lattices, both the lattice and the sublattice share the same translation vectors, and one obtains the usual relation, namely that for each energy $E$ at momentum $\vec k$ there has to be an energy $-E$ at momentum $-\vec k$. 
This is the case for lattices (10,3)b, (10,3)c, (10,3)d, (8,3)c, and (8,3)n -- which are precisely the five lattices that exhibit nodal lines as discussed in the following.

\begin{table}[b]
	\centering
	\begin{tabular}{c||c| c|c}
	lattice & \, \# NL \,& NL geometry & \,\, symmetry \,\, \\[-1mm]
    	& &  \\[-2mm]
	\hline \hline
    	& &   \\[-2mm]
	\,(10,3)b\, &\, 1 \, & --- & $PT$ \\
	(10,3)c & 1, 3 & separate lines & $S$ \\ 
	(10,3)d & 3 & chain geometry & $PT$ \\ 
    	& &  \\[-2mm]
	\hline
     	& &  \\[-2mm]
	(8,3)c & 1 & --- & $PT$  \\
	(8,3)n & 3, 5 & \, pairwise crossings \, & $PT$   \\
	\end{tabular}
	\caption{Overview of {\bf nodal-line semimetals}.
			The first column specifies the lattice, the second column provides the number of nodal lines (NLs).
			The third column specifies the mutual nodal line geometry, see also Fig.~\ref{Fig:OverviewNodalLines}.
			The last column lists the combination of symmetries that stabilize the nodal line(s),
			including time-reversal $T$, inversion $P$, and sublattice $S$  symmetries.
			\label{tab:symmetries}}
\end{table}

Upon close inspection of these symmetries for all lattices at hand, we find that in four cases -- lattices (10,3)b, (10,3)d, (8,3)c, and (8,3)n -- the nodal lines are protected by $PT$ symmetry. In one case, lattice (10,3)c, it is only sublattice symmetry $S$ that protects the  nodal line(s). This is summarized in Table \ref{tab:symmetries} below.

In addition, for some lattices further symmetries are at play. For lattice (10,3)d there is an additional glide symmetry, which pins one of the nodal lines (marked in red in Fig.~\ref{Fig:OverviewNodalLines}) to the $k_z=0$ plane.  For lattice (8,3)n, which exhibits three crossing nodal lines (see Fig.~\ref{Fig:OverviewNodalLines}), a mirror symmetry protects four of the six crossing points, while a four-fold rotation symmetry around the $z$-axis protects the remaining two crossing points.


\subsection{Longer-range exchange and particle-hole symmetry}
As discussed above, sublattice symmetry is a crucial symmetry for the nodal-line semimetals occurring in our three-dimensional honeycomb systems. If a system has both sublattice  and time-reversal symmetry, their product, denoted by $PS$, keeps the momentum invariant 
\footnote{Note that for the lattices (10,3)a, (8,3)a, and (8,3)b sublattice symmetry comes with a non-zero momentum $\vec{k}_0$ and therefore the product $PS$ of time-reversal and sublattice symmetry is {\em not} momentum invariant. It is for this reason that these lattices do not exhibit nodal lines.}
and defines an effective particle-hole symmetry. This in turn ensures that the nodal lines are located precisely at the Fermi energy (at half filling). 

\begin{figure}[b]
	\includegraphics[width=0.8\columnwidth]{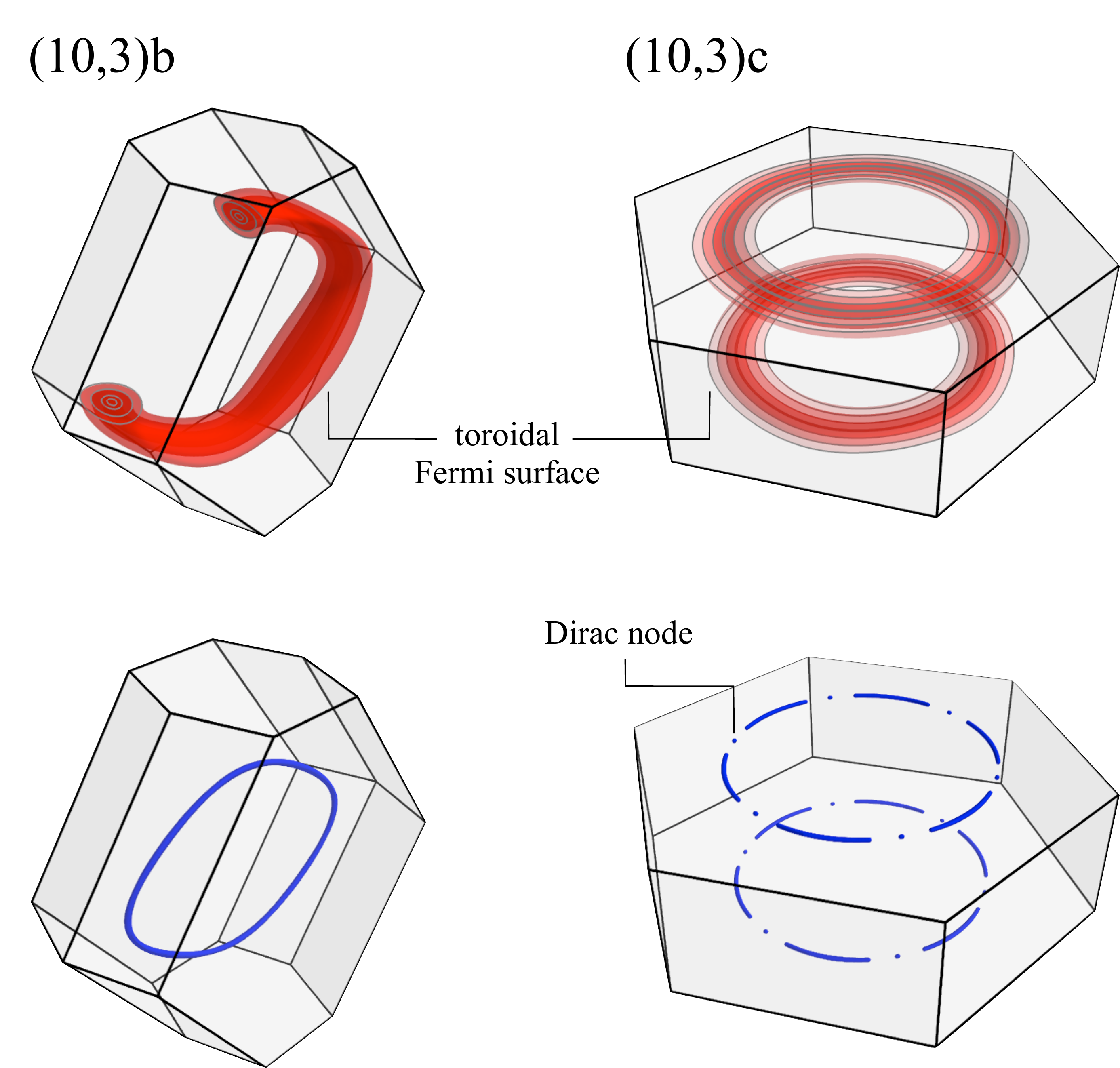}
	\caption{{\bf Toroidal Fermi surfaces} for lattices (10,3)b and (10,3)c 
		      upon inclusion of a (real) next-nearest neighbor hopping (top row).
		      In the bottom row we introduce an additional chemical potential that
		       shifts the Fermi energy such that the effect of the next-nearest neighbor hopping
		      is compensated. While for lattice (10,3)b the nodal line can be recovered this way
		      (see also the discussion in the main text),
		      a more complex situation arises for lattice (10,3)c where we find six isolated Dirac
		      nodes interlaced with small Fermi surface pockets (encompassing another set of six Dirac points).
			}
	\label{Fig:sausages_shifted}
\end{figure}

 However, sublattice symmetry is a rather fragile symmetry for the electronic systems considered here. Longer-range interactions (such as a next-nearest neighbor hopping), which are not symmetry-forbidden  and as such will be present in any realistic system, break sublattice symmetry and thereby immediately destroy particle-hole symmetry as well.  For those lattice where the nodal lines are protected by $PT$ symmetry, the nodal lines will shift away from the Fermi energy as a consequence. The result is a torroidal Fermi surface centered around the original nodal lines. This is illustrated for lattice (10,3)b in Fig.~\ref{Fig:sausages_shifted} below. One can compensate for this shift of the nodal line by introducing a chemical potential and thereby recover the nodal line close to the Fermi energy. (Note, however, that the Fermi line also generically tilts in energy upon breaking sublattice symmetry.) 
For lattice (10,3)c where the nodal line is protected by sublattice symmetry the inclusion of longer-range interactions, e.g. a next-nearest neighbor hopping, has a seemingly similar effect in that the resulting Fermi surface also exhibits a torroidal geometry. However, when introducing a chemical potential to compensate for this effect one finds, as illustrated in Fig.~\ref{Fig:sausages_shifted}, that the nodal line no longer exists anywhere in the energy spectrum, but has gapped out up to twelve symmetry-related Dirac points.


\subsection{Drumhead surface states}

The topological nature of a nodal-line semimetal manifests itself most directly in the occurrence of so-called drumhead surface states
\cite{Burkov2011,Phillips2014,Chiu2014}. For the sake of completeness, let us briefly recapitulate the elementary steps to see this in the context of the 3D honeycomb systems considered here.

\begin{figure}[b]
	\includegraphics[width=0.95\columnwidth]{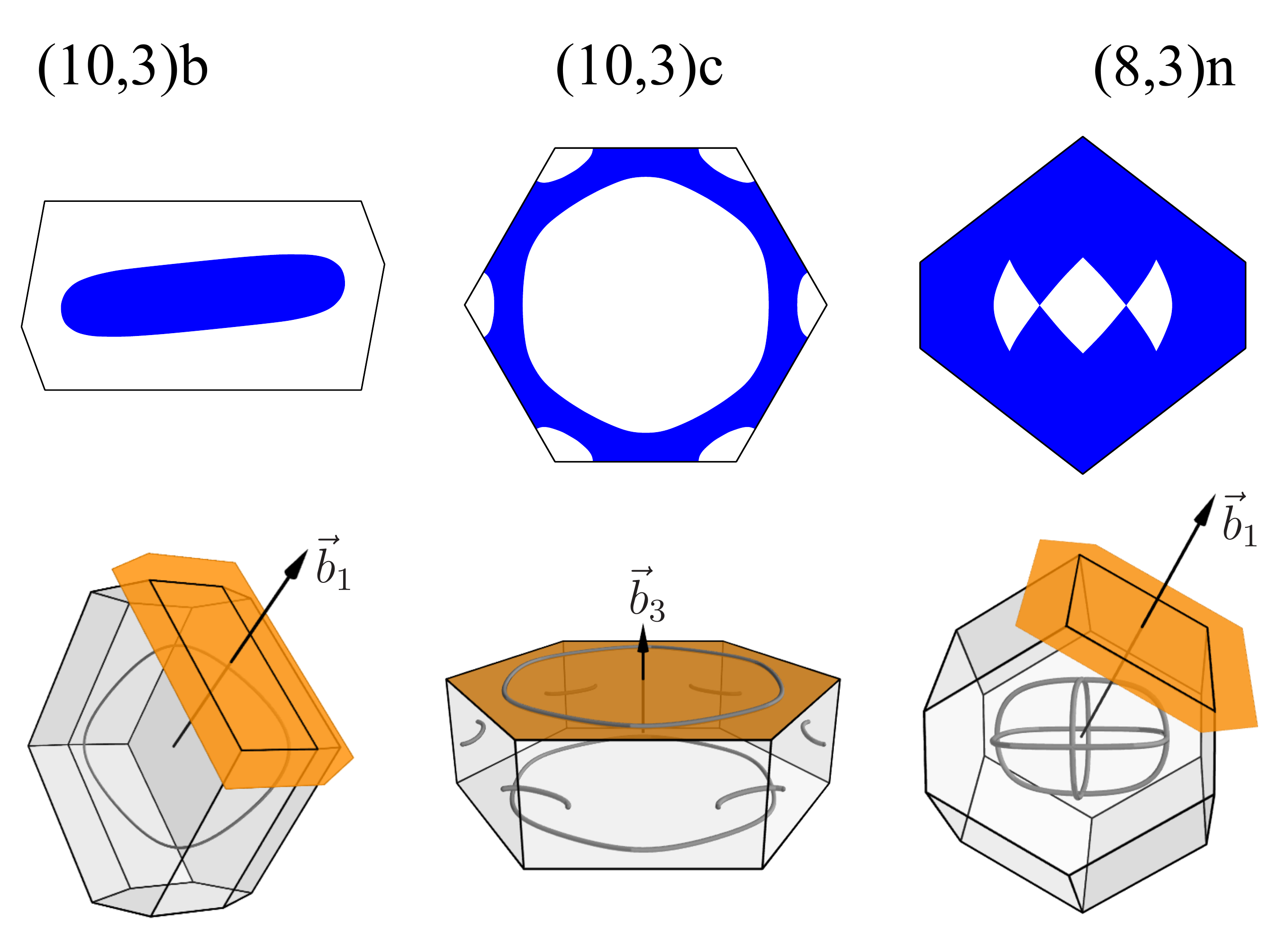}
	\caption{{\bf Drumhead surface states} for the nodal-line semimetals of lattices (10,3)b, (10,3)c, and (8,3)n.
			The bottom row shows the surface Brillouin zones along the projection used above.
			}
	\label{Fig:drumheads}
\end{figure}

Since all five lattices for which we find nodal lines exhibit sublattice symmetry, their Hamiltonian can be written in the off-diagonal form
\begin{align*}
\mathcal{H}\left(\vec{k}\right) = \begin{pmatrix}
0 & A^\dagger\left(\vec{k}\right)  \\
A\left(\vec{k}\right)  & 0
\end{pmatrix} \,,
\end{align*}
that is well known in the context of chiral symmetry classes \cite{Altland1997}.
For such off-diagonal Hamiltonians one can define a winding number
\begin{equation}
	W_\mathcal{C} = \frac{1}{2\pi} \oint_\mathcal{C} \text{d}\vec{k} \,\,\, \text{Im}\bracket{\text{tr}\bracket{A^{-1}\vec{\nabla}A}} 
\end{equation}
for any closed path $\mathcal{C}$ in momentum space. 
In particular, one can consider straight lines through the first Brillouin zone that either pierce through the interior of the nodal line or pass by them on the exterior. For any given nodal line, one of them must carry a non-trivial winding number.

If one now considers the projection of the nodal line to the surface Brillouin zone of a system in a slab geometry, one will find a band of flat surface states (taking the shape of a drumhead) for those momenta for which the corresponding winding number (modulo 2) is non-zero.
Examples for both scenarios are shown in Fig.~\ref{Fig:drumheads}.


\subsection{Landau level quantization and bulk flat bands}

The nodal line band structure can give rise to another topological phenomenon -- the formation of nearly dispersionless bands upon applying an external magnetic field parallel to the plane of the nodal line \cite{Rhim2015}. These bulk flat bands exhibit a Landau level quantization with an energy spacing that grows like $\sqrt{n}$  with the level index $n$. This square-root scaling traces back to the linear Dirac dispersion perpendicular to the nodal line and is familiar from the physics of graphene \cite{Li2007Observation}.
In the following, we briefly showcase how these generic features of nodal-line semimetals manifest themselvels in our three-dimensional honeycomb systems. Working directly with the microscopic model system (in lieu of a continuum approximation  \cite{Rhim2015})
we will focus our discussion on the (10,3)b hyperhoneycomb lattice.

\begin{figure}[b]
	\includegraphics[width=\columnwidth]{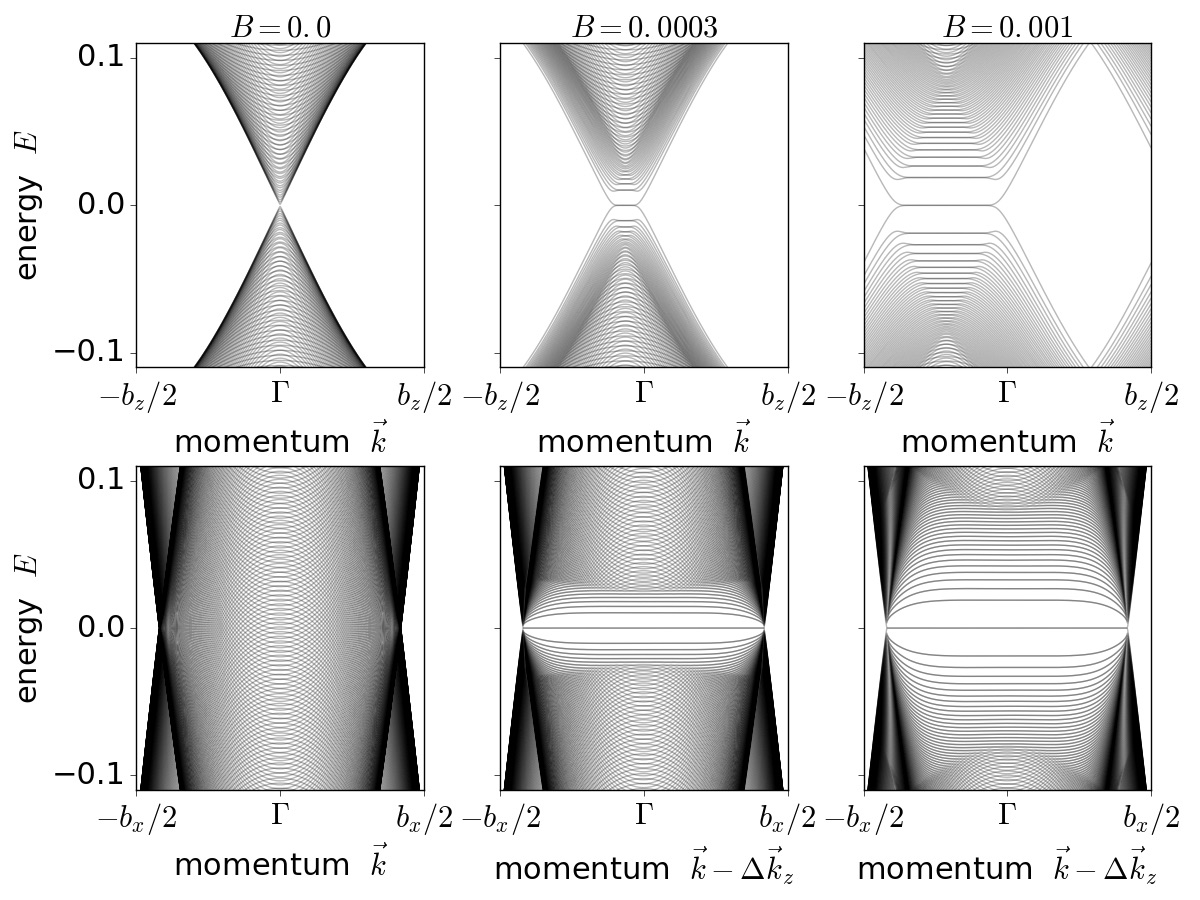}
	\caption{{\bf Formation of Landau levels} in the bulk energy spectrum for (10,3)b 
			for a magnetic field applied in plane of the nodal line.
			The different panels show varying magnetic field strength $B$.
	 		The top row shows the dispersion along the $k_z$ direction perpendicular to the nodal plane.
			The bottom row shows the dispersion along the $k_x$ direction 
			(where a field dependent offset $\Delta k_z$ is chosen such that the latter dispersion traverses the nodal line). 
		}
	\label{Fig:10b-LL}
\end{figure}

\begin{figure}[t]
	\includegraphics[width=\columnwidth]{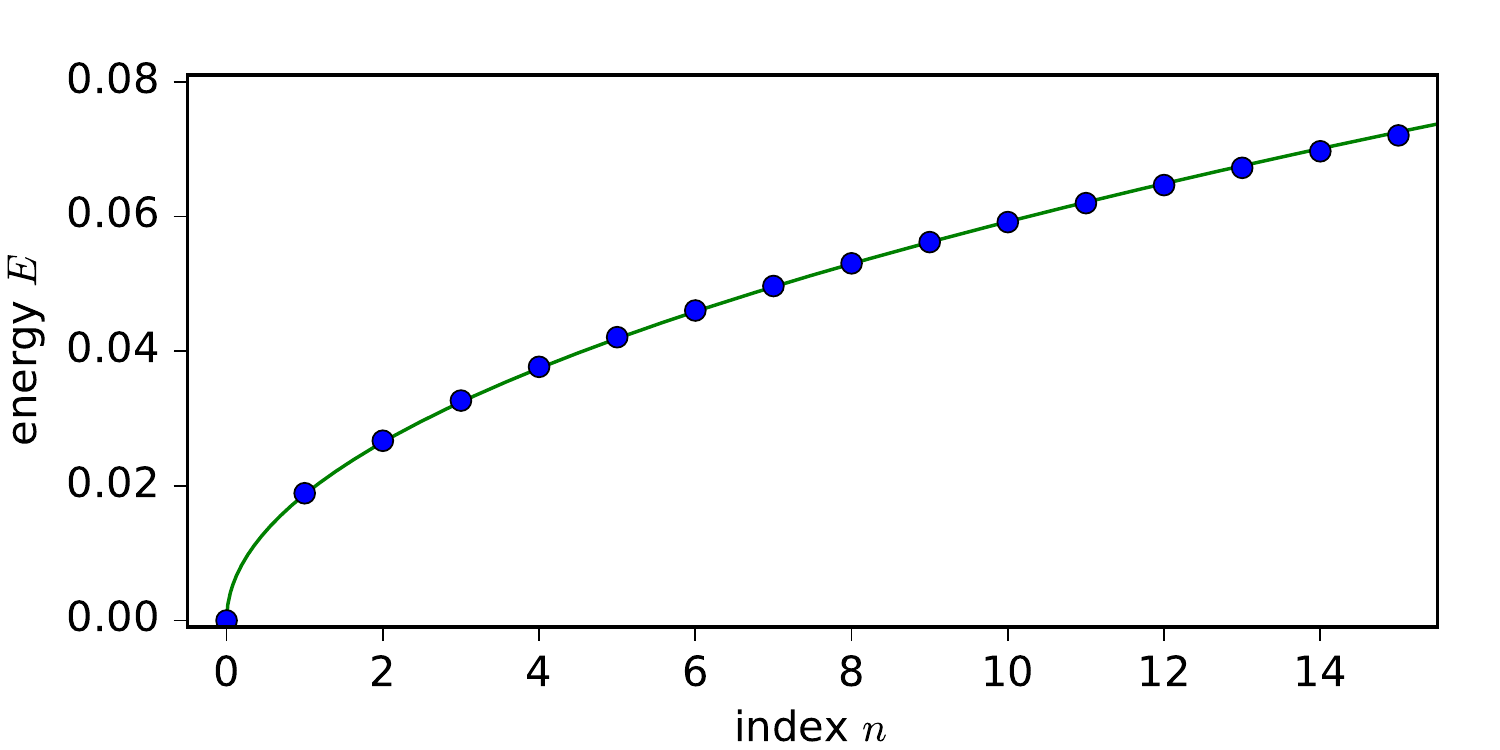}
	\caption{{\bf Spacing of Landau levels} in the energy spectrum for the (10,3)b hyperhoneycomb lattice.
		      The green line is a square root fit to the first 14 data points.}
	\label{Fig:10b-LL2}
\end{figure}

We consider a magnetic field 
\footnote{
We implement the magnetic field by assigning the phase $\phi=\pm n\pi \phi_F/\phi_{tot}$ to  the  $z$-bonds (see also the next footnote), 
with $n\vec a_y$ denoting the $y$-position of the unit cell and $\phi_{tot}$ denoting the total flux.  
This assigns a flux $2\phi$ to all the plaquettes that are penetrated by the magnetic field. 
Since we want to describe arbitrary flux through the plaquette, we use open boundary conditions in the $y$-direction. 
Note that $k_y$ is no longer a good quantum number. 
}
applied in the plane of the nodal line, e.g. along the $\hat x$-direction with the nodal line lying in the  $(k_x,k_y)$-plane
\footnote{To facilitate our analysis we work with the most symmetric representation of the (10,3)b hyperhoneycomb lattice. In this representation
the elementary zig-zag chains constituting the lattice, see Fig.~\ref{Fig:lattice_illustration}, are rotated by 90$^\circ$ with respect to each other. We further double the unit cell, see Appendix~\ref{App:Lattices}, which allows us to choose orthogonal lattice translation vectors (as compared with the minimal representation of the lattice). 
With these conventions the nodal line lies in the $(k_x,k_y)$-plane, see also Fig.~\ref{Fig:10b-rectangular} in Appendix~\ref{App:Lattices}. 
}.
Upon applying the magnetic field  almost dispersionless flat bands evolve in the middle of the spectrum and widen with increasing
 field strength, as can be seen in the upper panel of Fig.~\ref{Fig:10b-LL}, which shows the bulk energy dispersion along the $k_z$-direction perpendicular to the nodal plane.
In fact, the nodal line tilts upon applying a magnetic field away from the $k_x$-$k_y$ plane and acquires a finite extent also along the 
$k_z$-direction. It is precisely within this $k_z$ region that the nearly dispersionless bands develop. These flat bands constitute Landau levels
whose separation is plotted in Fig.~\ref{Fig:10b-LL2}, clearly revealing the $\sqrt{n}$ growth with the Landau level index $n$
expected for a linear Dirac dispersion \cite{McCann2006}.
The lower panel in Fig.~\ref{Fig:10b-LL} shows the dispersion along the $k_x$-direction revealing the flat bands precisely in the $k_x$ region inside the nodal line. Note that we need a field-dependent offset $\Delta k_z$ since the nodal line shifts in the $k_z$ direction. 

\begin{figure}[b]
	\includegraphics[width=\columnwidth]{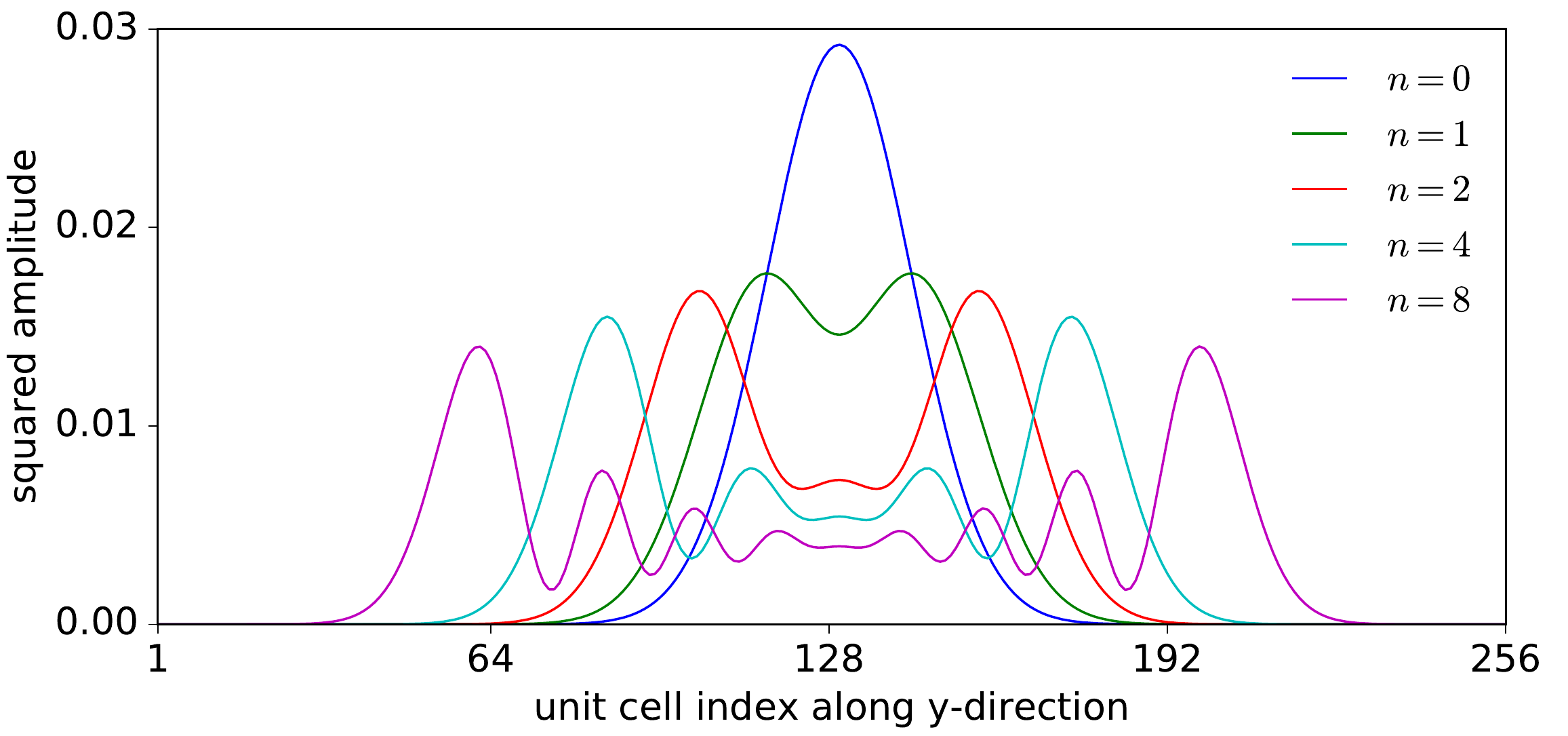}
	\caption{Squared amplitude of the {\bf wavefunction for the Landau levels} for a slab geometry 
			of the (10,3)b hyperhoneycomb lattice with open boundaries along the $y$-direction.
	}
	\label{Fig:10b-LL3}
\end{figure}

Each Landau level is four-fold degenerate, up to a very small Zeeman splitting of the order of $10^{-4}$ for the magnetic field strengths considered here.
This four-fold degeneracy can be explained by realizing that the system effectively sees, for any momentum within the nodal line,  two two-fold degenerate Dirac nodes (one from each side of the nodal line), similar to the physics of a graphene bilayer as nicely shown in Ref.~\cite{Rhim2015}. 

Finally, let us note that the Landau levels are not to be confused with the flat surface bands discussed earlier. 
In fact, the Landau levels discussed here are pure {\em bulk} features. 
This can be verified by plotting the spatial localization of their wavefunction as illustrated in Fig.~\ref{Fig:10b-LL3},
where the squared amplitude of the wavefunction (evaluated for a momentum in the middle of the Landau level) is plotted
against the $y$-position for a slab geometry.
As can be  seen, the wavefunction of the lowest Landau level ($n=0$) is clearly localized in the middle of the slab.
Going to higher Landau levels, one observes a spreading towards the edges of the slab with significant weight remaining in the bulk.


\subsection{Spin-orbit coupling and topological insulators}

\begin{figure}[b]
	\includegraphics[width=\columnwidth]{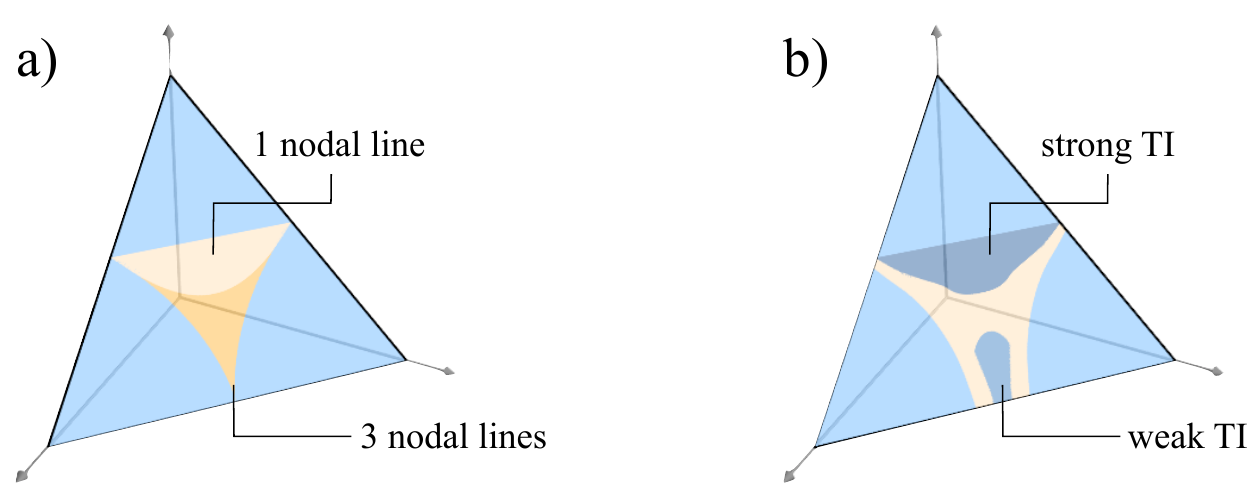}
	\caption{{\bf Schematic phase diagram} for (a) the pure tight-binding model and (b) in the presence of spin-orbit coupling.
			Generically, we find that around the isotropic coupling point ($t_x = t_y = t_z$) there is an extended semimetallic phase (indicated by the orange shading), whereas in the regimes where one of the three coupling
			strengths dominates one finds gapped band insulators (indicated by the light blue shading).
			The actual data shown here is for lattice (10,3)c, which in the absence of spin-orbit coupling exhibits
			two distinct semimetallic phases with one or three nodal lines, respectively. In the presence of spin-orbit
			coupling two topological insulator phases emerge that can be classified as strong/weak topological insulators
			as indicated.
			}
	\label{Fig:SchematicPhaseDiagram}
\end{figure}

Probably the most dramatic effect destabilizing the nodal-line band structures in the three-dimensional honeycomb systems at hand 
is the inclusion of spin-orbit coupling. We find that in the presence of spin-orbit coupling (implemented via a complex next-nearest neighbor hopping analogous to the one used in the work of Kane and Mele \cite{Kane2005}), all nodal-line semimetals generically gap out into topological insulators (TIs). This is illustrated in the exemplary phase diagram for lattice (10,3)c, which is shown in Fig.~\ref{Fig:SchematicPhaseDiagram}. 
Interestingly, for all systems considered here the arising topological insulator is in fact a {\em strong} topological insulator (which for some systems is flanked by a second parameter regime of a weak topological insulator as is the case for the phase diagram of lattice  (10,3)c in Fig.~\ref{Fig:SchematicPhaseDiagram}). A precise characterization of the TIs via their topological indices \cite{Fu2007} is given in the overview Table \ref{Tab:Overview}. For the example system (10,3)c the calculation of the topological indices via the surface energy spectrum  is illustrated in Fig.~\ref{Fig:10c-TI}, with further details provided in  Appendix \ref{app:TI}. This appendix also provides surface energy spectra and topological indices of the TI phases for all of the other honeycomb lattices.

\begin{figure}[t]
	\includegraphics[width=\columnwidth]{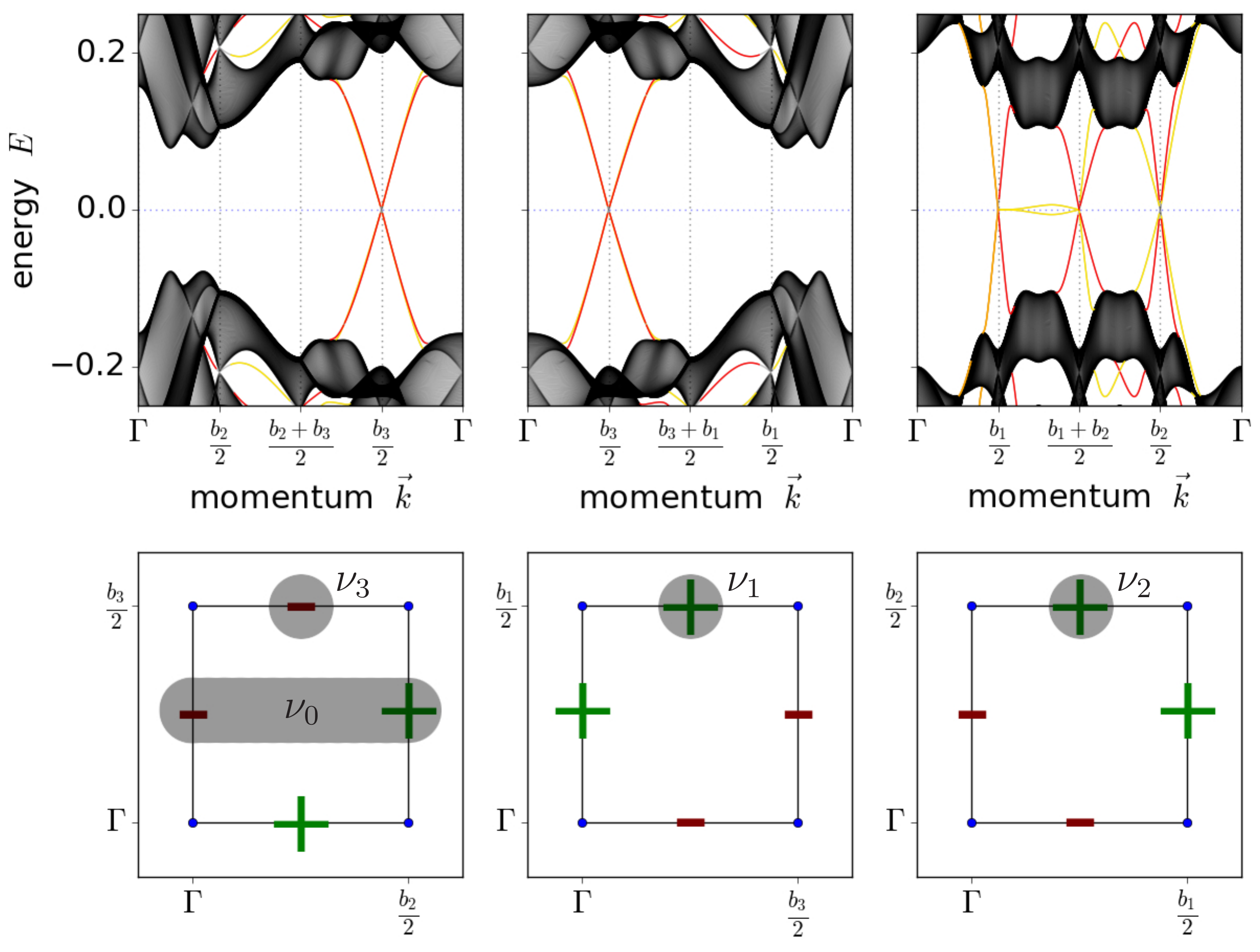}
	\caption{{\bf Strong topological insulator} for (10,3)c with indices 1;(001).
			The top row shows the surface energy spectrum along the high-symmetry path connecting the 
			time-reversal invariant momenta (TRIMs).
			 The colored bands indicate surface bands crossing the Fermi energy 
			 (with the yellow/red band corresponding to a band on the top/bottom surface).
			The bottom row shows the change of the time-reversal polarization for different paths between TRIMs
			(indicated by the blue dots at the corner of the squares).}
	\label{Fig:10c-TI}
\end{figure}
\begin{figure}[h]
	\includegraphics[width=\columnwidth]{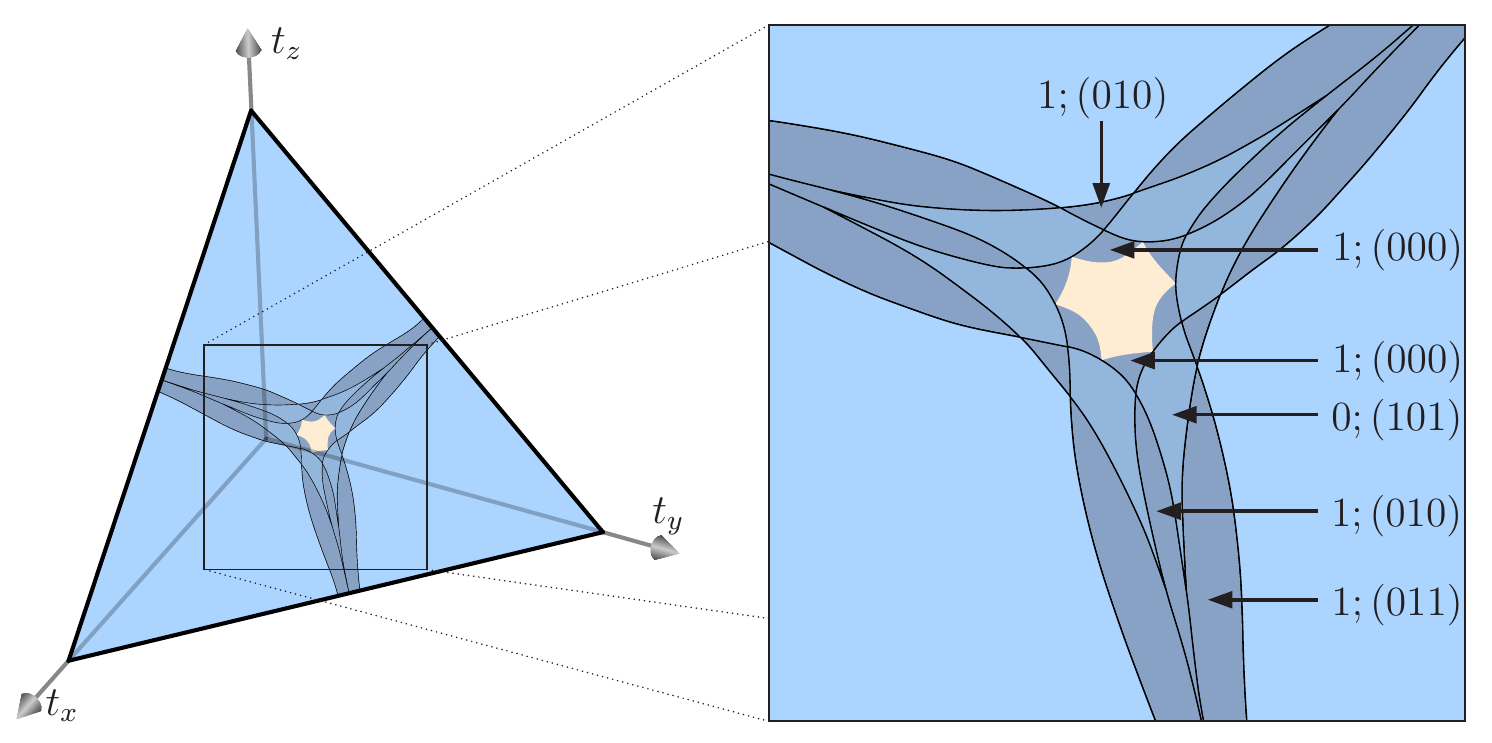}
	\caption{{\bf Kaleidoscope of topological phases} for lattice (8,3)c in the presence of spin-orbit coupling. 
			}
	\label{Fig:Overview8c}
\end{figure}

While most nodal-line systems exhibit a relatively simple phase diagram in the presence of spin-orbit coupling, similar to the one shown for 
lattice (10,3)c in Fig.~\ref{Fig:Overview8c}, there is one exception, lattice (8,3)c,  where the inclusion of spin-orbit coupling leads to a kaleidoscope of different topological phases as illustrated in Fig.~\ref{Fig:Overview8c}. Each line separating two TI phases in Fig.~\ref{Fig:Overview8c} corresponds to a closing of a gap at a {\em single} time-reversal invariant momentum (TRIM). Since this can change only one cone on the surface, the number of cones switches from even to odd or vice versa, so that every line separates a weak from a strong TI.

\begin{figure}[h]
	\includegraphics[width=\columnwidth]{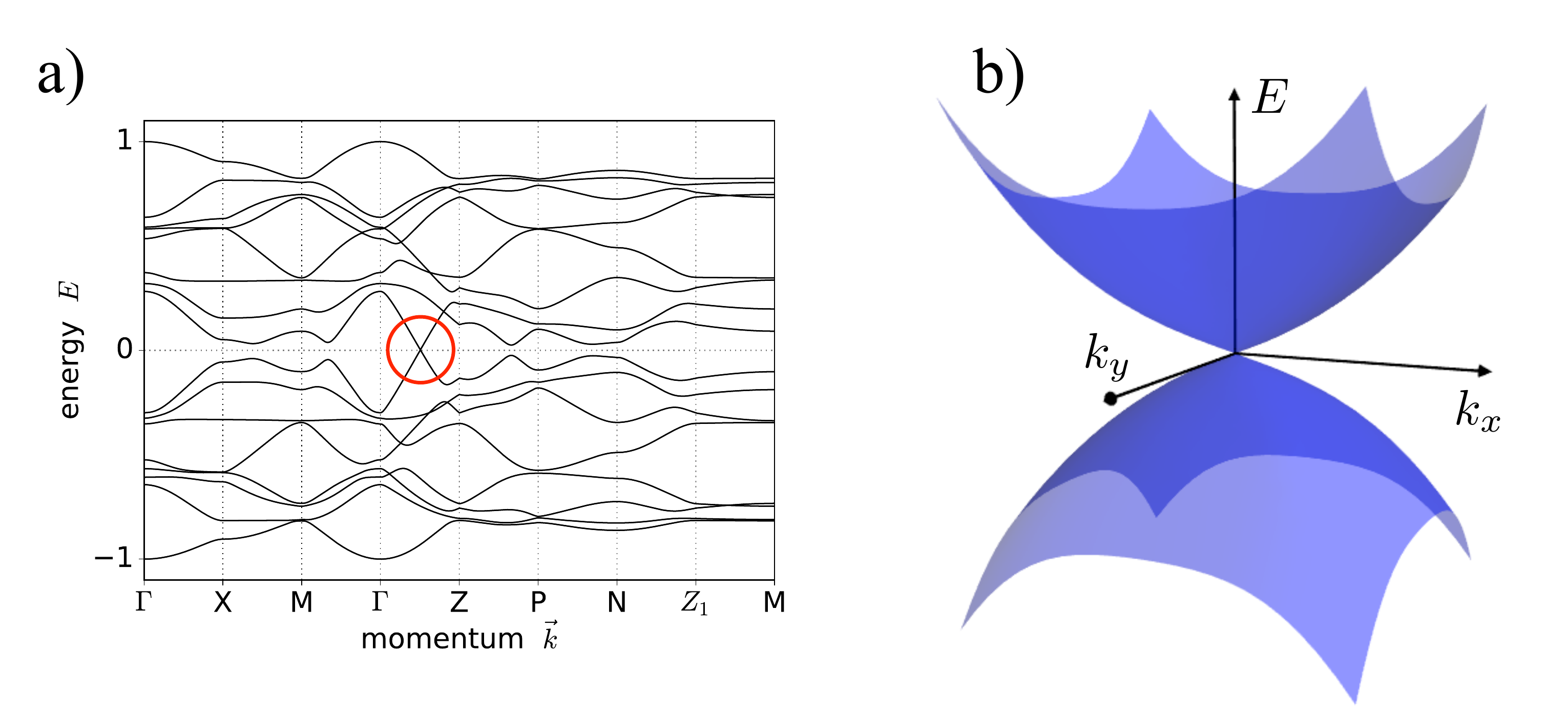}
	\caption{ {\bf Spectral features} for (8,3)n in the presence of spin-orbit coupling.
		      a) Bulk energy spectrum along a high-symmetry path with Dirac node highlighted by the red circle.
		      b) 3D energy spectrum around the Dirac node. 
			}
	\label{Fig:Overview8n}
\end{figure}

Lattices (8,3)c and (8,3)n exhibit another feature occurring in the presence of spin-orbit coupling (besides the formation of a strong TI) -- the emergence of (bulk) Dirac nodes in the spectrum for some parameter regime. For lattice (8,3)n, which exhibits three crossing nodal lines in the absence of spin-orbit coupling, the Dirac nodes emerge precisely at the crossing points (while the remaining nodal lines gap out). The bulk energy spectrum of lattice (8,3)n is illustrated, along a high-symmetry path, in Fig.~\ref{Fig:Overview8n}.

%


%
\begin{figure*}[t]
 	\includegraphics[width=.98\textwidth]{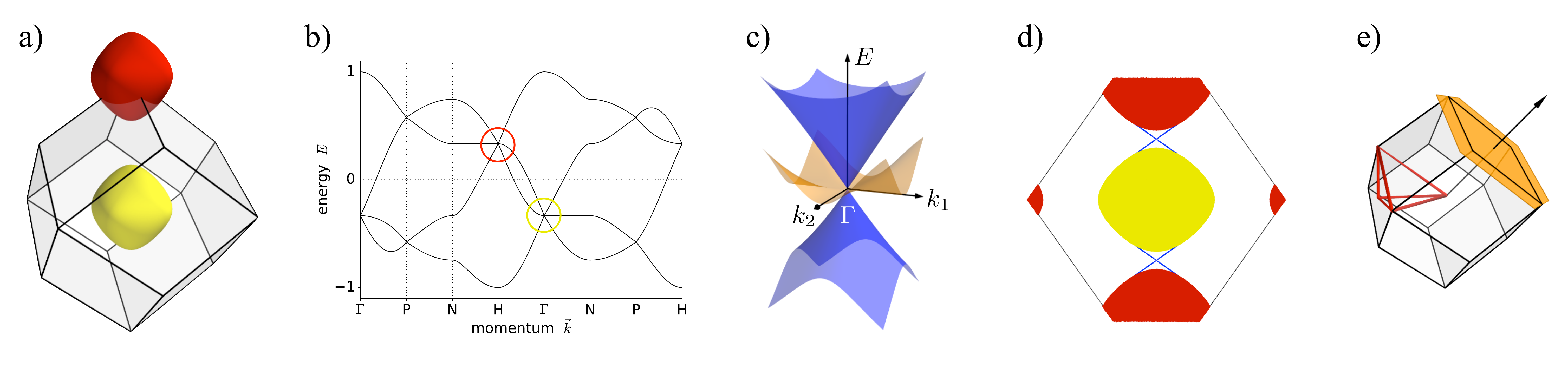}
	\caption{Overview of the {\bf spectral features for the (10,3)a hyperoctagon lattice}.
		      a) Fermi surfaces around the $\Gamma$ and $H$ points in the Brillouin zone.
		      b) Bulk energy spectrum along the high-symmetry path indicated in e).
		      c) Spin-1 Weyl point (below the Fermi surface) at the $\Gamma$ point.
		      d) Surface spectrum with Fermi arcs for the surface Brillouin zone indicated in e).
		      }
	\label{Fig:10a-SpectralFeatures}
\end{figure*}
\begin{figure*}[t]
 	\includegraphics[width=.98\textwidth]{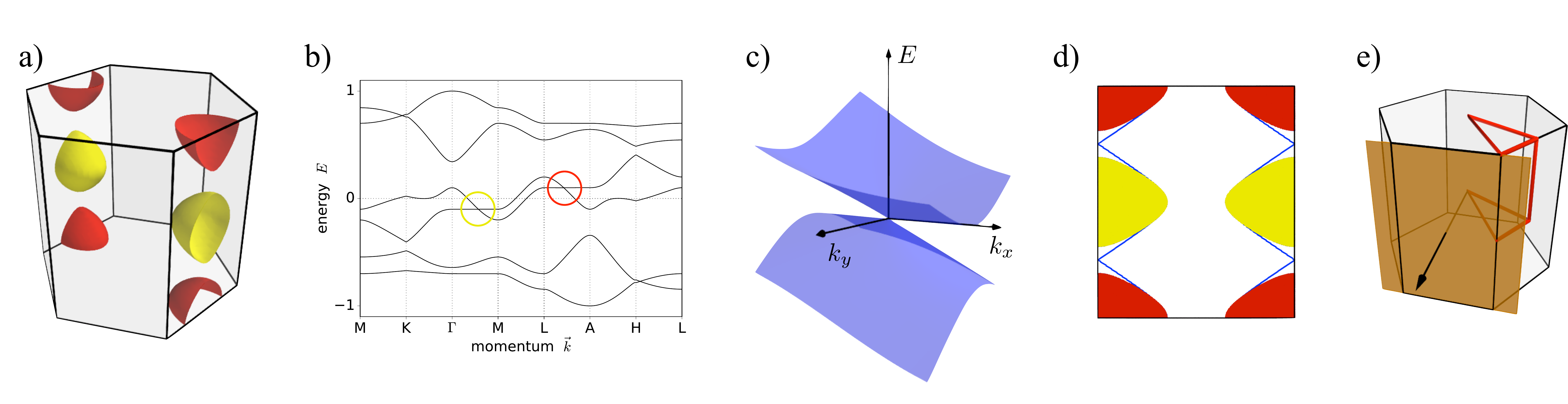}
	\caption{Overview of the {\bf spectral features for lattice (8,3)a}.
		      a) Fermi surfaces around the $M$ and $L$ points in the Brillouin zone.
		      b) Bulk energy spectrum along the high-symmetry path indicated in e).
		      c) Weyl point (below the Fermi surface) close to the $M$ point.
		      d) Surface spectrum with Fermi arcs for the surface Brillouin zone indicated in e).
		      }
	\label{Fig:8a-SpectralFeatures}
\end{figure*}

\section{Topological metals}
\label{sec:TopologicalMetals}

For the family of three-dimensional honeycomb models at the heart of this manuscript there are -- besides the ones that give rise to nodal-line semimetals -- a number of instances that form metals exhibiting ordinary two-dimensional Fermi surfaces, see the overview of Table \ref{Tab:Overview}. However, even in this seemingly conventional scenario, it turns out that topology is again at play and the metals are best characterized as {\em Weyl metals} or {\em Dirac metals} with a distinct topological feature (such as Weyl or Dirac cones) encompassed by the Fermi surfaces. As such the Fermi surfaces are characterized by a topological invariant (such as a Chern number), which again leads to the formation of distinct surface features such as Fermi arcs. We will concentrate our discussion on these topological features in the following, after providing some elementary symmetry considerations that motivate the formation of Fermi surfaces in the first place.

\subsection{Symmetry considerations}

For the 3D honeycomb systems at hand, we find a metallic behavior for the lattices (10,3)a, (9,3)a, (8,3)a, and (8,3)b as summarized in Table~\ref{Tab:Overview}. The formation of a metal in all these lattices is closely connected to the way  sublattice symmetry and inversion are implemented in these lattices. 

For the sublattice symmetry there are three cases to distinguish: (i) it might be entirely absent, as is the case for the non-bipartite lattice (9,3)a, (ii) it is implemented with an additional translation in momentum space (see Sec.~\ref{sec:sym}), such as for the lattices (10,3)a, (8,3)a, and (8,3)b of interest here, or (iii) its presence does not require an additional translation in momentum space. The last case is precisely the scenario needed for the formation of nodal lines as discussed in the previous Section, while the first two cases naturally lead to the formation of Fermi surfaces as discussed in the following.

Considering inversion symmetry, there are only two possible cases for the lattices at hand -- a lattice either breaks inversion symmetry and is chiral or not, see the overview of lattice properties in Table \ref{tab:lattice_overview}.
If the underlying lattice is inversion symmetric, as is the case for lattices (9,3)a and (8,3)b, then the combination of time-reversal and inversion symmetry, $PT$, leaves the Hamiltonian invariant and we would naively expect that the system should show nodal line physics. Indeed, looking at the band structure, one finds that for both lattices the  two bands in the middle of the energy spectrum touch along a closed line, which is protected by $PT$. However, due to the absence (or non-trivial implementation) of sublattice symmetry, the system is not particle-hole symmetric. 
Consequently, the nodal line(s) do not lie at constant energy (but are tilted) and are not pinned to the Fermi energy. As such the systems generically exhibit Fermi surfaces surrounding the nodal line(s). 
For the two chiral lattices at hand, lattices (10,3)a and (8,3)a, there is no symmetry protecting nodal lines in the first place, and one would expect these systems to form conventional Fermi surfaces.

\subsection{Hidden Weyl/Dirac nodes}
Weyl and Dirac semimetals have recently attracted much attention due to their unique combination of bulk and surface properties
\cite{Armitage2018}. 
It was already noted early on \cite{Herring1937} that band touchings at isolated points in the three-dimensional Brillouin zone can occur frequently and are in fact stable objects. 
The origin of this stability was later traced back to a topological invariant, the Chern number, that associates an integer charge with each one of them \cite{Wan2011}. 
Charge conservation immediately implies that these so-called Weyl nodes cannot be removed individually, but only in pairs with opposite charge \cite{Nielsen1981a,Nielsen1981b}. 
In order for the system to show semimetallic behavior, one needs additional symmetries that pin the energy of at least one pair of Weyl nodes to the Fermi energy. 
However, even in the absence of such symmetries, the system can show interesting features, particularly when the Weyl nodes are `hidden' within Fermi surfaces and thereby lend their topological properties to the metallic state. Such systems are referred to as Weyl metals.
Similarly, a Dirac node -- the composition of two Weyl nodes -- can be encompassed by a Fermi surface giving rise to what can analogously  be referred to as Dirac metal (as opposed to the commonly discussed Dirac {\em semi}metals).
It turns out that these latter scenarios are at play for some of the 3D honeycomb systems considered here.

Specifically, this is the case for lattices (10,3)a and (8,3)a whose spectral features (for the spinless case) are summarized in Figs.~\ref{Fig:10a-SpectralFeatures} and \ref{Fig:8a-SpectralFeatures}, respectively. 
Both lattices exhibit pair(s) of Weyl nodes located above/below the Fermi energy (as required by sublattice symmetry). 
The case of lattice (10,3)a is particularly interesting as the Weyl node is located at the intersection of {\em three} crossing bands, two forming the actual Weyl  cone and the third forming a flat band, see Figs.~\ref{Fig:10a-SpectralFeatures}~b) and c). This scenario can 
only play out at certain high-symmetry points in the Brillouin zone \cite{Bradlyn2016} and is referred to as spin-1 Weyl node \cite{Watanabe2011,Malcolm2014,Orlita2014,Bradlyn2016,Tsuchiizu2016,Fulga2017}. An additional consequence of this spin-1 scenario comes in the form of a higher integer Chern number of $\pm 2$ associated with these Weyl nodes in the bulk and a {\em pair} of Fermi arcs on the surface, indicated by the blue lines in Fig.~\ref{Fig:10a-SpectralFeatures}~d).

For lattice (8,3)a a more conventional scenario is found. The two Fermi surfaces around the $M$ and $L$ points each encompass two Weyl nodes (of which only one is shown along the high-symmetry path in Fig.~\ref{Fig:8a-SpectralFeatures}~b)). As such the Fermi surfaces again carry a total Chern number of $\pm 2$ and the surface energy spectrum shows two Fermi arcs, as shown in Fig.~\ref{Fig:8a-SpectralFeatures}~d).

It should be noted that the topological features of the Fermi surfaces typically come hand-in-hand with an instability. Since the Weyl nodes hidden within the Fermi surfaces always come in pairs \cite{Nielsen1981a,Nielsen1981b}, so do the topological Fermi surfaces which oftentimes leads to a perfect nesting situation between them. This certainly is the case for the two lattices (10,3)a and (8,3)a at hand. As a result, these systems are susceptible to show either (finite-momentum) Cooper pair formation leading to superconductivity \cite{Fulde1964,Larkin1964} or a Peierls instability leading to either charge or spin density formation \cite{Rosch2015}.

Finally, we briefly mention the two remaining lattices (9,3)a and (8,3)b, which both exhibit Fermi surfaces that enclose nodal lines (instead of Weyl nodes). But as such these Fermi surfaces are {\em not} endowed with any topological features (since the nodal lines in our models do not exhibit a non-trivial spherical Chern number \cite{Fang2015}).

\subsection{Spin-orbit coupling}
In the absence of spin-orbit coupling all spectral features discussed above are spin-degenerate. The inclusion of spin-orbit coupling will immediately lift this spin-degeneracy and the spin-degenerate Dirac cones will split into pairs of Weyl nodes. 

\begin{figure}[t]
	\includegraphics[width=\columnwidth]{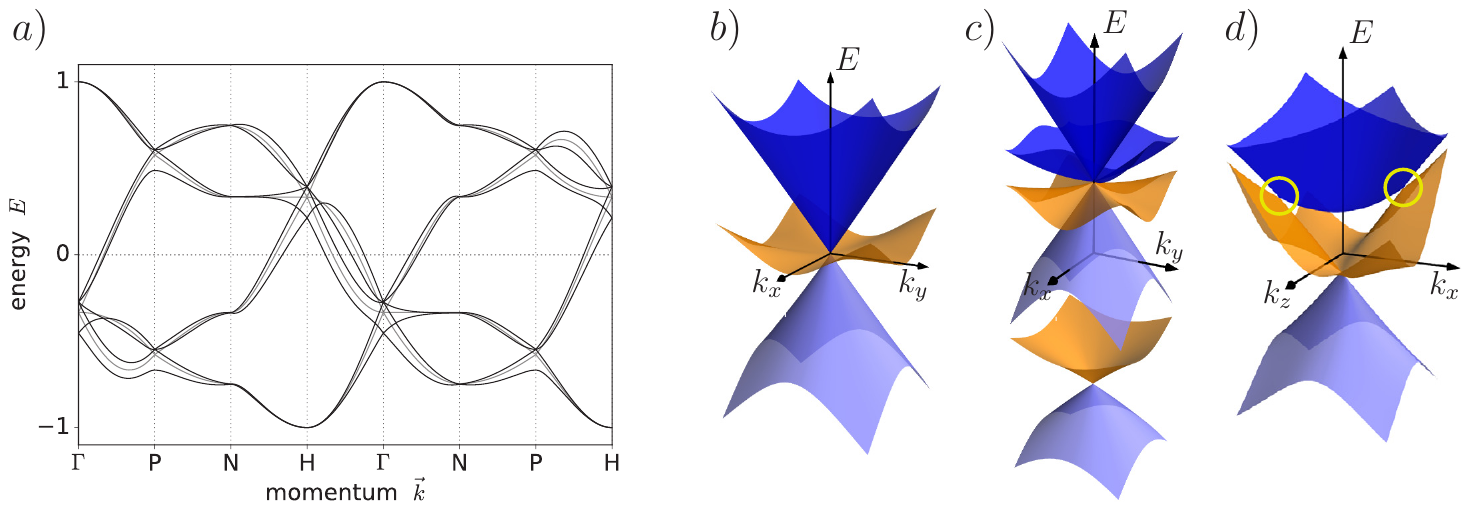}
	\caption{{\bf Effect of spin-orbit coupling on the spin-1 Weyl node} of the hyperoctagon lattice (10,3)a. 
	a) Band structure in the absence/presence of spin-orbit coupling (grey/black lines). 
	b) The six-fold degeneracy of the spin-1 Weyl node at the $\Gamma$ point is lifted in the presence of spin-orbit coupling
	to a two-fold and a four-fold band touching as shown in c). 
	d) Breaking the two-fold screw symmetry (in the absence of spin-orbit coupling), splits the original (charge 4) Weyl node 
	into two (charge 2) Weyl nodes (marked by the yellow circles), 	which remain spin-degenerate.
}
	\label{Fig:10a-GammaPoint}
\end{figure}

A slightly more subtle mechanism is at play for the hyperoctagon lattice (10,3)a where part of the degeneracy remains even in the presence of spin-orbit coupling (SOC).
Nonsymmorphic lattice symmetries, in particular a two-fold screw symmetry, pin the gapless modes to the high-symmetry points $\Gamma$ and $H$, so that the only allowed splitting is in energy. 
We find that SOC splits the original six-fold degenerate band touching, to a two-fold (with charge $\pm 1$) and a four-fold (with charge $\pm 3$) touching, as shown in Fig.~\ref{Fig:10a-GammaPoint}~b) and c). 
This degeneracy could be split further by also considering a breaking of the nonsymmorphic lattice symmetries. 
Fig.~\ref{Fig:10a-GammaPoint}~d) shows the effect of breaking the two-fold screw symmetry (in the absence of SOC), which is found to break the charge 4 Dirac node into two charge 2 Dirac nodes. The remaining spin degeneracy can then be destroyed by introducing SOC. 

\begin{figure}[t]
	\includegraphics[width=\columnwidth]{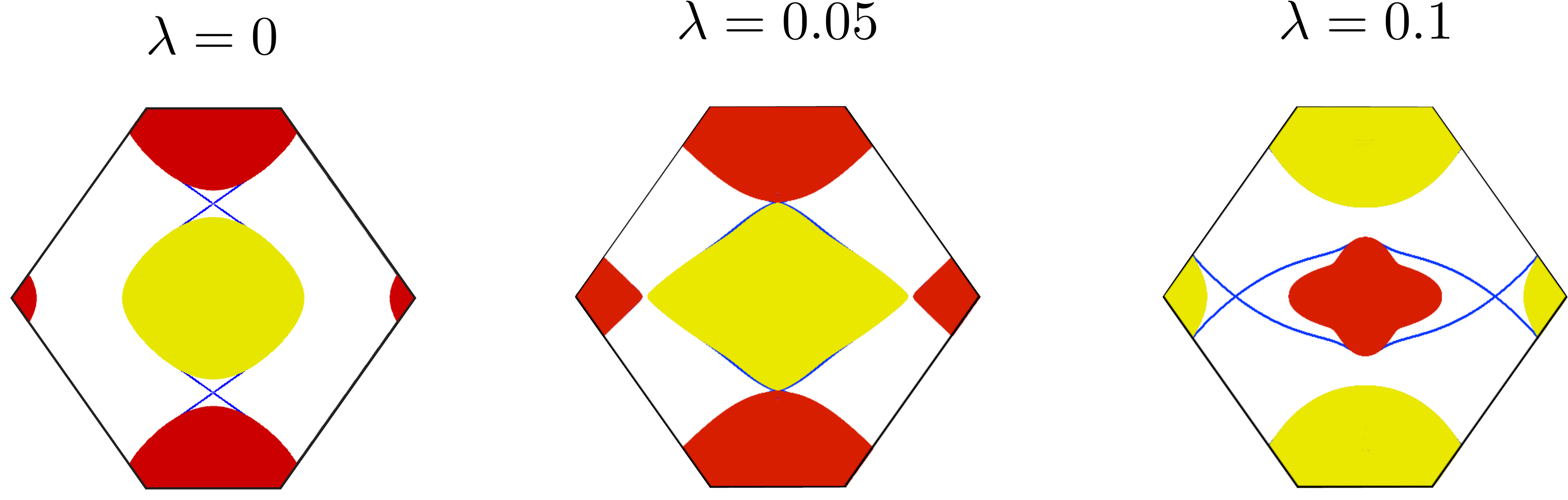}
	\caption{{\bf Evolution of surface states} for the (10,3)a hyperoctagon lattice 
		      for varying strength of the spin-orbit coupling $\lambda$. 
		      All spectra exhibit puddles, which originate from the projection of the Fermi surface onto the 
		      surface Brillouin zone. They are color-coded to reflect the Chern number associated with 
		      the Weyl nodes encompassed by the respective Fermi surface.
		      In addition, the surface spectrum exhibits a pair of Fermi arcs (colored in blue) that reflect the connection of the
		      hidden Weyl nodes. 
		      With increasing spin-orbit coupling the Weyl nodes reorganize underneath the Fermi surface 
		      and the Chern numbers associated with the Fermi surfaces change.}
	\label{Fig:10a-SurfaceStates_SOC}
\end{figure}

With increasing spin-orbit coupling the Weyl nodes encompassed by the Fermi surfaces move within the Brillouin zone. This reorganization of topological charges becomes most apparent in the surface spectrum, illustrated in Fig.~\ref{Fig:10a-SurfaceStates_SOC}. As can be seen in the evolution of these spectra, the Chern numbers associated with the individual Fermi surfaces change sign and the associated Fermi arcs flip their orientation.

%


\section{Summary}
\label{sec:Discussion}

Probably the most revealing result of our approach to systematically classify the band structures of three-dimensional honeycomb systems is the formation of topological features in almost all instances, as summarized in Table \ref{Tab:Overview}.
Amongst these topological band structures, an expansive family of Dirac nodal-line semimetals stands out. Depending on the underlying lattice geometry, these systems exhibit nodal lines forming a variety of mutual geometries, as summarized in Fig.~\ref{Fig:OverviewNodalLines}. 
They share many well understood common features such as the protection by $PT$ symmetry (with one exception), the formation of drumhead surface states, and a $\sqrt{n}$ Landau level quantization of bulk flat bands in the presence of an applied magnetic field. Other common features whose origin is less clear include the formation of an {\em odd} number of nodal lines for all lattice geometries and what appears to be the generic formation of a {\em strong} topological insulator in the presence of spin-orbit coupling.
Besides the formation of nodal-line semimetals we have discussed two instances of Weyl metals, topological metals in which (multiple) Weyl nodes are enclosed by the Fermi surface enriching these systems with a non-trivial bulk Weyl charge 
and Fermi arc surface states.

The current investigation of electronic band structures complements our recent work on classifying the band structures of real Majorana fermions relevant to the physics of gapless spin liquids in three-dimensional Kitaev models \cite{OBrien2016} defined  for the same set of elementary tricoordinated lattices. Interestingly,  there are a number of instances where the band structures of real and complex fermions differ substantially, see the comparison in Table \ref{Tab:Overview}. For certain lattice geometries, the Majorana fermions form Weyl semimetals \cite{Hermanns2015} in lieu of conventional (Fermi surface) metals in the corresponding electronic system. The deeper reason for this discrepancy is found in the way projective symmetries and in particular time-reversal symmetry act on the Majorana fermions in contrast to ordinary complex fermions as discussed in detail in Ref.~\onlinecite{OBrien2016}.

Beyond electronic and Majorana band structures, the emergence of nodal lines have long been discussed in the context of 
superconductors \cite{Schnyder2012} and more recently also in magnetic systems with non-trivial magnon bands 
\cite{Mook2017}. Likely, the 3D honeycomb lattices studied in the present work will also reveal non-trivial band structure phenomena for these alternate systems, but we will leave it to future studies to explore this. 

In the realm of materials synthesis, we hope that our work provides inspiration to explore candidate materials that could realize one of the 3D honeycomb systems discussed here, be it in the context of complex graphene networks \cite{Weng2015}  or in the form of a newly synthesized Kitaev material \cite{Trebst2017}.


\begin{acknowledgments}
We acknowledge insightful discussions with S. Parameswaran, K. O'Brien, M. G. Yamada, and V. Dwivedi.
This work was partially supported by the DFG within the CRC 1238 (project C02).
The numerical simulations were performed on the CHEOPS cluster at RRZK Cologne.
\end{acknowledgments}


\appendix


\section{Tricoordinated lattice geometries}
\label{App:Lattices}
To make the manuscript at hand fully self-contained, we provide in the following the details of all the tricoordinated lattices discussed in the main text. 
In particular, we provide the lattice vectors and their corresponding reciprocal lattice vectors, as well as the positions of the lattice sites within the unit cell. 
Details on the symmetry group for all lattices can be found in Table~\ref{tab:lattice_overview}.

\subsection{(10,3)a}
The lattice vectors of the chiral hyperoctagon or (10,3)a lattice  are given by 
\begin{align*}
\textbf{a}_1 &= \bracket{1,0,0},~~~~
\textbf{a}_2 = \bracket{\frac{1}{2}, \frac{1}{2},-\frac{1}{2}},\\
\textbf{a}_3 &= \bracket{\frac{1}{2},\frac{1}{2},\frac{1}{2}}
\end{align*}
and their corresponding reciprocal lattice vectors are
\begin{align*}
\textbf{b}_1 &= 2\pi \bracket{1,-1,0},~~~~
\textbf{b}_2 = 2\pi \bracket{0,1,-1},\\
\textbf{b}_3 &= 2\pi \bracket{0,1,1}. 
\end{align*}
This lattice has four sites per unit cell that are located at
\begin{align*}  
\textbf{r}_1 &= \frac{1}{8}\bracket{1,1,1},~~~~&&
\textbf{r}_2 = \frac{1}{8}\bracket{5,3,-1}\\
\textbf{r}_3 &= \frac{1}{8}\bracket{3,1,-1},~~~~&&
\textbf{r}_4 = \frac{1}{8}\bracket{7,3,1}.
\end{align*}

\subsection{(10,3)b}
The (10,3)b or hyperhoneycomb lattice has, in its original realization,  the translation vectors
\begin{align*}
\textbf{a}_1 &= \bracket{-1,1,-2},~~~~
\textbf{a}_2 = \bracket{-1,1,2},\\
\textbf{a}_3 &= \bracket{2,4,0}.
\end{align*}
Their corresponding  reciprocal lattice vectors are given by
\begin{align*}
\textbf{b}_1 &= 2\pi \bracket{-\frac{1}{3}, \frac{1}{6}, -\frac{1}{4}},~~~~
\textbf{b}_2 = 2\pi \bracket{-\frac{1}{3}, \frac{1}{6}, \frac{1}{4}},\\
\textbf{b}_3 &= 2\pi \bracket{\frac{1}{6}, \frac{1}{6}, 0}.
\end{align*}
This lattice has also four sites per unit cell that are located at 
\begin{align*}  
\textbf{r}_1 &= \bracket{0,0,0},~~~~&&
\textbf{r}_2 = \bracket{1,2,1}\\
\textbf{r}_3 &= \bracket{1,1,0},~~~~&&
\textbf{r}_4 = \bracket{2,3,1}.
\end{align*}

\begin{figure}[b]
	\includegraphics[width=\columnwidth]{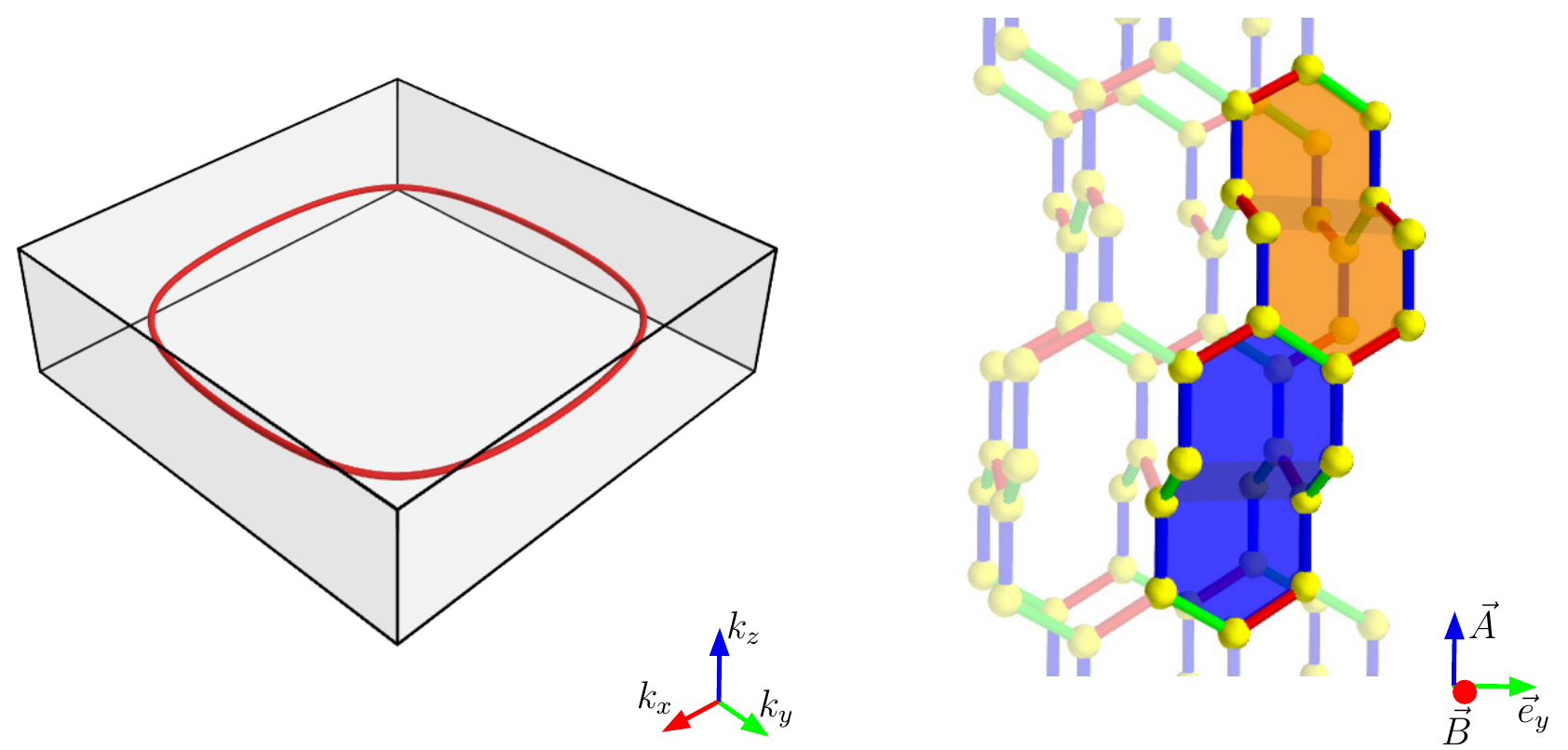}
	\caption{Left: Visualization of the nodal line of the {\bf deformed hyperhoneycomb lattice} in the rectangular Brillouin zone. Right: The deformed hyperhoneycomb lattice with enlarged unit cell and orthogonal lattice vectors. Of the 8 plaquettes per unit cell, only two enclose a magnetic flux (for $\mathbf{B}=B_0\hat{e}_x$), which we marked in orange and blue. 
	}
	\label{Fig:10b-rectangular}
\end{figure}

In order to include a magnetic field and still retain two good quantum numbers in reciprocal space , it is convenient to deform the hyperhoneycomb lattice. 
First we use that  the angle between the two different zig-zag chains in $\textbf{a}_1$ and $\textbf{a}_2$ direction can be chosen arbitraril. 
Thus to simplify the calculation we set them to be orthogonal, as e.g. done in Ref.~\cite{Ezawa2016}. 
By doubling the unit cell in the $z$-direction, we can furthermore set $\textbf{a}_3$ to be orthogonal to the first two lattice vectors, $\textbf{a}_1$ and $\textbf{a}_2$. 
The price we pay is that there are now eight sites per unit cell.
Solving the tight-binding model on the (deformed) hyperhoneycomb lattice, we find that the BZ is now rectangular and the nodal line is located in the $k_z=0$ plane, as shown on the left of Fig.~\ref{Fig:10b-rectangular}.
This deformation also ensures that upon applying a magnetic field in the $x$-direction --- i.e. $\textbf{B} = B_0 \hat{e}_x$ with vector potential given by $\textbf{A} = B_0y\hat{e}_z$ --- only two of the eight plaquettes per unit cell enclose a magnetic flux. 
These are shown on the right side of Fig.~\ref{Fig:10b-rectangular}, marked in orange and blue. For more details on the plaquettes per unit cell for the hyperhoneycomb lattice, we refer the reader to the discussion in Ref.~\cite{OBrien2016}. 
For the deformed lattice, the lattice vectors become 
\begin{align*}
\textbf{a}_1 &= \bracket{\sqrt{3}, 0, 0},~~~~
\textbf{a}_2 = \bracket{0, \sqrt{3}, 0},\\
\textbf{a}_3 &= \bracket{0, 0, 6},
\end{align*}
and the reciprocal lattice vectors
\begin{align*}
\textbf{b}_1 &= 2\pi \bracket{\frac{1}{\sqrt{3}}, 0, 0},~~~
\textbf{b}_2 = 2\pi \bracket{0, \frac{1}{\sqrt{3}}, 0},\\
\textbf{b}_3 &= 2\pi \bracket{0, 0, \frac{1}{6}}.
\end{align*}
The eight sites within the unit cell are localized at
\begin{align*}  
\textbf{r}_1 &= \bracket{-\frac{\sqrt{3}}{4}, -\frac{\sqrt{3}}{4}, -\frac{11}{4}}, &&
\textbf{r}_2 = \bracket{-\frac{\sqrt{3}}{4}, -\frac{\sqrt{3}}{4}, -\frac{7}{4}}&&\\
\textbf{r}_3 &= \bracket{\frac{\sqrt{3}}{4}, -\frac{\sqrt{3}}{4}, -\frac{5}{4}}, &&
\textbf{r}_4 = \bracket{\frac{\sqrt{3}}{4}, -\frac{\sqrt{3}}{4}, -\frac{1}{4}}&&\\ 
\textbf{r}_5 &= \bracket{\frac{\sqrt{3}}{4}, \frac{\sqrt{3}}{4}, \frac{1}{4}}, &&
\textbf{r}_6 = \bracket{\frac{\sqrt{3}}{4}, \frac{\sqrt{3}}{4}, \frac{5}{4}}&&\\ 
\textbf{r}_7 &= \bracket{-\frac{\sqrt{3}}{4}, \frac{\sqrt{3}}{4}, \frac{7}{4}}, &&
\textbf{r}_8 = \bracket{-\frac{\sqrt{3}}{4}, \frac{\sqrt{3}}{4}, \frac{11}{4}} .
\end{align*}
\subsection{(10,3)c}
The (10,3)c lattice is closely related to the hyperhoneycomb lattice, except that its zig-zag chains run along three different directions that are 120$^\circ$ rotated against each other (see Fig.~\ref{Fig:lattice_illustration}). Its lattice vectors are given by
\begin{align*}
\textbf{a}_1 &= \bracket{1,0,0},~~~~
\textbf{a}_2 = \bracket{-\frac{1}{2}, \frac{\sqrt{3}}{2}, 0},\\
\textbf{a}_3 &= \bracket{0,0,\frac{3\sqrt{3}}{21}},
\end{align*}
and the corresponding reciprocal lattice vectors are
\begin{align*}
\textbf{b}_1 &= 2\pi \bracket{1, \frac{1}{\sqrt{3}}, 0},~~~~
\textbf{b}_2 = 2\pi \bracket{0, \frac{2}{\sqrt{3}}, 0},\\
\textbf{b}_3 &= 2\pi \bracket{0, 0, \frac{2}{3\sqrt{3}}}.
\end{align*}
This lattice has six sites per unit cell, which are located at
\begin{align*}  
\textbf{r}_1 &= \frac{1}{4\sqrt{3}}\bracket{\sqrt{3}, 1, 2},~~~~&&
\textbf{r}_2 = \frac{1}{4\sqrt{3}}\bracket{3\sqrt{3}, 1, 8}\\
\textbf{r}_3 &= \frac{1}{4\sqrt{3}}\bracket{2\sqrt{3}, 4, 14},~~~~&&
\textbf{r}_4 = \frac{1}{4\sqrt{3}}\bracket{3\sqrt{3}, 1, 4}\\
\textbf{r}_5 &= \frac{1}{4\sqrt{3}}\bracket{2\sqrt{3}, 4, 10},~~~~&&
\textbf{r}_6 = \frac{1}{4\sqrt{3}}\bracket{\sqrt{3}, 1, 16}.
\end{align*}

\subsection{(10,3)d}
The inversion-symmetric (10,3)d lattice is closely related to the chiral hyperoctagon (10,3)a lattice, except that the rotation of the `square' spirals is alternating as shown in Fig.~\ref{Fig:lattice_illustration}.
Its lattice vectors can be chosen as
\begin{align*}
\textbf{a}_1 &= \bracket{1, -1, 0},~~~~
\textbf{a}_2 = \bracket{1, 1, 0},\\
\textbf{a}_3 &= \bracket{0, 0, 1}
\end{align*}
with the reciprocal lattice vectors given by
\begin{align*}
\textbf{b}_1 &= \pi \bracket{1, -1, 0},~~~~
\textbf{b}_2 = \pi \bracket{1, 1, 0},\\
\textbf{b}_3 &= \pi \bracket{0, 0,2}.
\end{align*}
The (10,3)d lattice has eight sites per unit cell that are located at
\begin{align*}  
\textbf{r}_1 &= \frac{1}{4}\bracket{a, b, 1},~~~~&&
\textbf{r}_2 = \frac{1}{4}\bracket{0, a+b, 2}\\
\textbf{r}_3 &= \frac{1}{4}\bracket{-a,b,3},~~~~&&
\textbf{r}_4 = \frac{1}{4}\bracket{0, -a+b, 4}\\
\textbf{r}_5 &= \frac{1}{4}\bracket{0, a-b, 3},~~~~&&
\textbf{r}_6 = \frac{1}{4}\bracket{-a, -b, 2}\\
\textbf{r}_7 &= \frac{1}{4}\bracket{0, -a-b, 1},~~~~&&
\textbf{r}_8 = \frac{1}{4}\bracket{a, -b, 4},
\end{align*}
where  we used the same conventions, 
\begin{align*}
a = 4-2\sqrt{2}m\qquad b = 2,
\end{align*}
as  in Ref.~\onlinecite{Yamada2017b}. 

\subsection{(9,3)a}
The (9,3)a lattice is the only example of the lattices at hand that is not bipartite. 
In its most symmetric form, its lattice vectors are given by 
\begin{align*}
\textbf{a}_1 &= \frac{1}{3}\bracket{-\textbf{a}+\textbf{b}+\textbf{c}},~~~~
\textbf{a}_2 = \frac{1}{3}\bracket{-\textbf{a}-2\textbf{b}+\textbf{c}},\\
\textbf{a}_3 &= \frac{1}{3}\bracket{2\textbf{a}+\textbf{b}+\textbf{c}}
\end{align*}
where 
\begin{align*}
\textbf{a} &= (1,0,0),~~
\textbf{b} = \bracket{-\frac{1}{2}, \frac{\sqrt{3}}{2}, 0},~~
\textbf{c} = \bracket{0,0, \alpha}
\end{align*}
with
\begin{align*}
\alpha &= \frac{\sqrt{6\bracket{4+\sqrt{3}}}}{1+2\sqrt{3}},~~
\delta_f = \frac{\sqrt{3}}{1+2\sqrt{3}},~~
\delta_h = \frac{29-3\sqrt{3}}{132}.
\end{align*}
The corresponding reciprocal lattice vectors then become
\begin{align*}
\textbf{b}_1 &= 2\pi \bracket{-1, \frac{1}{\sqrt{3}}, \frac{1}{\alpha}},~~~~
\textbf{b}_2 = 2\pi \bracket{0, -\frac{2}{\sqrt{3}}, \frac{1}{\alpha}},\\
\textbf{b}_3 &= 2\pi \bracket{1, \frac{1}{\sqrt{3}}, \frac{1}{\alpha}}
\end{align*}
It has twelve sites per unit cell which are located at
\begin{align*}
&\textbf{r}_1 = \delta_f \textbf{a},~~~~&&
\textbf{r}_2 = 2\delta_h\textbf{a} + \delta_h\textbf{b} + \frac{1}{12}\textbf{c}\\
&\textbf{r}_3 = \delta_h\bracket{\textbf{a} + \textbf{b}},~~~~&&
\textbf{r}_4 = \delta_h\textbf{a} +2\delta_h\textbf{b} - \frac{1}{12}\textbf{c}\\
&\textbf{r}_5 = \delta_f\textbf{b},~~~~&&
\textbf{r}_6 = -\delta_h\textbf{a} + \delta_h\textbf{b} + \frac{1}{12\textbf{c}}\\
&\textbf{r}_7 = -\delta_f\textbf{a},~~~~&&
\textbf{r}_8 = -2\delta_h\textbf{a} - \delta_h\textbf{b} - \frac{1}{12}\textbf{c}\\
&\textbf{r}_9 = -\delta_f\bracket{\textbf{a} + \textbf{b}},~~~~&&
\textbf{r}_{10} = -\delta_h\textbf{a} - 2\delta_h\textbf{b} + \frac{1}{12}\textbf{c}\\
&\textbf{r}_{11} = -\delta_f\textbf{b},~~~~&&
\textbf{r}_{12} = \delta_h\textbf{a} - \delta_h\textbf{b} - \frac{1}{12}\textbf{c}.
\end{align*}
A simpler, but also less symmetric version of the lattice can be found in Ref.~\onlinecite{OBrien2016}. 

\subsection{(8,3)a}
The (8,3)a lattice is a chiral lattice. 
For all computations shown in this manuscript, we use the version where all triangular spirals are  rotating clockwise
\footnote{Note that the (8,3)a lattice shown in Fig.~\ref{Fig:lattice_illustration} has counter-clockwise rotating spirals.}, as in Ref.~\cite{OBrien2016}. Its lattice vectors are given by
\begin{align*}
\textbf{a}_1 &= (1,0,0),~~~~
\textbf{a}_2 = \left(-\frac{1}{2}, \frac{\sqrt{3}}{2}, 0\right),\\
\textbf{a}_3 &= \left(0,0,\frac{3\sqrt{2}}{5}\right)
\end{align*}
and its reciprocal lattice vectors by
\begin{align*}
\textbf{b}_1 &= 2\pi\left(1, \frac{1}{\sqrt{3}}, 0 \right),~~~~
\textbf{b}_2 = 2\pi\left(0, \frac{2}{\sqrt{3}}, 0 \right),\\
\textbf{b}_3 &= 2\pi\left(0,0,\frac{5\sqrt{2}}{6} \right).
\end{align*}
The lattice has six sites per unit cell at positions
\begin{align*}
\textbf{r}_1 &= \left(\frac{1}{2}, \frac{\sqrt{3}}{10}, 0 \right),~~~~&&
\textbf{r}_2 = \left(\frac{3}{5} , \frac{\sqrt{3}}{5},  \frac{2\sqrt{2}}{5}\right)\\
\textbf{r}_3 &= \left(\frac{1}{10} , \frac{3\sqrt{3}}{10},\frac{\sqrt{2}}{5} \right),~~&&
\textbf{r}_4 = \left(\frac{2}{5} , \frac{\sqrt{3}}{5},  \frac{\sqrt{2}}{5} \right)\\
\textbf{r}_5 &= \left(0, \frac{2\sqrt{3}}{5}, 0 \right),~~~~&&
\textbf{r}_6 = \left(-\frac{1}{10} , \frac{3\sqrt{3}}{10},\frac{2\sqrt{2}}{5} \right).
\end{align*}

\subsection{(8,3)b}
The (8,3)b lattice is closely related to the (8,3)a lattice, except that it is inversion symmetric with alternating rotation directions of the triangular spirals, see Fig.~\ref{Fig:lattice_illustration}.
Its lattice translation vectors are given by
\begin{align*}
\textbf{a}_1 &= \left(\frac{1}{2}, \frac{1}{2\sqrt{3}}, \frac{\sqrt{2}}{5\sqrt{3}} \right),~~~~
\textbf{a}_2 = \left(0, \frac{1}{\sqrt{3}}, \frac{2\sqrt{2}}{5\sqrt{3}} \right),\\
\textbf{a}_3 &= \left(0, 0, \frac{\sqrt{6}}{5} \right),
\end{align*}
and the corresponding reciprocal lattice vectors are
\begin{align*}
\textbf{b}_1 &= 2\pi\left(2, 0, 0 \right),~~~~
\textbf{b}_2 = 2\pi\left(-1, \sqrt{3}, 0 \right),\\
\textbf{b}_3 &= 2\pi\left(0, -\frac{2}{\sqrt{3}}, \frac{5}{\sqrt{6}} \right).
\end{align*}
It also has six sites per unit cell at positions
\begin{align*}
\textbf{r}_1 &= \left(\frac{1}{10}, \frac{1}{2\sqrt{3}}, \frac{\sqrt{2}}{5\sqrt{3}}\right),~~~~&&
\textbf{r}_2 = \left(\frac{1}{5}, \frac{\sqrt{3}}{5}, \frac{\sqrt{6}}{5} \right)\\
\textbf{r}_3 &= \left(\frac{3}{10}, \frac{11}{10\sqrt{3}}, \frac{4\sqrt{2}}{5\sqrt{3}} \right),~~~~&&
\textbf{r}_4 = \left(\frac{1}{5}, \frac{2}{5\sqrt{3}}, \frac{2\sqrt{2}}{5\sqrt{3}} \right)\\
\textbf{r}_5 &= \left(\frac{3}{10}, \frac{3\sqrt{3}}{10}, \frac{\sqrt{6}}{5} \right),~~~~&&
\textbf{r}_6 = \left(\frac{2}{5}, \frac{1}{\sqrt{3}}, \sqrt{\frac{2}{3}} \right).
\end{align*}

\subsection{(8,3)c}
The lattice vectors of the inversion-symmetric (8,3)c lattice are chosen as
\begin{align*}
\textbf{a}_1 &= \left(1, 0, 0 \right),~~~~
\textbf{a}_2 = \left(-\frac{1}{2}, \frac{\sqrt{3}}{2}, 0 \right),\\
\textbf{a}_3 &= \left(0, 0, \frac{2}{5} \right)
\end{align*}
and the reciprocal lattice vectors become
\begin{align*}
\textbf{b}_1 &= 2\pi\left(1, \frac{1}{\sqrt{3}}, 0 \right),~~~~
\textbf{b}_2 = 2\pi\left(0, \frac{2}{\sqrt{3}}, 0 \right),\\
\textbf{b}_3 &= 2\pi\left(0, 0, \frac{5}{2} \right).
\end{align*}
It has eight sites per unit cell that are located at
\begin{align*}
\textbf{r}_1 &= \left(-\frac{1}{5}, \frac{4}{5\sqrt{3}}, \frac{1}{10} \right),~~~~&&
\textbf{r}_2 = \left(0, \frac{1}{\sqrt{3}}, \frac{1}{10} \right)\\
\textbf{r}_3 &= \left(\frac{3}{10}, \frac{7}{10\sqrt{3}}, \frac{3}{10} \right),~~~~&&
\textbf{r}_4 = \left(\frac{1}{2}, \frac{1}{10\sqrt{3}}, \frac{3}{10} \right)\\
\textbf{r}_5 &= \left(0, \frac{7}{5\sqrt{3}}, \frac{1}{10} \right),~~~~&&
\textbf{r}_6 = \left(\frac{1}{5}, \frac{4}{5\sqrt{3}}, \frac{1}{10} \right)\\
\textbf{r}_7 &= \left(\frac{1}{2}, \frac{1}{2\sqrt{3}}, \frac{3}{10} \right),~~~~&&
\textbf{r}_8 = \left(\frac{7}{10}, \frac{7}{10\sqrt{3}}, \frac{3}{10} \right).
\end{align*}

\subsection{(8,3)n}
The (8,3)n lattice is again an inversion-symmetric lattice. 
Using 
\begin{align*}
\textbf{a} &= (1,0,0),~~~~
\textbf{b} = (0,1,0),~~~~
\textbf{c} = \left( 0,0, \alpha \right)\\
\alpha &= \frac{4}{2\sqrt{3}+\sqrt{2}},~~~~
x = \frac{\sqrt{3}+\sqrt{2}}{2\left(2\sqrt{3}+\sqrt{2}\right)},~~~~
z = \frac{1}{8},
\end{align*}
we can express the lattice translation vectors as 
\begin{align*}
\textbf{a}_1 &= \textbf{a},~~~~
\textbf{a}_2 = \textbf{b},~~~~
\textbf{a}_3 = \frac{1}{2} \left( \textbf{a}+\textbf{b}+\textbf{c} \right)
\end{align*}
while the corresponding reciprocal lattice vectors become
\begin{align*}
\textbf{b}_1 &= 2\pi\left( 1, 0, -\frac{1}{\alpha} \right),~~~~
\textbf{b}_2 = 2\pi\left( 0, 1, -\frac{1}{\alpha} \right),\\
\textbf{b}_3 &= 2\pi\left( 0, 0, \frac{2}{\alpha}\right).
\end{align*}
This lattice has a rather large unit cell consisting of  16 sites that are located at 
\allowdisplaybreaks
\begin{align*}
&\textbf{r}_1 = x\textbf{a} + \bracket{\frac{1}{2}-x}\textbf{b} + \frac{1}{4}\textbf{c}\\
&\textbf{r}_2 = \bracket{1-x}\textbf{a} + \bracket{\frac{1}{2}-x}\textbf{b} + \frac{1}{4}\textbf{c}\\
&\textbf{r}_3 = \bracket{\frac{1}{2}+x}\textbf{a} + \frac{1}{2}\textbf{b} + \bracket{\frac{1}{2}-z}\textbf{c}\\
&\textbf{r}_4 = \bracket{1-x}\textbf{a} + \bracket{\frac{1}{2}+x}\textbf{b} + \frac{1}{4}\textbf{c}\\
&\textbf{r}_5 = x\textbf{a} + \bracket{\frac{1}{2}+x}\textbf{b} + \frac{1}{4}\textbf{c}\\
&\textbf{r}_6 = \bracket{\frac{1}{2}-x}\textbf{a} + \frac{1}{2}\textbf{b} + \bracket{\frac{1}{2}-z}\textbf{c}\\
&\textbf{r}_7 = \bracket{1-x}\textbf{b} + z\textbf{c}\\
&\textbf{r}_8 = x\textbf{b} + z\textbf{c}\\
&\textbf{r}_9 = \bracket{\frac{1}{2}-x}\textbf{a} + x\textbf{b} + \frac{1}{4}\textbf{c}\\
&\textbf{r}_{10} = \frac{1}{2}\textbf{a} + \bracket{\frac{1}{2}-x}\textbf{b} + \bracket{\frac{1}{2}-z}\textbf{c}\\
&\textbf{r}_{11} = \bracket{\frac{1}{2}+x}\textbf{a} + x\textbf{b} + \frac{1}{4}\textbf{c}\\
&\textbf{r}_{12} = \bracket{\frac{1}{2}+x}\textbf{a} + \bracket{1-x}\textbf{b} + \frac{1}{4}\textbf{c}\\
&\textbf{r}_{13} = \frac{1}{2}\textbf{a} + \bracket{\frac{1}{2}+x}\textbf{b} + \bracket{\frac{1}{2}-z}\textbf{c}\\
&\textbf{r}_{14} = \bracket{\frac{1}{2}-x}\textbf{a} + \bracket{1-x}\textbf{b} + \frac{1}{4}\textbf{c}\\
&\textbf{r}_{15} = x\textbf{a} + z\textbf{c}\\
&\textbf{r}_{16} = \bracket{1-x}\textbf{a} + z\textbf{c}.
\end{align*}
It has a four-fold rotation symmetry around the $z$-axis, as well as several mirror symmetries. 


\section{Topological insulator classification}
\label{app:TI}
In three spatial dimensions, topological insulators (TIs) can be classified \cite{Fu2007} by   four $\mathbb{Z}_2$ topological indices 
$\nu_0;(\nu_1 \nu_2 \nu_3)$. 
The first index $\nu_0$ differentiates between strong ($\nu_0=1$) and weak ($\nu_0=0$) topological insulators. 
Technically, it states whether the TI exhibits an odd (strong) or even (weak) number of Dirac cones on the surface. 
As such, a strong TI enjoys a higher level of topological protection, since in contrast to their weak counterparts their surface Dirac cones cannot be  gapped out by simply merging pairs of Dirac cones.
In more general terms, a strong TI is an intrinsically three-dimensional topological state, while a weak TI can be seen as layered composite of two-dimensional topological states (such as the quantum spin Hall states) \cite{Moore2007,Fu2007,Roy2009}.
The remaining indices $(\nu_1 \nu_2 \nu_3)$ are the so-called weak indices, which further specify the positions of the Dirac cones on the various surfaces of the TI.

Topologically protected Dirac cones on the surface of a three-dimensional TI cannot appear at arbitrary momenta in the surface Brillouin zone, but are fixed to certain high-symmetry points \cite{Moore2007,Fu2007,Roy2009}, i.e.~the set of time-reversal invariant momenta (TRIMs).
 These TRIMs can be found at linear combinations of half reciprocal lattice vectors $\vec{b}_i$ , i.e. 
\[
   \Lambda_{n_1n_2n_3} = \frac{1}{2} \sum n_i \vec{b}_i \quad {\rm with} \quad n_i \in \{0,1\} \,.
\]
The relevance of the TRIMs here is that they can be used to calculate the topological indices $\nu_0;(\nu_1 \nu_2 \nu_3)$ mentioned above.
Specifically, we have 
\begin{align*}
(-1)^{\nu_0} &= \prod_{n_1,n_2,n_3 = 0}^1 \delta_{n_1n_2n_3}\\
(-1)^{\nu_i} &= \prod_{\substack{n_{j\neq i} = 0,1\\n_i = 1}} \delta_{n_1n_2n_3}
\end{align*}
where $\delta = \pm 1$ is  the time-reversal polarization (TRP)   at the TRIM defined by $n_1$, $n_2$, $n_3$. 
To determine the indices, it is in fact not necessary  to calculate this time-reversal polarization explicitly, but it suffices to determine its change between two TRIMs. 
Following Ref.~\onlinecite{Fu2007}, the simplest way to do so is to count the number of surface bands crossing the Fermi energy between two TRIMs. 
If this number is odd, the corresponding  TRP changes sign, otherwise not.
To determine the topological indices, it then suffices to calculate the TRP change for a fixed path between two TRIMs to obtain the weak topological indices and to calculate the product of TRP changes on opposite paths to obtain the strong index.

In the following, we document the determination of topological indices for the topological insulators found in the various 3D honeycomb systems at the heart of the current study. 
To do so, we provide an illustration of the surface energy spectra along a high-symmetry path connecting the TRIMs and indicate the change of the TRP along this path. 
The surface bands are marked in red or yellow, depending on whether they are located at the upper or lower surface. 
If the bands are degenerated they are marked in orange.

\begin{figure}
	\includegraphics[width=.85\columnwidth]{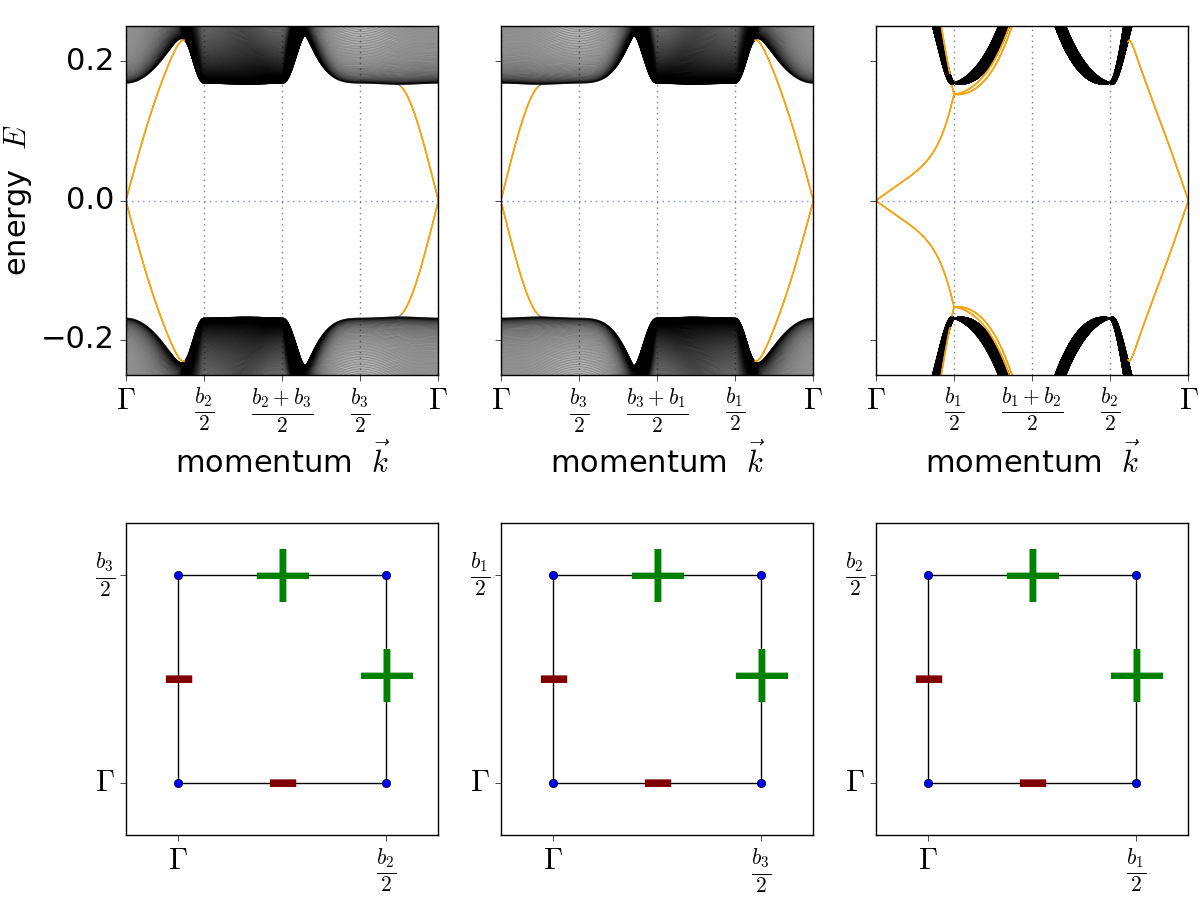}
	\caption{Surface energy spectra along high-symmetry paths for the 
	strong topological insulator $1;(000)$ for {\bf (10,3)b} at the isotropic point $t_x=t_y=t_z=1/3$ with $\lambda = 0.1$.}
	\label{Fig:10b-TI}
\end{figure}
\begin{figure}
	\includegraphics[width=.85\columnwidth]{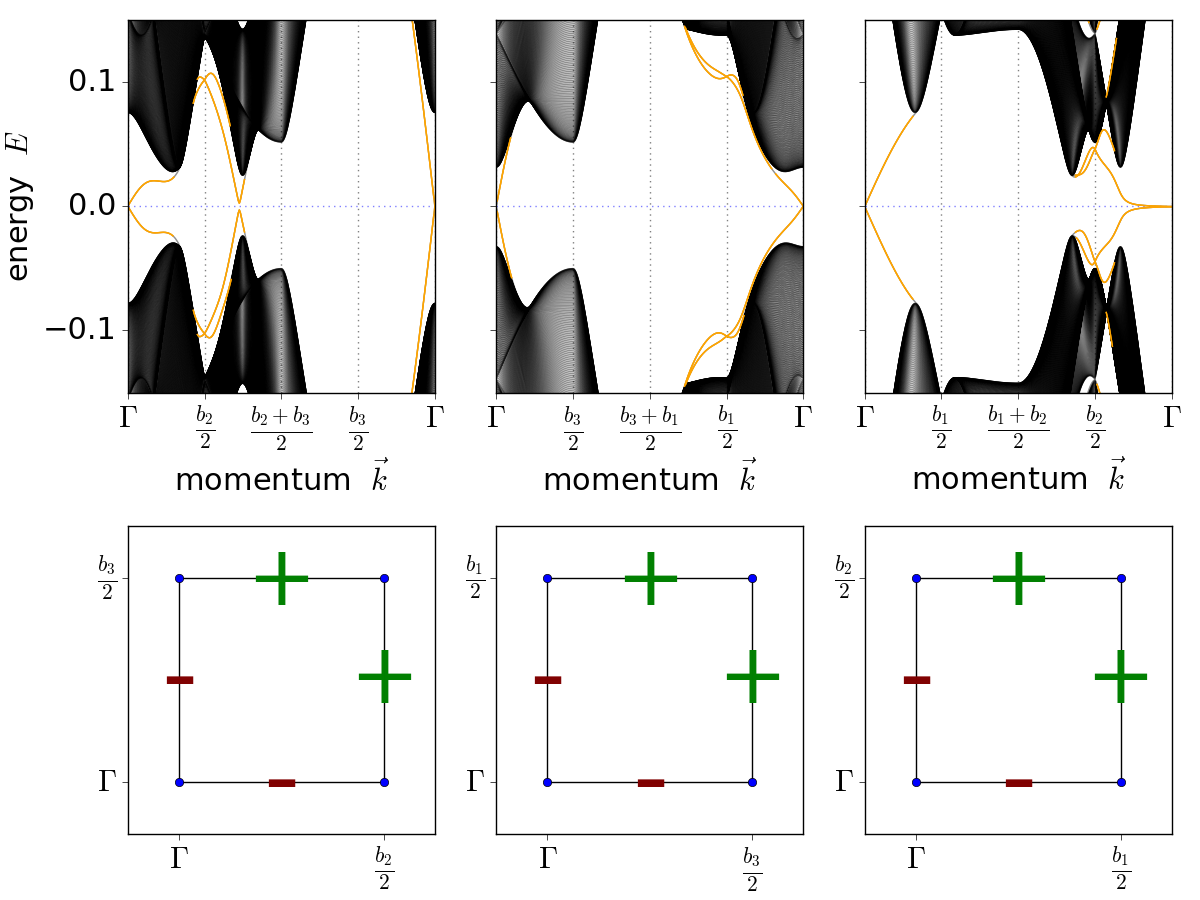}
	\caption{Surface energy spectra along high-symmetry paths for the 
	strong topological insulator 1;(000) for {\bf (10,3)d} at the isotropic point with $\lambda = 0.03$.}
	\label{Fig:10d-TI}
\end{figure}
\begin{figure}
	\includegraphics[width=.85\columnwidth]{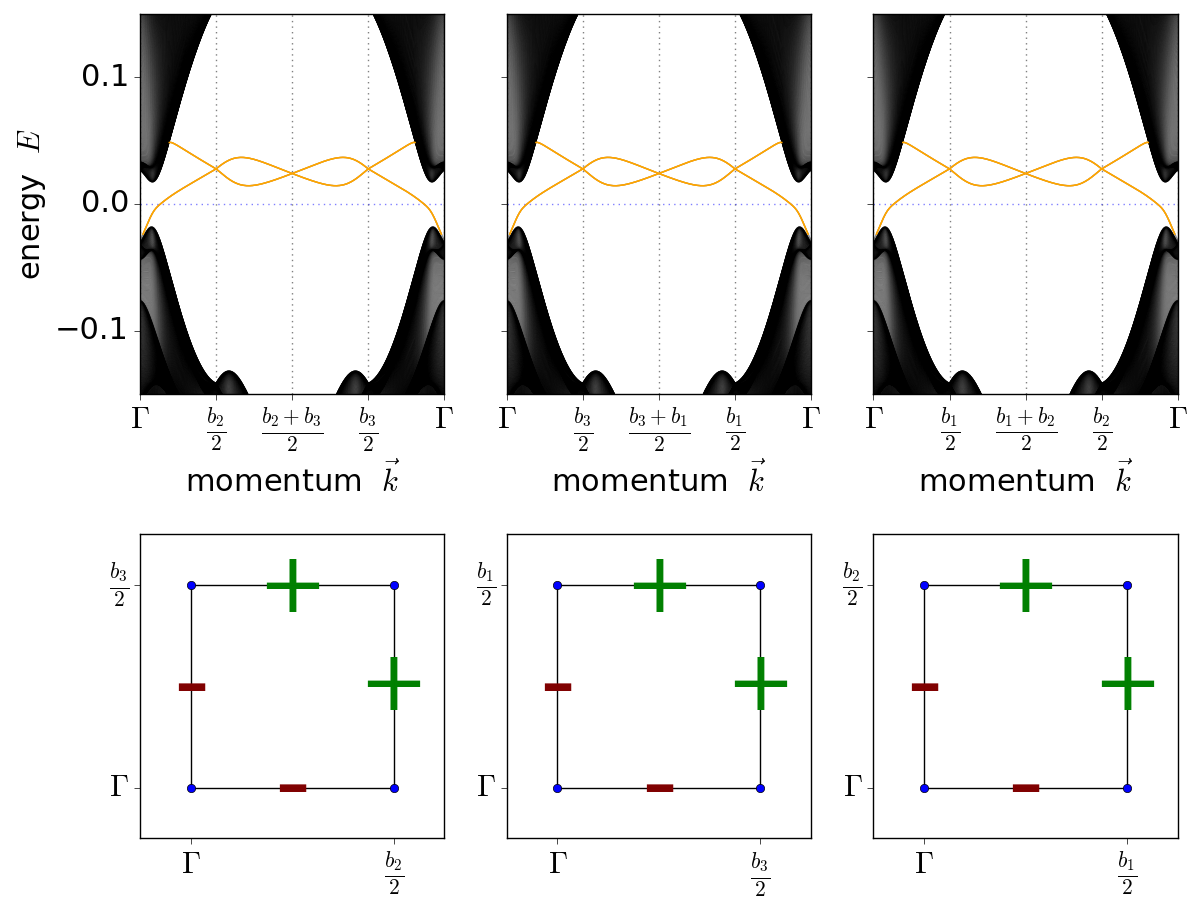}
	\caption{Surface energy spectra along high-symmetry paths for the 
	strong topological insulator 1;(000) for {\bf (9,3)a} at $t_z=0.475$, $t_x = t_y = 0.2625$ and $\lambda = 0.03$.}
	\label{Fig:9a-TI}
\end{figure}
\begin{figure}
	\includegraphics[width=.85\columnwidth]{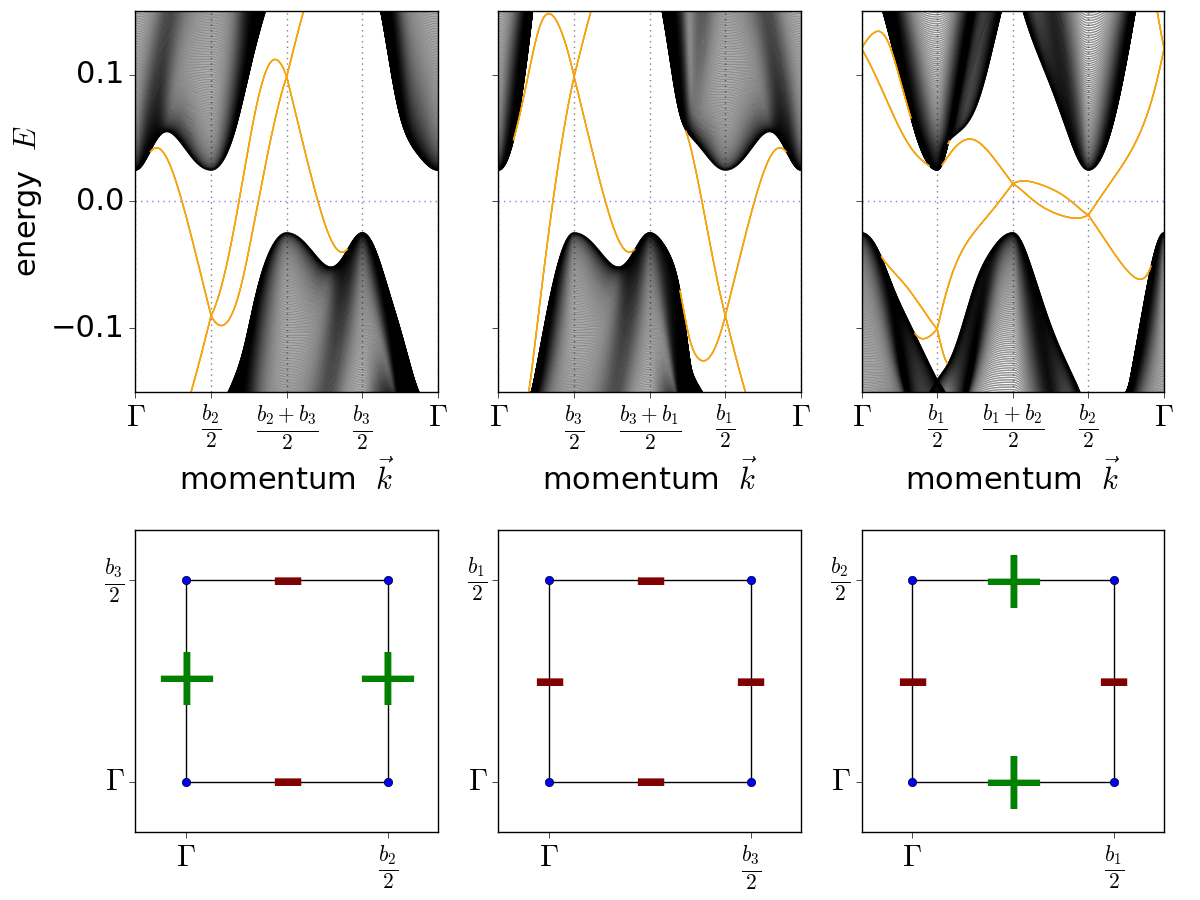}
	\caption{Surface energy spectra along high-symmetry paths for the 
	weak topological insulator 0;(101) for {\bf (8,3)b} at $t_z=0.2$, $t_x = t_y = 0.4$ and $\lambda = 0.1$.}
	\label{Fig:8b-TI}
\end{figure}
\begin{figure}
	\includegraphics[width=.85\columnwidth]{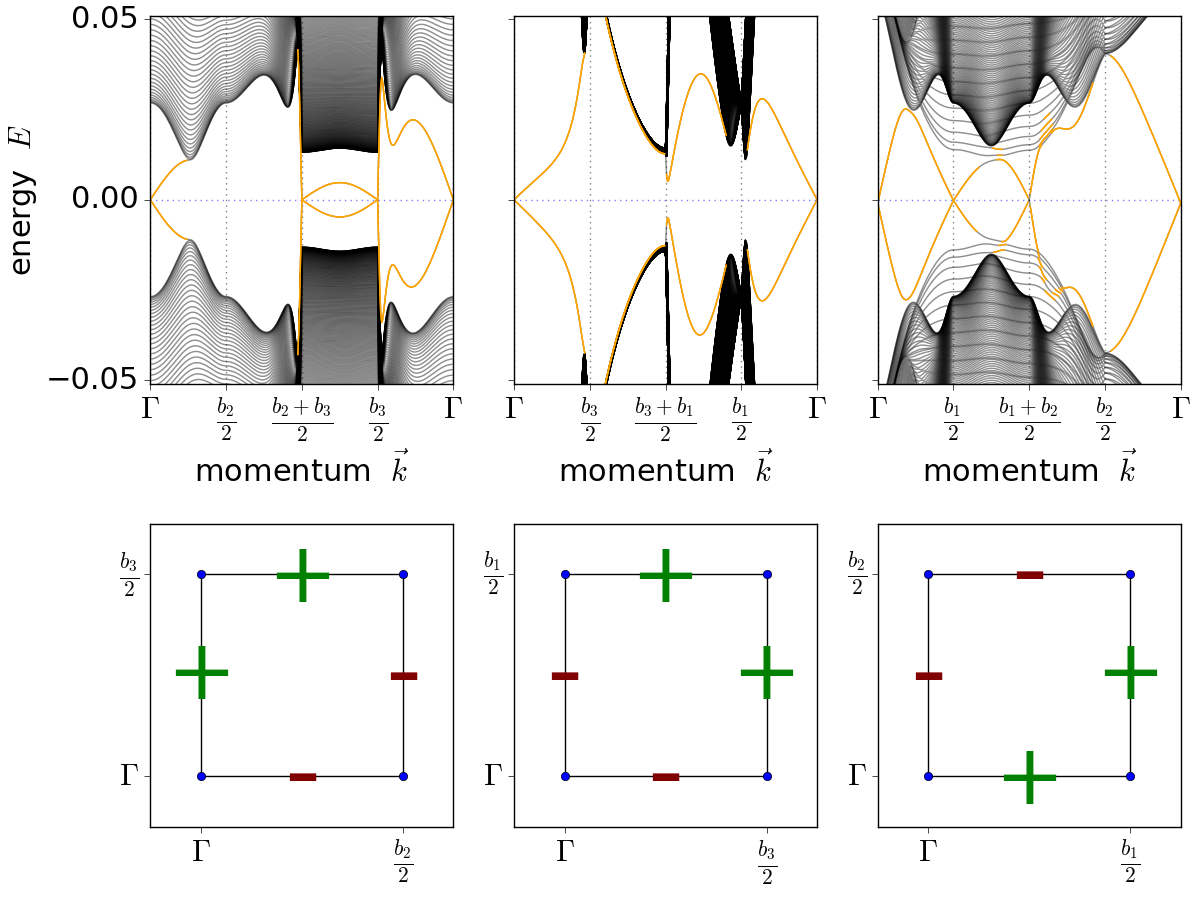}
	\caption{Surface energy spectra along high-symmetry paths for the 
	strong topological insulator 1;(010) for {\bf (8,3)c} at $t_z=0.2$, $t_x = t_y = 0.4$ and $\lambda = 0.03$.}
	\label{Fig:8c-TI}
\end{figure}
\begin{figure}
	\includegraphics[width=.85\columnwidth]{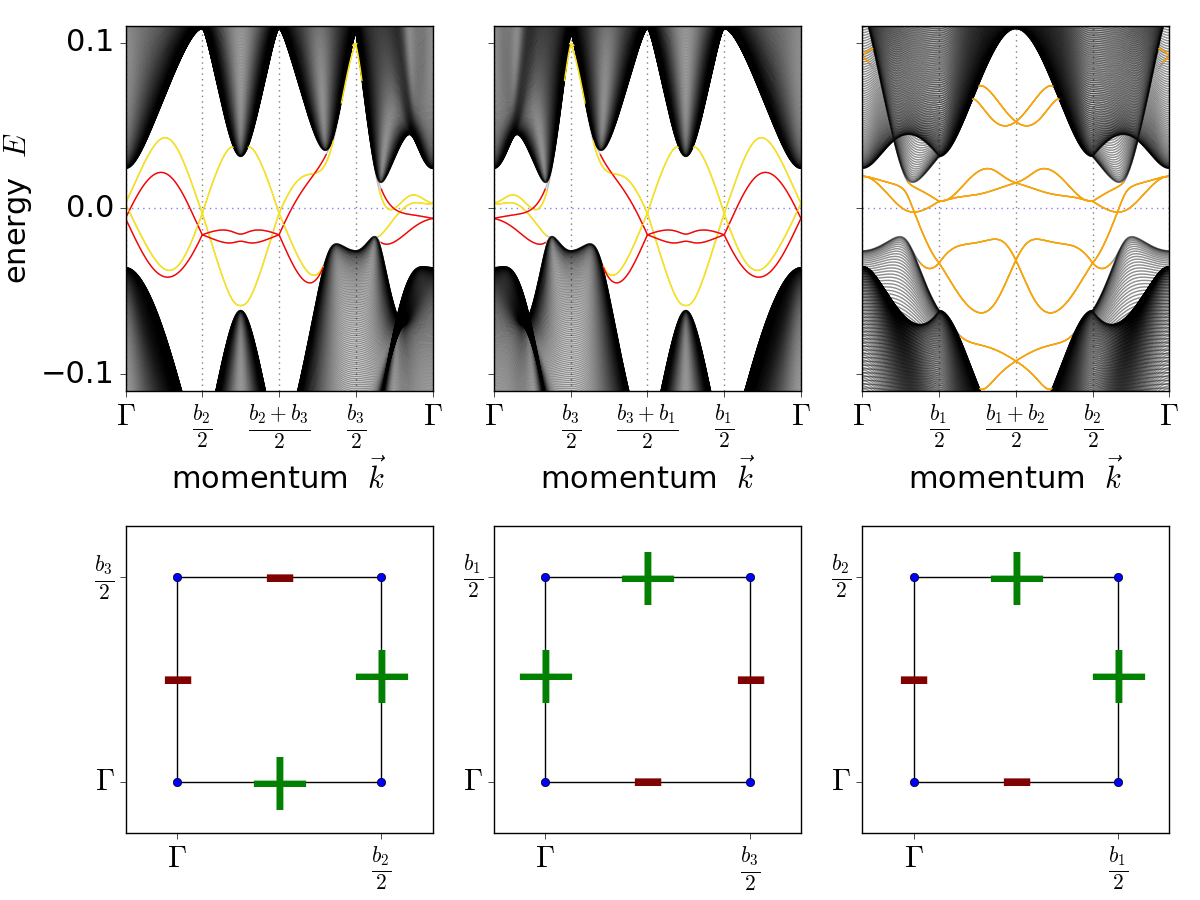}
	\caption{Surface energy spectra along high-symmetry paths for the 
	strong topological insulator 1;(001) for {\bf (8,3)n} at $t_z=0.51$, $t_x = t_y = 0.245$ and $\lambda = 0.1$.}
	\label{Fig:8n-TI}
\end{figure}
%

\newpage

\bibliography{Dirac}

\begin{thebibliography}{58}%
\makeatletter
\providecommand \@ifxundefined [1]{%
 \@ifx{#1\undefined}
}%
\providecommand \@ifnum [1]{%
 \ifnum #1\expandafter \@firstoftwo
 \else \expandafter \@secondoftwo
 \fi
}%
\providecommand \@ifx [1]{%
 \ifx #1\expandafter \@firstoftwo
 \else \expandafter \@secondoftwo
 \fi
}%
\providecommand \natexlab [1]{#1}%
\providecommand \enquote  [1]{``#1''}%
\providecommand \bibnamefont  [1]{#1}%
\providecommand \bibfnamefont [1]{#1}%
\providecommand \citenamefont [1]{#1}%
\providecommand \href@noop [0]{\@secondoftwo}%
\providecommand \href [0]{\begingroup \@sanitize@url \@href}%
\providecommand \@href[1]{\@@startlink{#1}\@@href}%
\providecommand \@@href[1]{\endgroup#1\@@endlink}%
\providecommand \@sanitize@url [0]{\catcode `\\12\catcode `\$12\catcode
  `\&12\catcode `\#12\catcode `\^12\catcode `\_12\catcode `\%12\relax}%
\providecommand \@@startlink[1]{}%
\providecommand \@@endlink[0]{}%
\providecommand \url  [0]{\begingroup\@sanitize@url \@url }%
\providecommand \@url [1]{\endgroup\@href {#1}{\urlprefix }}%
\providecommand \urlprefix  [0]{URL }%
\providecommand \Eprint [0]{\href }%
\providecommand \doibase [0]{http://dx.doi.org/}%
\providecommand \selectlanguage [0]{\@gobble}%
\providecommand \bibinfo  [0]{\@secondoftwo}%
\providecommand \bibfield  [0]{\@secondoftwo}%
\providecommand \translation [1]{[#1]}%
\providecommand \BibitemOpen [0]{}%
\providecommand \bibitemStop [0]{}%
\providecommand \bibitemNoStop [0]{.\EOS\space}%
\providecommand \EOS [0]{\spacefactor3000\relax}%
\providecommand \BibitemShut  [1]{\csname bibitem#1\endcsname}%
\let\auto@bib@innerbib\@empty
\bibitem [{\citenamefont {Herring}(1937)}]{Herring1937}%
  \BibitemOpen
  \bibfield  {author} {\bibinfo {author} {\bibfnamefont {C.}~\bibnamefont
  {Herring}},\ }\bibfield  {title} {{\color{Gray}\small \bibinfo {title}
  {{Accidental Degeneracy in the Energy Bands of Crystals}},\ }}\href {\doibase
  10.1103/PhysRev.52.365} {\bibfield  {journal} {\bibinfo  {journal} {Phys.
  Rev.}\ }\textbf {\bibinfo {volume} {52}},\ \bibinfo {pages} {365} (\bibinfo
  {year} {1937})}\BibitemShut {NoStop}%
\bibitem [{\citenamefont {Volovik}(2003)}]{Volovik2003}%
  \BibitemOpen
  \bibfield  {author} {\bibinfo {author} {\bibfnamefont {G.~E.}\ \bibnamefont
  {Volovik}},\ }\href@noop {} {\emph {\bibinfo {title} {{The Universe in a
  Helium Droplet}}}}\ (\bibinfo  {publisher} {Clarendon,Oxford},\ \bibinfo
  {year} {2003})\BibitemShut {NoStop}%
\bibitem [{\citenamefont {Volovik}(2007)}]{Volovik2007}%
  \BibitemOpen
  \bibfield  {author} {\bibinfo {author} {\bibfnamefont {G.~E.}\ \bibnamefont
  {Volovik}},\ }\bibinfo {title} {Quantum phase transitions from topology in
  momentum space},\ in\ \href {\doibase 10.1007/3-540-70859-6_3} {\emph
  {\bibinfo {booktitle} {Quantum Analogues: From Phase Transitions to Black
  Holes and Cosmology}}},\ \bibinfo {editor} {edited by\ \bibinfo {editor}
  {\bibfnamefont {W.~G.}\ \bibnamefont {Unruh}}\ and\ \bibinfo {editor}
  {\bibfnamefont {R.}~\bibnamefont {Sch{\"u}tzhold}}}\ (\bibinfo  {publisher}
  {Springer Berlin Heidelberg},\ \bibinfo {address} {Berlin, Heidelberg},\
  \bibinfo {year} {2007})\ pp.\ \bibinfo {pages} {31--73}\BibitemShut {NoStop}%
\bibitem [{\citenamefont {Wan}\ \emph {et~al.}(2011)\citenamefont {Wan},
  \citenamefont {Turner}, \citenamefont {Vishwanath},\ and\ \citenamefont
  {Savrasov}}]{Wan2011}%
  \BibitemOpen
  \bibfield  {author} {\bibinfo {author} {\bibfnamefont {X.}~\bibnamefont
  {Wan}}, \bibinfo {author} {\bibfnamefont {A.~M.}\ \bibnamefont {Turner}},
  \bibinfo {author} {\bibfnamefont {A.}~\bibnamefont {Vishwanath}}, \ and\
  \bibinfo {author} {\bibfnamefont {S.~Y.}\ \bibnamefont {Savrasov}},\
  }\bibfield  {title} {{\color{Gray}\small \bibinfo {title} {{Topological
  semimetal and Fermi-arc surface states in the electronic structure of
  pyrochlore iridates}},\ }}\href {\doibase 10.1103/PhysRevB.83.205101}
  {\bibfield  {journal} {\bibinfo  {journal} {Phys. Rev. B}\ }\textbf {\bibinfo
  {volume} {83}},\ \bibinfo {pages} {205101} (\bibinfo {year}
  {2011})}\BibitemShut {NoStop}%
\bibitem [{\citenamefont {Armitage}\ \emph {et~al.}(2018)\citenamefont
  {Armitage}, \citenamefont {Mele},\ and\ \citenamefont
  {Vishwanath}}]{Armitage2018}%
  \BibitemOpen
  \bibfield  {author} {\bibinfo {author} {\bibfnamefont {N.~P.}\ \bibnamefont
  {Armitage}}, \bibinfo {author} {\bibfnamefont {E.~J.}\ \bibnamefont {Mele}},
  \ and\ \bibinfo {author} {\bibfnamefont {A.}~\bibnamefont {Vishwanath}},\
  }\bibfield  {title} {{\color{Gray}\small \bibinfo {title} {{Weyl and Dirac
  semimetals in three-dimensional solids}},\ }}\href {\doibase
  10.1103/RevModPhys.90.015001} {\bibfield  {journal} {\bibinfo  {journal}
  {Rev. Mod. Phys.}\ }\textbf {\bibinfo {volume} {90}},\ \bibinfo {pages}
  {015001} (\bibinfo {year} {2018})}\BibitemShut {NoStop}%
\bibitem [{\citenamefont {Burkov}\ \emph {et~al.}(2011)\citenamefont {Burkov},
  \citenamefont {Hook},\ and\ \citenamefont {Balents}}]{Burkov2011}%
  \BibitemOpen
  \bibfield  {author} {\bibinfo {author} {\bibfnamefont {A.~A.}\ \bibnamefont
  {Burkov}}, \bibinfo {author} {\bibfnamefont {M.~D.}\ \bibnamefont {Hook}}, \
  and\ \bibinfo {author} {\bibfnamefont {L.}~\bibnamefont {Balents}},\
  }\bibfield  {title} {{\color{Gray}\small \bibinfo {title} {Topological nodal
  semimetals},\ }}\href {\doibase 10.1103/PhysRevB.84.235126} {\bibfield
  {journal} {\bibinfo  {journal} {Phys. Rev. B}\ }\textbf {\bibinfo {volume}
  {84}},\ \bibinfo {pages} {235126} (\bibinfo {year} {2011})}\BibitemShut
  {NoStop}%
\bibitem [{\citenamefont {Phillips}\ and\ \citenamefont
  {Aji}(2014)}]{Phillips2014}%
  \BibitemOpen
  \bibfield  {author} {\bibinfo {author} {\bibfnamefont {M.}~\bibnamefont
  {Phillips}}\ and\ \bibinfo {author} {\bibfnamefont {V.}~\bibnamefont {Aji}},\
  }\bibfield  {title} {{\color{Gray}\small \bibinfo {title} {Tunable line node
  semimetals},\ }}\href {\doibase 10.1103/PhysRevB.90.115111} {\bibfield
  {journal} {\bibinfo  {journal} {Phys. Rev. B}\ }\textbf {\bibinfo {volume}
  {90}},\ \bibinfo {pages} {115111} (\bibinfo {year} {2014})}\BibitemShut
  {NoStop}%
\bibitem [{\citenamefont {Chiu}\ and\ \citenamefont
  {Schnyder}(2014)}]{Chiu2014}%
  \BibitemOpen
  \bibfield  {author} {\bibinfo {author} {\bibfnamefont {C.-K.}\ \bibnamefont
  {Chiu}}\ and\ \bibinfo {author} {\bibfnamefont {A.~P.}\ \bibnamefont
  {Schnyder}},\ }\bibfield  {title} {{\color{Gray}\small \bibinfo {title}
  {Classification of reflection-symmetry-protected topological semimetals and
  nodal superconductors},\ }}\href {\doibase 10.1103/PhysRevB.90.205136}
  {\bibfield  {journal} {\bibinfo  {journal} {Phys. Rev. B}\ }\textbf {\bibinfo
  {volume} {90}},\ \bibinfo {pages} {205136} (\bibinfo {year}
  {2014})}\BibitemShut {NoStop}%
\bibitem [{\citenamefont {Bzdu{\v s}ek}\ \emph {et~al.}(2016)\citenamefont
  {Bzdu{\v s}ek}, \citenamefont {Wu}, \citenamefont {R{\"u}egg}, \citenamefont
  {Sigrist},\ and\ \citenamefont {Soluyanov}}]{Bzdusek2016}%
  \BibitemOpen
  \bibfield  {author} {\bibinfo {author} {\bibfnamefont {T.}~\bibnamefont
  {Bzdu{\v s}ek}}, \bibinfo {author} {\bibfnamefont {Q.}~\bibnamefont {Wu}},
  \bibinfo {author} {\bibfnamefont {A.}~\bibnamefont {R{\"u}egg}}, \bibinfo
  {author} {\bibfnamefont {M.}~\bibnamefont {Sigrist}}, \ and\ \bibinfo
  {author} {\bibfnamefont {A.~A.}\ \bibnamefont {Soluyanov}},\ }\bibfield
  {title} {{\color{Gray}\small \bibinfo {title} {Nodal-chain metals},\ }}\href
  {http://dx.doi.org/10.1038/nature19099} {\bibfield  {journal} {\bibinfo
  {journal} {Nature}\ }\textbf {\bibinfo {volume} {538}},\ \bibinfo {pages}
  {75} (\bibinfo {year} {2016})}\BibitemShut {NoStop}%
\bibitem [{\citenamefont {Rhim}\ and\ \citenamefont {Kim}(2015)}]{Rhim2015}%
  \BibitemOpen
  \bibfield  {author} {\bibinfo {author} {\bibfnamefont {J.-W.}\ \bibnamefont
  {Rhim}}\ and\ \bibinfo {author} {\bibfnamefont {Y.~B.}\ \bibnamefont {Kim}},\
  }\bibfield  {title} {{\color{Gray}\small \bibinfo {title} {Landau level
  quantization and almost flat modes in three-dimensional semimetals with nodal
  ring spectra},\ }}\href {\doibase 10.1103/PhysRevB.92.045126} {\bibfield
  {journal} {\bibinfo  {journal} {Phys. Rev. B}\ }\textbf {\bibinfo {volume}
  {92}},\ \bibinfo {pages} {045126} (\bibinfo {year} {2015})}\BibitemShut
  {NoStop}%
\bibitem [{\citenamefont {Kane}\ and\ \citenamefont {Mele}(2005)}]{Kane2005}%
  \BibitemOpen
  \bibfield  {author} {\bibinfo {author} {\bibfnamefont {C.~L.}\ \bibnamefont
  {Kane}}\ and\ \bibinfo {author} {\bibfnamefont {E.~J.}\ \bibnamefont
  {Mele}},\ }\bibfield  {title} {{\color{Gray}\small \bibinfo {title}
  {{${Z}_{2}$ Topological Order and the Quantum Spin Hall Effect}},\ }}\href
  {\doibase 10.1103/PhysRevLett.95.146802} {\bibfield  {journal} {\bibinfo
  {journal} {Phys. Rev. Lett.}\ }\textbf {\bibinfo {volume} {95}},\ \bibinfo
  {pages} {146802} (\bibinfo {year} {2005})}\BibitemShut {NoStop}%
\bibitem [{\citenamefont {Bernevig}\ \emph {et~al.}(2006)\citenamefont
  {Bernevig}, \citenamefont {Hughes},\ and\ \citenamefont
  {Zhang}}]{Bernevig2006}%
  \BibitemOpen
  \bibfield  {author} {\bibinfo {author} {\bibfnamefont {B.~A.}\ \bibnamefont
  {Bernevig}}, \bibinfo {author} {\bibfnamefont {T.~L.}\ \bibnamefont
  {Hughes}}, \ and\ \bibinfo {author} {\bibfnamefont {S.-C.}\ \bibnamefont
  {Zhang}},\ }\bibfield  {title} {{\color{Gray}\small \bibinfo {title}
  {{Quantum Spin Hall Effect and Topological Phase Transition in HgTe Quantum
  Wells}},\ }}\href {\doibase 10.1126/science.1133734} {\bibfield  {journal}
  {\bibinfo  {journal} {Science}\ }\textbf {\bibinfo {volume} {314}},\ \bibinfo
  {pages} {1757} (\bibinfo {year} {2006})}\BibitemShut {NoStop}%
\bibitem [{\citenamefont {Moore}\ and\ \citenamefont
  {Balents}(2007)}]{Moore2007}%
  \BibitemOpen
  \bibfield  {author} {\bibinfo {author} {\bibfnamefont {J.~E.}\ \bibnamefont
  {Moore}}\ and\ \bibinfo {author} {\bibfnamefont {L.}~\bibnamefont
  {Balents}},\ }\bibfield  {title} {{\color{Gray}\small \bibinfo {title}
  {Topological invariants of time-reversal-invariant band structures},\ }}\href
  {\doibase 10.1103/PhysRevB.75.121306} {\bibfield  {journal} {\bibinfo
  {journal} {Phys. Rev. B}\ }\textbf {\bibinfo {volume} {75}},\ \bibinfo
  {pages} {121306} (\bibinfo {year} {2007})}\BibitemShut {NoStop}%
\bibitem [{\citenamefont {Fu}\ \emph {et~al.}(2007)\citenamefont {Fu},
  \citenamefont {Kane},\ and\ \citenamefont {Mele}}]{Fu2007}%
  \BibitemOpen
  \bibfield  {author} {\bibinfo {author} {\bibfnamefont {L.}~\bibnamefont
  {Fu}}, \bibinfo {author} {\bibfnamefont {C.~L.}\ \bibnamefont {Kane}}, \ and\
  \bibinfo {author} {\bibfnamefont {E.~J.}\ \bibnamefont {Mele}},\ }\bibfield
  {title} {{\color{Gray}\small \bibinfo {title} {{Topological Insulators in
  Three Dimensions}},\ }}\href {\doibase 10.1103/PhysRevLett.98.106803}
  {\bibfield  {journal} {\bibinfo  {journal} {Phys. Rev. Lett.}\ }\textbf
  {\bibinfo {volume} {98}},\ \bibinfo {pages} {106803} (\bibinfo {year}
  {2007})}\BibitemShut {NoStop}%
\bibitem [{\citenamefont {Roy}(2009)}]{Roy2009}%
  \BibitemOpen
  \bibfield  {author} {\bibinfo {author} {\bibfnamefont {R.}~\bibnamefont
  {Roy}},\ }\bibfield  {title} {{\color{Gray}\small \bibinfo {title}
  {{Topological phases and the quantum spin Hall effect in three dimensions}},\
  }}\href {\doibase 10.1103/PhysRevB.79.195322} {\bibfield  {journal} {\bibinfo
   {journal} {Phys. Rev. B}\ }\textbf {\bibinfo {volume} {79}},\ \bibinfo
  {pages} {195322} (\bibinfo {year} {2009})}\BibitemShut {NoStop}%
\bibitem [{\citenamefont {K{\"o}nig}\ \emph {et~al.}(2007)\citenamefont
  {K{\"o}nig}, \citenamefont {Wiedmann}, \citenamefont {Br{\"u}ne},
  \citenamefont {Roth}, \citenamefont {Buhmann}, \citenamefont {Molenkamp},
  \citenamefont {Qi},\ and\ \citenamefont {Zhang}}]{Koenig2007}%
  \BibitemOpen
  \bibfield  {author} {\bibinfo {author} {\bibfnamefont {M.}~\bibnamefont
  {K{\"o}nig}}, \bibinfo {author} {\bibfnamefont {S.}~\bibnamefont {Wiedmann}},
  \bibinfo {author} {\bibfnamefont {C.}~\bibnamefont {Br{\"u}ne}}, \bibinfo
  {author} {\bibfnamefont {A.}~\bibnamefont {Roth}}, \bibinfo {author}
  {\bibfnamefont {H.}~\bibnamefont {Buhmann}}, \bibinfo {author} {\bibfnamefont
  {L.~W.}\ \bibnamefont {Molenkamp}}, \bibinfo {author} {\bibfnamefont {X.-L.}\
  \bibnamefont {Qi}}, \ and\ \bibinfo {author} {\bibfnamefont {S.-C.}\
  \bibnamefont {Zhang}},\ }\bibfield  {title} {{\color{Gray}\small \bibinfo
  {title} {{Quantum Spin Hall Insulator State in HgTe Quantum Wells}},\ }}\href
  {\doibase 10.1126/science.1148047} {\bibfield  {journal} {\bibinfo  {journal}
  {Science}\ }\textbf {\bibinfo {volume} {318}},\ \bibinfo {pages} {766}
  (\bibinfo {year} {2007})}\BibitemShut {NoStop}%
\bibitem [{\citenamefont {Hasan}\ and\ \citenamefont {Kane}(2010)}]{Hasan2010}%
  \BibitemOpen
  \bibfield  {author} {\bibinfo {author} {\bibfnamefont {M.~Z.}\ \bibnamefont
  {Hasan}}\ and\ \bibinfo {author} {\bibfnamefont {C.~L.}\ \bibnamefont
  {Kane}},\ }\bibfield  {title} {{\color{Gray}\small \bibinfo {title}
  {{Colloquium: Topological insulators}},\ }}\href {\doibase
  10.1103/RevModPhys.82.3045} {\bibfield  {journal} {\bibinfo  {journal} {Rev.
  Mod. Phys.}\ }\textbf {\bibinfo {volume} {82}},\ \bibinfo {pages} {3045}
  (\bibinfo {year} {2010})}\BibitemShut {NoStop}%
\bibitem [{\citenamefont {Qi}\ and\ \citenamefont {Zhang}(2011)}]{Qi2011}%
  \BibitemOpen
  \bibfield  {author} {\bibinfo {author} {\bibfnamefont {X.-L.}\ \bibnamefont
  {Qi}}\ and\ \bibinfo {author} {\bibfnamefont {S.-C.}\ \bibnamefont {Zhang}},\
  }\bibfield  {title} {{\color{Gray}\small \bibinfo {title} {Topological
  insulators and superconductors},\ }}\href {\doibase
  10.1103/RevModPhys.83.1057} {\bibfield  {journal} {\bibinfo  {journal} {Rev.
  Mod. Phys.}\ }\textbf {\bibinfo {volume} {83}},\ \bibinfo {pages} {1057}
  (\bibinfo {year} {2011})}\BibitemShut {NoStop}%
\bibitem [{\citenamefont {Kim}\ \emph {et~al.}(2015)\citenamefont {Kim},
  \citenamefont {Wieder}, \citenamefont {Kane},\ and\ \citenamefont
  {Rappe}}]{Kim2015}%
  \BibitemOpen
  \bibfield  {author} {\bibinfo {author} {\bibfnamefont {Y.}~\bibnamefont
  {Kim}}, \bibinfo {author} {\bibfnamefont {B.~J.}\ \bibnamefont {Wieder}},
  \bibinfo {author} {\bibfnamefont {C.~L.}\ \bibnamefont {Kane}}, \ and\
  \bibinfo {author} {\bibfnamefont {A.~M.}\ \bibnamefont {Rappe}},\ }\bibfield
  {title} {{\color{Gray}\small \bibinfo {title} {{Dirac Line Nodes in
  Inversion-Symmetric Crystals}},\ }}\href {\doibase
  10.1103/PhysRevLett.115.036806} {\bibfield  {journal} {\bibinfo  {journal}
  {Phys. Rev. Lett.}\ }\textbf {\bibinfo {volume} {115}},\ \bibinfo {pages}
  {036806} (\bibinfo {year} {2015})}\BibitemShut {NoStop}%
\bibitem [{\citenamefont {Fang}\ \emph {et~al.}(2015)\citenamefont {Fang},
  \citenamefont {Chen}, \citenamefont {Kee},\ and\ \citenamefont
  {Fu}}]{Fang2015}%
  \BibitemOpen
  \bibfield  {author} {\bibinfo {author} {\bibfnamefont {C.}~\bibnamefont
  {Fang}}, \bibinfo {author} {\bibfnamefont {Y.}~\bibnamefont {Chen}}, \bibinfo
  {author} {\bibfnamefont {H.-Y.}\ \bibnamefont {Kee}}, \ and\ \bibinfo
  {author} {\bibfnamefont {L.}~\bibnamefont {Fu}},\ }\bibfield  {title}
  {{\color{Gray}\small \bibinfo {title} {Topological nodal line semimetals with
  and without spin-orbital coupling},\ }}\href {\doibase
  10.1103/PhysRevB.92.081201} {\bibfield  {journal} {\bibinfo  {journal} {Phys.
  Rev. B}\ }\textbf {\bibinfo {volume} {92}},\ \bibinfo {pages} {081201}
  (\bibinfo {year} {2015})}\BibitemShut {NoStop}%
\bibitem [{\citenamefont {Chan}\ \emph {et~al.}(2016)\citenamefont {Chan},
  \citenamefont {Chiu}, \citenamefont {Chou},\ and\ \citenamefont
  {Schnyder}}]{Chan2016}%
  \BibitemOpen
  \bibfield  {author} {\bibinfo {author} {\bibfnamefont {Y.-H.}\ \bibnamefont
  {Chan}}, \bibinfo {author} {\bibfnamefont {C.-K.}\ \bibnamefont {Chiu}},
  \bibinfo {author} {\bibfnamefont {M.~Y.}\ \bibnamefont {Chou}}, \ and\
  \bibinfo {author} {\bibfnamefont {A.~P.}\ \bibnamefont {Schnyder}},\
  }\bibfield  {title} {{\color{Gray}\small \bibinfo {title}
  {{${\mathrm{Ca}}_{3}{\mathrm{P}}_{2}$ and other topological semimetals with
  line nodes and drumhead surface states}},\ }}\href {\doibase
  10.1103/PhysRevB.93.205132} {\bibfield  {journal} {\bibinfo  {journal} {Phys.
  Rev. B}\ }\textbf {\bibinfo {volume} {93}},\ \bibinfo {pages} {205132}
  (\bibinfo {year} {2016})}\BibitemShut {NoStop}%
\bibitem [{\citenamefont {Zhao}\ and\ \citenamefont {Wang}(2013)}]{Zhao2013}%
  \BibitemOpen
  \bibfield  {author} {\bibinfo {author} {\bibfnamefont {Y.~X.}\ \bibnamefont
  {Zhao}}\ and\ \bibinfo {author} {\bibfnamefont {Z.~D.}\ \bibnamefont
  {Wang}},\ }\bibfield  {title} {{\color{Gray}\small \bibinfo {title}
  {{Topological Classification and Stability of Fermi Surfaces}},\ }}\href
  {\doibase 10.1103/PhysRevLett.110.240404} {\bibfield  {journal} {\bibinfo
  {journal} {Phys. Rev. Lett.}\ }\textbf {\bibinfo {volume} {110}},\ \bibinfo
  {pages} {240404} (\bibinfo {year} {2013})}\BibitemShut {NoStop}%
\bibitem [{\citenamefont {Yamakage}\ \emph {et~al.}(2016)\citenamefont
  {Yamakage}, \citenamefont {Yamakawa}, \citenamefont {Tanaka},\ and\
  \citenamefont {Okamoto}}]{Yamakage2016}%
  \BibitemOpen
  \bibfield  {author} {\bibinfo {author} {\bibfnamefont {A.}~\bibnamefont
  {Yamakage}}, \bibinfo {author} {\bibfnamefont {Y.}~\bibnamefont {Yamakawa}},
  \bibinfo {author} {\bibfnamefont {Y.}~\bibnamefont {Tanaka}}, \ and\ \bibinfo
  {author} {\bibfnamefont {Y.}~\bibnamefont {Okamoto}},\ }\bibfield  {title}
  {{\color{Gray}\small \bibinfo {title} {{Line-Node Dirac Semimetal and
  Topological Insulating Phase in Noncentrosymmetric Pnictides CaAgX (X = P,
  As)}},\ }}\href {\doibase 10.7566/JPSJ.85.013708} {\bibfield  {journal}
  {\bibinfo  {journal} {Journal of the Physical Society of Japan}\ }\textbf
  {\bibinfo {volume} {85}},\ \bibinfo {pages} {013708} (\bibinfo {year}
  {2016})}\BibitemShut {NoStop}%
\bibitem [{\citenamefont {Yang}\ \emph {et~al.}(2017)\citenamefont {Yang},
  \citenamefont {Bojesen}, \citenamefont {Morimoto},\ and\ \citenamefont
  {Furusaki}}]{Yang2017}%
  \BibitemOpen
  \bibfield  {author} {\bibinfo {author} {\bibfnamefont {B.-J.}\ \bibnamefont
  {Yang}}, \bibinfo {author} {\bibfnamefont {T.~A.}\ \bibnamefont {Bojesen}},
  \bibinfo {author} {\bibfnamefont {T.}~\bibnamefont {Morimoto}}, \ and\
  \bibinfo {author} {\bibfnamefont {A.}~\bibnamefont {Furusaki}},\ }\bibfield
  {title} {{\color{Gray}\small \bibinfo {title} {Topological semimetals
  protected by off-centered symmetries in nonsymmorphic crystals},\ }}\href
  {\doibase 10.1103/PhysRevB.95.075135} {\bibfield  {journal} {\bibinfo
  {journal} {Phys. Rev. B}\ }\textbf {\bibinfo {volume} {95}},\ \bibinfo
  {pages} {075135} (\bibinfo {year} {2017})}\BibitemShut {NoStop}%
\bibitem [{\citenamefont {O'Brien}\ \emph {et~al.}(2016)\citenamefont
  {O'Brien}, \citenamefont {Hermanns},\ and\ \citenamefont
  {Trebst}}]{OBrien2016}%
  \BibitemOpen
  \bibfield  {author} {\bibinfo {author} {\bibfnamefont {K.}~\bibnamefont
  {O'Brien}}, \bibinfo {author} {\bibfnamefont {M.}~\bibnamefont {Hermanns}}, \
  and\ \bibinfo {author} {\bibfnamefont {S.}~\bibnamefont {Trebst}},\
  }\bibfield  {title} {{\color{Gray}\small \bibinfo {title} {{Classification of
  gapless ${\mathbb{Z}}_{2}$ spin liquids in three-dimensional Kitaev
  models}},\ }}\href {\doibase 10.1103/PhysRevB.93.085101} {\bibfield
  {journal} {\bibinfo  {journal} {Phys. Rev. B}\ }\textbf {\bibinfo {volume}
  {93}},\ \bibinfo {pages} {085101} (\bibinfo {year} {2016})}\BibitemShut
  {NoStop}%
\bibitem [{\citenamefont {Mullen}\ \emph {et~al.}(2015)\citenamefont {Mullen},
  \citenamefont {Uchoa},\ and\ \citenamefont {Glatzhofer}}]{Mullen2015}%
  \BibitemOpen
  \bibfield  {author} {\bibinfo {author} {\bibfnamefont {K.}~\bibnamefont
  {Mullen}}, \bibinfo {author} {\bibfnamefont {B.}~\bibnamefont {Uchoa}}, \
  and\ \bibinfo {author} {\bibfnamefont {D.~T.}\ \bibnamefont {Glatzhofer}},\
  }\bibfield  {title} {{\color{Gray}\small \bibinfo {title} {{Line of Dirac
  Nodes in Hyperhoneycomb Lattices}},\ }}\href {\doibase
  10.1103/PhysRevLett.115.026403} {\bibfield  {journal} {\bibinfo  {journal}
  {Phys. Rev. Lett.}\ }\textbf {\bibinfo {volume} {115}},\ \bibinfo {pages}
  {026403} (\bibinfo {year} {2015})}\BibitemShut {NoStop}%
\bibitem [{\citenamefont {Ezawa}(2016)}]{Ezawa2016}%
  \BibitemOpen
  \bibfield  {author} {\bibinfo {author} {\bibfnamefont {M.}~\bibnamefont
  {Ezawa}},\ }\bibfield  {title} {{\color{Gray}\small \bibinfo {title}
  {{Loop-Nodal and Point-Nodal Semimetals in Three-Dimensional Honeycomb
  Lattices}},\ }}\href {\doibase 10.1103/PhysRevLett.116.127202} {\bibfield
  {journal} {\bibinfo  {journal} {Phys. Rev. Lett.}\ }\textbf {\bibinfo
  {volume} {116}},\ \bibinfo {pages} {127202} (\bibinfo {year}
  {2016})}\BibitemShut {NoStop}%
\bibitem [{\citenamefont {Tsuchiizu}(2016)}]{Tsuchiizu2016}%
  \BibitemOpen
  \bibfield  {author} {\bibinfo {author} {\bibfnamefont {M.}~\bibnamefont
  {Tsuchiizu}},\ }\bibfield  {title} {{\color{Gray}\small \bibinfo {title}
  {{Three-dimensional higher-spin Dirac and Weyl dispersions in the strongly
  isotropic ${K}_{4}$ crystal}},\ }}\href {\doibase 10.1103/PhysRevB.94.195426}
  {\bibfield  {journal} {\bibinfo  {journal} {Phys. Rev. B}\ }\textbf {\bibinfo
  {volume} {94}},\ \bibinfo {pages} {195426} (\bibinfo {year}
  {2016})}\BibitemShut {NoStop}%
\bibitem [{\citenamefont {Yamada}\ \emph
  {et~al.}(2017{\natexlab{a}})\citenamefont {Yamada}, \citenamefont {Dwivedi},\
  and\ \citenamefont {Hermanns}}]{Yamada2017b}%
  \BibitemOpen
  \bibfield  {author} {\bibinfo {author} {\bibfnamefont {M.~G.}\ \bibnamefont
  {Yamada}}, \bibinfo {author} {\bibfnamefont {V.}~\bibnamefont {Dwivedi}}, \
  and\ \bibinfo {author} {\bibfnamefont {M.}~\bibnamefont {Hermanns}},\
  }\bibfield  {title} {{\color{Gray}\small \bibinfo {title} {{Crystalline
  Kitaev spin liquids}},\ }}\href {\doibase 10.1103/PhysRevB.96.155107}
  {\bibfield  {journal} {\bibinfo  {journal} {Phys. Rev. B}\ }\textbf {\bibinfo
  {volume} {96}},\ \bibinfo {pages} {155107} (\bibinfo {year}
  {2017}{\natexlab{a}})}\BibitemShut {NoStop}%
\bibitem [{\citenamefont {Weng}\ \emph {et~al.}(2015)\citenamefont {Weng},
  \citenamefont {Liang}, \citenamefont {Xu}, \citenamefont {Yu}, \citenamefont
  {Fang}, \citenamefont {Dai},\ and\ \citenamefont {Kawazoe}}]{Weng2015}%
  \BibitemOpen
  \bibfield  {author} {\bibinfo {author} {\bibfnamefont {H.}~\bibnamefont
  {Weng}}, \bibinfo {author} {\bibfnamefont {Y.}~\bibnamefont {Liang}},
  \bibinfo {author} {\bibfnamefont {Q.}~\bibnamefont {Xu}}, \bibinfo {author}
  {\bibfnamefont {R.}~\bibnamefont {Yu}}, \bibinfo {author} {\bibfnamefont
  {Z.}~\bibnamefont {Fang}}, \bibinfo {author} {\bibfnamefont {X.}~\bibnamefont
  {Dai}}, \ and\ \bibinfo {author} {\bibfnamefont {Y.}~\bibnamefont
  {Kawazoe}},\ }\bibfield  {title} {{\color{Gray}\small \bibinfo {title}
  {Topological node-line semimetal in three-dimensional graphene networks},\
  }}\href {\doibase 10.1103/PhysRevB.92.045108} {\bibfield  {journal} {\bibinfo
   {journal} {Phys. Rev. B}\ }\textbf {\bibinfo {volume} {92}},\ \bibinfo
  {pages} {045108} (\bibinfo {year} {2015})}\BibitemShut {NoStop}%
\bibitem [{\citenamefont {Takayama}\ \emph {et~al.}(2015)\citenamefont
  {Takayama}, \citenamefont {Kato}, \citenamefont {Dinnebier}, \citenamefont
  {Nuss}, \citenamefont {Kono}, \citenamefont {Veiga}, \citenamefont {Fabbris},
  \citenamefont {Haskel},\ and\ \citenamefont {Takagi}}]{Takayama2015}%
  \BibitemOpen
  \bibfield  {author} {\bibinfo {author} {\bibfnamefont {T.}~\bibnamefont
  {Takayama}}, \bibinfo {author} {\bibfnamefont {A.}~\bibnamefont {Kato}},
  \bibinfo {author} {\bibfnamefont {R.}~\bibnamefont {Dinnebier}}, \bibinfo
  {author} {\bibfnamefont {J.}~\bibnamefont {Nuss}}, \bibinfo {author}
  {\bibfnamefont {H.}~\bibnamefont {Kono}}, \bibinfo {author} {\bibfnamefont
  {L.~S.~I.}\ \bibnamefont {Veiga}}, \bibinfo {author} {\bibfnamefont
  {G.}~\bibnamefont {Fabbris}}, \bibinfo {author} {\bibfnamefont
  {D.}~\bibnamefont {Haskel}}, \ and\ \bibinfo {author} {\bibfnamefont
  {H.}~\bibnamefont {Takagi}},\ }\bibfield  {title} {{\color{Gray}\small
  \bibinfo {title} {{Hyperhoneycomb Iridate
  $\ensuremath{\beta}\text{\ensuremath{-}}{\mathrm{Li}}_{2}{\mathrm{IrO}}_{3}$
  as a Platform for Kitaev Magnetism}},\ }}\href {\doibase
  10.1103/PhysRevLett.114.077202} {\bibfield  {journal} {\bibinfo  {journal}
  {Phys. Rev. Lett.}\ }\textbf {\bibinfo {volume} {114}},\ \bibinfo {pages}
  {077202} (\bibinfo {year} {2015})}\BibitemShut {NoStop}%
\bibitem [{\citenamefont {Modic}\ \emph {et~al.}(2014)\citenamefont {Modic},
  \citenamefont {Smidt}, \citenamefont {Kimchi}, \citenamefont {Breznay},
  \citenamefont {Biffin}, \citenamefont {Choi}, \citenamefont {Johnson},
  \citenamefont {Coldea}, \citenamefont {Watkins-Curry}, \citenamefont
  {McCandless}, \citenamefont {Chan}, \citenamefont {Gandara}, \citenamefont
  {Islam}, \citenamefont {Vishwanath}, \citenamefont {Shekhter}, \citenamefont
  {McDonald},\ and\ \citenamefont {Analytis}}]{Modic2014}%
  \BibitemOpen
  \bibfield  {author} {\bibinfo {author} {\bibfnamefont {K.~A.}\ \bibnamefont
  {Modic}}, \bibinfo {author} {\bibfnamefont {T.~E.}\ \bibnamefont {Smidt}},
  \bibinfo {author} {\bibfnamefont {I.}~\bibnamefont {Kimchi}}, \bibinfo
  {author} {\bibfnamefont {N.~P.}\ \bibnamefont {Breznay}}, \bibinfo {author}
  {\bibfnamefont {A.}~\bibnamefont {Biffin}}, \bibinfo {author} {\bibfnamefont
  {S.}~\bibnamefont {Choi}}, \bibinfo {author} {\bibfnamefont {R.~D.}\
  \bibnamefont {Johnson}}, \bibinfo {author} {\bibfnamefont {R.}~\bibnamefont
  {Coldea}}, \bibinfo {author} {\bibfnamefont {P.}~\bibnamefont
  {Watkins-Curry}}, \bibinfo {author} {\bibfnamefont {G.~T.}\ \bibnamefont
  {McCandless}}, \bibinfo {author} {\bibfnamefont {J.~Y.}\ \bibnamefont
  {Chan}}, \bibinfo {author} {\bibfnamefont {F.}~\bibnamefont {Gandara}},
  \bibinfo {author} {\bibfnamefont {Z.}~\bibnamefont {Islam}}, \bibinfo
  {author} {\bibfnamefont {A.}~\bibnamefont {Vishwanath}}, \bibinfo {author}
  {\bibfnamefont {A.}~\bibnamefont {Shekhter}}, \bibinfo {author}
  {\bibfnamefont {R.~D.}\ \bibnamefont {McDonald}}, \ and\ \bibinfo {author}
  {\bibfnamefont {J.~G.}\ \bibnamefont {Analytis}},\ }\bibfield  {title}
  {{\color{Gray}\small \bibinfo {title} {{Realization of a three-dimensional
  spin--anisotropic harmonic honeycomb iridate}},\ }}\href
  {http://dx.doi.org/10.1038/ncomms5203} {\bibfield  {journal} {\bibinfo
  {journal} {Nature Communications}\ }\textbf {\bibinfo {volume} {5}},\
  \bibinfo {pages} {4203} (\bibinfo {year} {2014})}\BibitemShut {NoStop}%
\bibitem [{\citenamefont {Yamada}\ \emph
  {et~al.}(2017{\natexlab{b}})\citenamefont {Yamada}, \citenamefont {Fujita},\
  and\ \citenamefont {Oshikawa}}]{Yamada2017a}%
  \BibitemOpen
  \bibfield  {author} {\bibinfo {author} {\bibfnamefont {M.~G.}\ \bibnamefont
  {Yamada}}, \bibinfo {author} {\bibfnamefont {H.}~\bibnamefont {Fujita}}, \
  and\ \bibinfo {author} {\bibfnamefont {M.}~\bibnamefont {Oshikawa}},\
  }\bibfield  {title} {{\color{Gray}\small \bibinfo {title} {{Designing Kitaev
  Spin Liquids in Metal-Organic Frameworks}},\ }}\href {\doibase
  10.1103/PhysRevLett.119.057202} {\bibfield  {journal} {\bibinfo  {journal}
  {Phys. Rev. Lett.}\ }\textbf {\bibinfo {volume} {119}},\ \bibinfo {pages}
  {057202} (\bibinfo {year} {2017}{\natexlab{b}})}\BibitemShut {NoStop}%
\bibitem [{\citenamefont {Wells}(1977)}]{Wells1977}%
  \BibitemOpen
  \bibfield  {author} {\bibinfo {author} {\bibfnamefont {A.~F.}\ \bibnamefont
  {Wells}},\ }\href@noop {} {\emph {\bibinfo {title} {{Three-Dimensional Nets
  and Polyhedra}}}}\ (\bibinfo  {publisher} {Wiley, New York},\ \bibinfo {year}
  {1977})\BibitemShut {NoStop}%
\bibitem [{\citenamefont {Hermanns}\ and\ \citenamefont
  {Trebst}(2014)}]{Hermanns2014}%
  \BibitemOpen
  \bibfield  {author} {\bibinfo {author} {\bibfnamefont {M.}~\bibnamefont
  {Hermanns}}\ and\ \bibinfo {author} {\bibfnamefont {S.}~\bibnamefont
  {Trebst}},\ }\bibfield  {title} {{\color{Gray}\small \bibinfo {title}
  {{Quantum spin liquid with a Majorana Fermi surface on the three-dimensional
  hyperoctagon lattice}},\ }}\href {\doibase 10.1103/PhysRevB.89.235102}
  {\bibfield  {journal} {\bibinfo  {journal} {Phys. Rev. B}\ }\textbf {\bibinfo
  {volume} {89}},\ \bibinfo {pages} {235102} (\bibinfo {year}
  {2014})}\BibitemShut {NoStop}%
\bibitem [{\citenamefont {Heesch}\ and\ \citenamefont
  {Laves}(1993)}]{Laves1993}%
  \BibitemOpen
  \bibfield  {author} {\bibinfo {author} {\bibfnamefont {H.}~\bibnamefont
  {Heesch}}\ and\ \bibinfo {author} {\bibfnamefont {F.}~\bibnamefont {Laves}},\
  }\bibfield  {title} {{\color{Gray}\small \bibinfo {title} {{\"Uber d\"unne
  Kugelpackungen}},\ }}\href@noop {} {\bibfield  {journal} {\bibinfo  {journal}
  {Z. Kristallogr.}\ }\textbf {\bibinfo {volume} {85}},\ \bibinfo {pages} {443}
  (\bibinfo {year} {1993})}\BibitemShut {NoStop}%
\bibitem [{\citenamefont {Sunada}(2008)}]{K4}%
  \BibitemOpen
  \bibfield  {author} {\bibinfo {author} {\bibfnamefont {T.}~\bibnamefont
  {Sunada}},\ }\bibfield  {title} {{\color{Gray}\small \bibinfo {title}
  {{Crystals That Nature Might Miss Creating}},\ }}\href@noop {} {\bibfield
  {journal} {\bibinfo  {journal} {Notices of the AMS}\ }\textbf {\bibinfo
  {volume} {55}},\ \bibinfo {pages} {208} (\bibinfo {year} {2008})}\BibitemShut
  {NoStop}%
\bibitem [{Note1()}]{Note1}%
  \BibitemOpen
  \bibinfo {note} {Note that for the lattices (10,3)a, (8,3)a, and (8,3)b
  sublattice symmetry comes with a non-zero momentum $\protect \mathaccentV
  {vec}17E{k}_0$ and therefore the product $PS$ of time-reversal and sublattice
  symmetry is {\protect \em not} momentum invariant. It is for this reason that
  these lattices do not exhibit nodal lines.}\BibitemShut {Stop}%
\bibitem [{\citenamefont {Altland}\ and\ \citenamefont
  {Zirnbauer}(1997)}]{Altland1997}%
  \BibitemOpen
  \bibfield  {author} {\bibinfo {author} {\bibfnamefont {A.}~\bibnamefont
  {Altland}}\ and\ \bibinfo {author} {\bibfnamefont {M.~R.}\ \bibnamefont
  {Zirnbauer}},\ }\bibfield  {title} {{\color{Gray}\small \bibinfo {title}
  {{Nonstandard symmetry classes in mesoscopic normal-superconducting hybrid
  structures}},\ }}\href {\doibase 10.1103/PhysRevB.55.1142} {\bibfield
  {journal} {\bibinfo  {journal} {Phys. Rev. B}\ }\textbf {\bibinfo {volume}
  {55}},\ \bibinfo {pages} {1142} (\bibinfo {year} {1997})}\BibitemShut
  {NoStop}%
\bibitem [{\citenamefont {Li}\ and\ \citenamefont
  {Andrei}(2007)}]{Li2007Observation}%
  \BibitemOpen
  \bibfield  {author} {\bibinfo {author} {\bibfnamefont {G.}~\bibnamefont
  {Li}}\ and\ \bibinfo {author} {\bibfnamefont {E.~Y.}\ \bibnamefont
  {Andrei}},\ }\bibfield  {title} {{\color{Gray}\small \bibinfo {title}
  {{Observation of Landau levels of Dirac fermions in graphite}},\ }}\href@noop
  {} {\bibfield  {journal} {\bibinfo  {journal} {Nature Physics}\ }\textbf
  {\bibinfo {volume} {3}},\ \bibinfo {pages} {623} (\bibinfo {year}
  {2007})}\BibitemShut {NoStop}%
\bibitem [{Note2()}]{Note2}%
  \BibitemOpen
  \bibinfo {note} {We implement the magnetic field by assigning the phase $\phi
  =\pm n\pi \phi _F/\phi _{tot}$ to the $z$-bonds (see also the next footnote),
  with $n\protect \mathaccentV {vec}17Ea_y$ denoting the $y$-position of the
  unit cell and $\phi _{tot}$ denoting the total flux. This assigns a flux
  $2\phi $ to all the plaquettes that are penetrated by the magnetic field.
  Since we want to describe arbitrary flux through the plaquette, we use open
  boundary conditions in the $y$-direction. Note that $k_y$ is no longer a good
  quantum number.}\BibitemShut {Stop}%
\bibitem [{Note3()}]{Note3}%
  \BibitemOpen
  \bibinfo {note} {To facilitate our analysis we work with the most symmetric
  representation of the (10,3)b hyperhoneycomb lattice. In this representation
  the elementary zig-zag chains constituting the lattice, see Fig.~\ref
  {Fig:lattice_illustration}, are rotated by 90$^\circ $ with respect to each
  other. We further double the unit cell, see Appendix~\ref {App:Lattices},
  which allows us to choose orthogonal lattice translation vectors (as compared
  with the minimal representation of the lattice). With these conventions the
  nodal line lies in the $(k_x,k_y)$-plane, see also Fig.~\ref
  {Fig:10b-rectangular} in Appendix~\ref {App:Lattices}.}\BibitemShut {Stop}%
\bibitem [{\citenamefont {McCann}\ and\ \citenamefont
  {Fal'ko}(2006)}]{McCann2006}%
  \BibitemOpen
  \bibfield  {author} {\bibinfo {author} {\bibfnamefont {E.}~\bibnamefont
  {McCann}}\ and\ \bibinfo {author} {\bibfnamefont {V.~I.}\ \bibnamefont
  {Fal'ko}},\ }\bibfield  {title} {{\color{Gray}\small \bibinfo {title}
  {{Landau-Level Degeneracy and Quantum Hall Effect in a Graphite Bilayer}},\
  }}\href {\doibase 10.1103/PhysRevLett.96.086805} {\bibfield  {journal}
  {\bibinfo  {journal} {Phys. Rev. Lett.}\ }\textbf {\bibinfo {volume} {96}},\
  \bibinfo {pages} {086805} (\bibinfo {year} {2006})}\BibitemShut {NoStop}%
\bibitem [{\citenamefont {Nielsen}\ and\ \citenamefont
  {Ninomiya}(1981{\natexlab{a}})}]{Nielsen1981a}%
  \BibitemOpen
  \bibfield  {author} {\bibinfo {author} {\bibfnamefont {H.}~\bibnamefont
  {Nielsen}}\ and\ \bibinfo {author} {\bibfnamefont {M.}~\bibnamefont
  {Ninomiya}},\ }\bibfield  {title} {{\color{Gray}\small \bibinfo {title}
  {{Absence of neutrinos on a lattice: (II). Intuitive topological proof}},\
  }}\href {\doibase https://doi.org/10.1016/0550-3213(81)90524-1} {\bibfield
  {journal} {\bibinfo  {journal} {Nuclear Physics B}\ }\textbf {\bibinfo
  {volume} {193}},\ \bibinfo {pages} {173 } (\bibinfo {year}
  {1981}{\natexlab{a}})}\BibitemShut {NoStop}%
\bibitem [{\citenamefont {Nielsen}\ and\ \citenamefont
  {Ninomiya}(1981{\natexlab{b}})}]{Nielsen1981b}%
  \BibitemOpen
  \bibfield  {author} {\bibinfo {author} {\bibfnamefont {H.}~\bibnamefont
  {Nielsen}}\ and\ \bibinfo {author} {\bibfnamefont {M.}~\bibnamefont
  {Ninomiya}},\ }\bibfield  {title} {{\color{Gray}\small \bibinfo {title}
  {{Absence of neutrinos on a lattice: (I). Proof by homotopy theory}},\
  }}\href {\doibase https://doi.org/10.1016/0550-3213(81)90361-8} {\bibfield
  {journal} {\bibinfo  {journal} {Nuclear Physics B}\ }\textbf {\bibinfo
  {volume} {185}},\ \bibinfo {pages} {20 } (\bibinfo {year}
  {1981}{\natexlab{b}})}\BibitemShut {NoStop}%
\bibitem [{\citenamefont {Bradlyn}\ \emph {et~al.}(2016)\citenamefont
  {Bradlyn}, \citenamefont {Cano}, \citenamefont {Wang}, \citenamefont
  {Vergniory}, \citenamefont {Felser}, \citenamefont {Cava},\ and\
  \citenamefont {Bernevig}}]{Bradlyn2016}%
  \BibitemOpen
  \bibfield  {author} {\bibinfo {author} {\bibfnamefont {B.}~\bibnamefont
  {Bradlyn}}, \bibinfo {author} {\bibfnamefont {J.}~\bibnamefont {Cano}},
  \bibinfo {author} {\bibfnamefont {Z.}~\bibnamefont {Wang}}, \bibinfo {author}
  {\bibfnamefont {M.~G.}\ \bibnamefont {Vergniory}}, \bibinfo {author}
  {\bibfnamefont {C.}~\bibnamefont {Felser}}, \bibinfo {author} {\bibfnamefont
  {R.~J.}\ \bibnamefont {Cava}}, \ and\ \bibinfo {author} {\bibfnamefont
  {B.~A.}\ \bibnamefont {Bernevig}},\ }\bibfield  {title} {{\color{Gray}\small
  \bibinfo {title} {{Beyond Dirac and Weyl fermions: Unconventional
  quasiparticles in conventional crystals}},\ }}\href
  {http://science.sciencemag.org/content/353/6299/aaf5037} {\bibfield
  {journal} {\bibinfo  {journal} {Science}\ }\textbf {\bibinfo {volume}
  {353}},\ \bibinfo {pages} {5037} (\bibinfo {year} {2016})}\BibitemShut
  {NoStop}%
\bibitem [{\citenamefont {Watanabe}\ \emph {et~al.}(2011)\citenamefont
  {Watanabe}, \citenamefont {Hatsugai},\ and\ \citenamefont
  {Aoki}}]{Watanabe2011}%
  \BibitemOpen
  \bibfield  {author} {\bibinfo {author} {\bibfnamefont {H.}~\bibnamefont
  {Watanabe}}, \bibinfo {author} {\bibfnamefont {Y.}~\bibnamefont {Hatsugai}},
  \ and\ \bibinfo {author} {\bibfnamefont {H.}~\bibnamefont {Aoki}},\
  }\bibfield  {title} {{\color{Gray}\small \bibinfo {title} {{Manipulation of
  the Dirac cones and the anomaly in the graphene related quantum Hall
  effect}},\ }}\href {http://stacks.iop.org/1742-6596/334/i=1/a=012044}
  {\bibfield  {journal} {\bibinfo  {journal} {Journal of Physics: Conference
  Series}\ }\textbf {\bibinfo {volume} {334}},\ \bibinfo {pages} {012044}
  (\bibinfo {year} {2011})}\BibitemShut {NoStop}%
\bibitem [{\citenamefont {Malcolm}\ and\ \citenamefont
  {Nicol}(2014)}]{Malcolm2014}%
  \BibitemOpen
  \bibfield  {author} {\bibinfo {author} {\bibfnamefont {J.~D.}\ \bibnamefont
  {Malcolm}}\ and\ \bibinfo {author} {\bibfnamefont {E.~J.}\ \bibnamefont
  {Nicol}},\ }\bibfield  {title} {{\color{Gray}\small \bibinfo {title}
  {{Magneto-optics of general pseudospin-$s$ two-dimensional Dirac-Weyl
  fermions}},\ }}\href {\doibase 10.1103/PhysRevB.90.035405} {\bibfield
  {journal} {\bibinfo  {journal} {Phys. Rev. B}\ }\textbf {\bibinfo {volume}
  {90}},\ \bibinfo {pages} {035405} (\bibinfo {year} {2014})}\BibitemShut
  {NoStop}%
\bibitem [{\citenamefont {Orlita}\ \emph {et~al.}(2014)\citenamefont {Orlita},
  \citenamefont {Basko}, \citenamefont {Zholudev}, \citenamefont {Teppe},
  \citenamefont {Knap}, \citenamefont {Gavrilenko}, \citenamefont {Mikhailov},
  \citenamefont {Dvoretskii}, \citenamefont {Neugebauer}, \citenamefont
  {Faugeras}, \citenamefont {Barra}, \citenamefont {Martinez},\ and\
  \citenamefont {Potemski}}]{Orlita2014}%
  \BibitemOpen
  \bibfield  {author} {\bibinfo {author} {\bibfnamefont {M.}~\bibnamefont
  {Orlita}}, \bibinfo {author} {\bibfnamefont {D.~M.}\ \bibnamefont {Basko}},
  \bibinfo {author} {\bibfnamefont {M.~S.}\ \bibnamefont {Zholudev}}, \bibinfo
  {author} {\bibfnamefont {F.}~\bibnamefont {Teppe}}, \bibinfo {author}
  {\bibfnamefont {W.}~\bibnamefont {Knap}}, \bibinfo {author} {\bibfnamefont
  {V.~I.}\ \bibnamefont {Gavrilenko}}, \bibinfo {author} {\bibfnamefont
  {N.~N.}\ \bibnamefont {Mikhailov}}, \bibinfo {author} {\bibfnamefont {S.~A.}\
  \bibnamefont {Dvoretskii}}, \bibinfo {author} {\bibfnamefont
  {P.}~\bibnamefont {Neugebauer}}, \bibinfo {author} {\bibfnamefont
  {C.}~\bibnamefont {Faugeras}}, \bibinfo {author} {\bibfnamefont {A.-L.}\
  \bibnamefont {Barra}}, \bibinfo {author} {\bibfnamefont {G.}~\bibnamefont
  {Martinez}}, \ and\ \bibinfo {author} {\bibfnamefont {M.}~\bibnamefont
  {Potemski}},\ }\bibfield  {title} {{\color{Gray}\small \bibinfo {title}
  {{Observation of three-dimensional massless Kane fermions in a zinc-blende
  crystal}},\ }}\href {http://dx.doi.org/10.1038/nphys2857} {\bibfield
  {journal} {\bibinfo  {journal} {Nature Physics}\ }\textbf {\bibinfo {volume}
  {10}},\ \bibinfo {pages} {233 EP } (\bibinfo {year} {2014})}\BibitemShut
  {NoStop}%
\bibitem [{\citenamefont {Fulga}\ and\ \citenamefont
  {Stern}(2017)}]{Fulga2017}%
  \BibitemOpen
  \bibfield  {author} {\bibinfo {author} {\bibfnamefont {I.~C.}\ \bibnamefont
  {Fulga}}\ and\ \bibinfo {author} {\bibfnamefont {A.}~\bibnamefont {Stern}},\
  }\bibfield  {title} {{\color{Gray}\small \bibinfo {title} {Triple point
  fermions in a minimal symmorphic model},\ }}\href {\doibase
  10.1103/PhysRevB.95.241116} {\bibfield  {journal} {\bibinfo  {journal} {Phys.
  Rev. B}\ }\textbf {\bibinfo {volume} {95}},\ \bibinfo {pages} {241116}
  (\bibinfo {year} {2017})}\BibitemShut {NoStop}%
\bibitem [{\citenamefont {Fulde}\ and\ \citenamefont
  {Ferrell}(1964)}]{Fulde1964}%
  \BibitemOpen
  \bibfield  {author} {\bibinfo {author} {\bibfnamefont {P.}~\bibnamefont
  {Fulde}}\ and\ \bibinfo {author} {\bibfnamefont {R.~A.}\ \bibnamefont
  {Ferrell}},\ }\bibfield  {title} {{\color{Gray}\small \bibinfo {title}
  {{Superconductivity in a Strong Spin-Exchange Field}},\ }}\href {\doibase
  10.1103/PhysRev.135.A550} {\bibfield  {journal} {\bibinfo  {journal} {Phys.
  Rev.}\ }\textbf {\bibinfo {volume} {135}},\ \bibinfo {pages} {A550} (\bibinfo
  {year} {1964})}\BibitemShut {NoStop}%
\bibitem [{\citenamefont {Larkin}\ and\ \citenamefont
  {Ovchinnikov}(1965)}]{Larkin1964}%
  \BibitemOpen
  \bibfield  {author} {\bibinfo {author} {\bibfnamefont {A.~I.}\ \bibnamefont
  {Larkin}}\ and\ \bibinfo {author} {\bibfnamefont {Y.~N.}\ \bibnamefont
  {Ovchinnikov}},\ }\bibfield  {title} {{\color{Gray}\small \bibinfo {title}
  {Nonuniform state of superconductors},\ }}\href@noop {} {\bibfield  {journal}
  {\bibinfo  {journal} {Sov. Phys. JETP}\ }\textbf {\bibinfo {volume} {20}},\
  \bibinfo {pages} {762} (\bibinfo {year} {1965})}\BibitemShut {NoStop}%
\bibitem [{\citenamefont {Hermanns}\ \emph
  {et~al.}(2015{\natexlab{a}})\citenamefont {Hermanns}, \citenamefont
  {Trebst},\ and\ \citenamefont {Rosch}}]{Rosch2015}%
  \BibitemOpen
  \bibfield  {author} {\bibinfo {author} {\bibfnamefont {M.}~\bibnamefont
  {Hermanns}}, \bibinfo {author} {\bibfnamefont {S.}~\bibnamefont {Trebst}}, \
  and\ \bibinfo {author} {\bibfnamefont {A.}~\bibnamefont {Rosch}},\ }\bibfield
   {title} {{\color{Gray}\small \bibinfo {title} {{Spin-Peierls Instability of
  Three-Dimensional Spin Liquids with Majorana Fermi Surfaces}},\ }}\href
  {\doibase 10.1103/PhysRevLett.115.177205} {\bibfield  {journal} {\bibinfo
  {journal} {Phys. Rev. Lett.}\ }\textbf {\bibinfo {volume} {115}},\ \bibinfo
  {pages} {177205} (\bibinfo {year} {2015}{\natexlab{a}})}\BibitemShut
  {NoStop}%
\bibitem [{\citenamefont {Hermanns}\ \emph
  {et~al.}(2015{\natexlab{b}})\citenamefont {Hermanns}, \citenamefont
  {O'Brien},\ and\ \citenamefont {Trebst}}]{Hermanns2015}%
  \BibitemOpen
  \bibfield  {author} {\bibinfo {author} {\bibfnamefont {M.}~\bibnamefont
  {Hermanns}}, \bibinfo {author} {\bibfnamefont {K.}~\bibnamefont {O'Brien}}, \
  and\ \bibinfo {author} {\bibfnamefont {S.}~\bibnamefont {Trebst}},\
  }\bibfield  {title} {{\color{Gray}\small \bibinfo {title} {{Weyl Spin
  Liquids}},\ }}\href {\doibase 10.1103/PhysRevLett.114.157202} {\bibfield
  {journal} {\bibinfo  {journal} {Phys. Rev. Lett.}\ }\textbf {\bibinfo
  {volume} {114}},\ \bibinfo {pages} {157202} (\bibinfo {year}
  {2015}{\natexlab{b}})}\BibitemShut {NoStop}%
\bibitem [{\citenamefont {Schnyder}\ \emph {et~al.}(2012)\citenamefont
  {Schnyder}, \citenamefont {Brydon},\ and\ \citenamefont
  {Timm}}]{Schnyder2012}%
  \BibitemOpen
  \bibfield  {author} {\bibinfo {author} {\bibfnamefont {A.~P.}\ \bibnamefont
  {Schnyder}}, \bibinfo {author} {\bibfnamefont {P.~M.~R.}\ \bibnamefont
  {Brydon}}, \ and\ \bibinfo {author} {\bibfnamefont {C.}~\bibnamefont
  {Timm}},\ }\bibfield  {title} {{\color{Gray}\small \bibinfo {title} {Types of
  topological surface states in nodal noncentrosymmetric superconductors},\
  }}\href {\doibase 10.1103/PhysRevB.85.024522} {\bibfield  {journal} {\bibinfo
   {journal} {Phys. Rev. B}\ }\textbf {\bibinfo {volume} {85}},\ \bibinfo
  {pages} {024522} (\bibinfo {year} {2012})}\BibitemShut {NoStop}%
\bibitem [{\citenamefont {Mook}\ \emph {et~al.}(2017)\citenamefont {Mook},
  \citenamefont {Henk},\ and\ \citenamefont {Mertig}}]{Mook2017}%
  \BibitemOpen
  \bibfield  {author} {\bibinfo {author} {\bibfnamefont {A.}~\bibnamefont
  {Mook}}, \bibinfo {author} {\bibfnamefont {J.}~\bibnamefont {Henk}}, \ and\
  \bibinfo {author} {\bibfnamefont {I.}~\bibnamefont {Mertig}},\ }\bibfield
  {title} {{\color{Gray}\small \bibinfo {title} {Magnon nodal-line semimetals
  and drumhead surface states in anisotropic pyrochlore ferromagnets},\ }}\href
  {\doibase 10.1103/PhysRevB.95.014418} {\bibfield  {journal} {\bibinfo
  {journal} {Phys. Rev. B}\ }\textbf {\bibinfo {volume} {95}},\ \bibinfo
  {pages} {014418} (\bibinfo {year} {2017})}\BibitemShut {NoStop}%
\bibitem [{\citenamefont {Trebst}()}]{Trebst2017}%
  \BibitemOpen
  \bibfield  {author} {\bibinfo {author} {\bibfnamefont {S.}~\bibnamefont
  {Trebst}},\ }\bibfield  {title} {{\color{Gray}\small \bibinfo {title}
  {{Kitaev Materials}},\ }}\href@noop {} {\ }\Eprint
  {http://arxiv.org/abs/arXiv:1701.07056} {arXiv:1701.07056} \BibitemShut
  {NoStop}%
\bibitem [{Note4()}]{Note4}%
  \BibitemOpen
  \bibinfo {note} {Note that the (8,3)a lattice shown in Fig.~\ref
  {Fig:lattice_illustration} has counter-clockwise rotating
  spirals.}\BibitemShut {Stop}%
\end{thebibliography}%

\end{document}